\documentclass{article}
\usepackage[T1]{fontenc}
\usepackage[a4paper,top=3cm,bottom=3cm,left=3.5cm,right=3.5cm]{geometry}
\usepackage{graphicx} 
\usepackage{amsmath,amssymb,amsfonts}  
\usepackage[colorlinks=true, citecolor=blue, urlcolor=blue, linkcolor=blue]{hyperref}
\usepackage{soul}
\usepackage{natbib}
\usepackage{comment}
\usepackage{longtable}
\usepackage{adjustbox}  
\usepackage{booktabs}
\usepackage{soul}

\usepackage[nameinlink]{cleveref}
\crefname{figure}{figure}{figures}
\Crefname{figure}{Figure}{Figures}
\crefrangelabelformat{figure}{#3#1#4--#5#2#6}  

\crefname{table}{table}{tables}
\Crefname{table}{Table}{Tables}
\crefrangelabelformat{table}{#3#1#4--#5#2#6}

\crefname{section}{Section}{Sections}
\Crefname{section}{Section}{Sections}

\crefname{subsection}{Subsection}{Subsection}
\Crefname{subsection}{Subsection}{Subsections}

\newcommand{\aref}[1]{\hyperref[#1]{Appendix~\ref{#1}}} 
\usepackage{chngcntr}  
\usepackage{float}
\usepackage{caption}
\usepackage{subfigure}

\usepackage{eso-pic}
\usepackage{fancyhdr}

\fancypagestyle{plain}{%
    \fancyhf{}

    \fancyfoot[C]{\thepage}
}

\definecolor{asparagus}{rgb}{0.55, 0.71, 0.0}

\title{VOLatility Archive for Realized Estimates \\(VOLARE)}

\author{Fabrizio Cipollini\textsuperscript{a},
        Giulia Cruciani\textsuperscript{b},
        Giampiero M. Gallo\textsuperscript{c, e},\\
        Alessandra Insana\textsuperscript{b},
        Edoardo Otranto\textsuperscript{d, e},
        Fabio Spagnolo\textsuperscript{b}}

\date{}

\begin{document}

\maketitle

\renewcommand{\thefootnote}{\alph{footnote}}
\footnotetext[1]{DiSIA, University of Florence, Italy}
\footnotetext[2]{Department of Economics, University of Messina, Italy}
\footnotetext[3]{NYU in Florence}
\footnotetext[4]{Department of Social and Economic Sciences, Sapienza University of Rome, Italy}
\footnotetext[5]{CRENoS}

\renewcommand{\thefootnote}{\arabic{footnote}}
\setcounter{footnote}{0}

\tableofcontents
\clearpage

\begin{abstract}
VOLARE (VOLatility Archive for Realized Estimates --  \url{https://volare.unime.it}) is an open research infrastructure providing standardized realized 
volatility and covariance measures constructed from ultra-high-frequency financial data. The platform processes tick-level observations across equities, exchange rates, and futures using an asset-specific pipeline that addresses heterogeneous trading calendars, microstructure noise, and timestamp precision. For equities, price series are cleaned using a documented outlier detection procedure and sampled at regular intervals. 

VOLARE delivers a comprehensive set of realized estimators, including realized variance, range-based measures, bipower variation, semivariances, realized quarticity, realized kernels, and multivariate covariance measures, ensuring methodological consistency and cross-asset comparability. In addition to bulk dataset download, the platform supports interactive 
visualization and real-time estimation of established volatility models such as HAR and MEM specifications. 

\end{abstract}


\section{Introduction}
Measuring and modelling market volatility is fundamental to financial econometrics, as it is essential for forecasting market movements, quantifying tail risks, and supporting portfolio allocation decisions. Understanding how volatility evolves over time and across assets provides essential insights into market dynamics, liquidity conditions, and systemic risk. Since the introduction of the Autoregressive Conditional Heteroskedasticity (ARCH) model by \citet {engle1982autoregressive} and its generalization (GARCH) by \citet{bollerslev1986generalized}, volatility modelling has been an essential component of empirical finance, forming the basis for derivative pricing, risk management, and financial stability analysis.

A major step forward in this field came with the development of realized volatility measures.  \citet{engle2000econometrics}, introduces the notion of \emph{ultra} high-frequency data (UHFD) as recording detailed information about market activity across financial instruments.

Introduced by \citet{andersen1998answering}, realized variance computes the ex-post variation in prices as the sum of squared intraday returns. Compared with daily GARCH-type models-which infer latent conditional variance from squared daily returns, realized measures offer a more accurate and less noisy assessment of market variability. Moreover, realized volatility retains key empirical properties such as volatility clustering and long memory \citep{andersen2003modeling}, while serving as a benchmark for evaluating the forecasting performance of econometric models.

Subsequent methodological advances have broadened the scope of realized measures, extending the original plain vanilla realized variance, allowing the incorporation of various features of the return series, such as microstructure noise, jump components, and asymmetries in price movements. In particular, the realized kernel \citep{barndorff2008designing} provides robust estimators under noisy conditions, while the realized semivariances \citep{barndorff2010semivariance} capture asymmetric responses to positive and negative returns. These developments have made realized measures the foundation for modern volatility modelling and forecasting frameworks, including both univariate and multivariate settings.

UHFD are inherently complex: they comprise millions of tick-level observations per asset per year and require significant computational and methodological effort to clean, filter, and aggregate these data. Moreover, for many a researcher, the usage is hindered by the cost of access. Historically, the Realized Library of the Oxford-Man Institute \citep{heber2009oxford} represented a valuable reference resource, offering various daily measures of realized volatility, along with open, high, low, and close prices related to several market indices from North American, European, and Asian exchanges. However, since its discontinuation in mid-2022, no open-access databases providing comparable data coverage or regular updates are available, despite their theoretical and practical importance. As a result, researchers and practitioners face significant barriers when attempting to replicate empirical findings, test new models, or conduct volatility forecasting exercises using realized data.

VOLARE (VOLatility Archive for Realized Estimates) aims at filling this gap by developing a comprehensive and openly accessible database of realized volatility and covariance measures derived from UHFD (source: \url{http://www.kibot.com}). Within VOLARE, we will refer to:
\begin{itemize}
    \item A \textit{library} where we process tick-level observations for stocks, exchange rates and futures, applying rigorous filtering, aggregation, and outlier detection procedures to produce clean and accurate series. Here, we integrate advanced econometric methodologies -- including the \citet{brownlees2006financial} outlier detection procedure and the \citet{barndorff2009realized} realized kernel framework -- ensuring methodological consistency and comparability across assets and time.
    \item A \textit{research infrastructure} where we estimate and compare a wide range of volatility models, such as the Heterogeneous Autoregressive (HAR) model \citep{corsi2009simple}, its quarticity-augmented extension HAR-Q \citep{bollerslev2016exploiting}, and the Multiplicative Error Model (MEM) family \citep{engle2002new, engle2006multiple}, including asymmetric specifications (AMEM). The platform enables consistent estimation, validation, and forecasting of volatility dynamics, similar to some of the features of the volatility analysis page in Engle’s V-Lab at NYU Stern (\url{https://vlab.stern.nyu.edu/volatility}).
    \item A data visualization facility (\textit{platform}) with an innovative user-oriented design. The platform integrates a back-end pipeline for large-scale data acquisition and cleaning with a web-based front-end for visualization, download, model estimation, and forecast dissemination. Users can access realized volatility and covariance time series, perform econometric modelling directly within the interface, and generate short- and medium-term forecasts, thereby bridging the gap between methodological research and applied financial analysis.
\end{itemize}

This paper provides a technical documentation and a justification for some of the choices made, both from the point of view of the econometric measures computed and of the architecture of data organization and back-end to front-end interaction. 

The rest of this paper is structured as follows. \autoref{sec:Kibot} describes the characteristics of the raw data used in the project, focusing on the Kibot database and the procedures adopted to assess data quality. \autoref{sec:oddlots} examines the impact of odd-lot trades on UHFD. \autoref{sec:constructingVolare} details the construction of the VOLARE library, covering the downloading, cleaning, and 
aggregation procedures: that includes the outlier detection procedure and the conversion of unevenly spaced tick data into regularly sampled series. \autoref{sec:variance} presents the realized measures computed within VOLARE, encompassing both univariate series (\autoref{sec:univariate}) and multivariate covariance 
measures (\autoref{sec:covariance}). \autoref{sec:frontend} illustrates the system integration between the back-end and the user interface. Overall, VOLARE makes advanced volatility analysis operational: its research infrastructure comprises a real-time capability of estimating, evaluating, and forecasting several popular univariate econometric models, discussed in detail in   \autoref{sec:volatilitymodels}. Some concluding remarks synthesize the contribution and lie the ground for future developments. The final appendix provides the complete list of assets available on the VOLARE platform and presents an empirical application of the volatility models implemented therein.

\section{The Raw Data from Kibot} \label{sec:Kibot}


The raw data used to construct VOLARE were obtained from Kibot, a provider offering data for stocks, ETFs, futures, exchange rates, indices, and OTC securities at various levels of 
temporal aggregation: from end-of-day (EOD) series, to intraday minute bars, down to tick-level data. For transaction data, Kibot offers two tick formats: \texttt{tickms}, with millisecond-precision timestamps, and \texttt{tick}, with second-level timestamps; the former is the closest available approximation to the original Trades and Quotes (TAQ) data, which carry nanosecond-precision timestamps.\footnote{When comparing TAQ and \texttt{tickms}, despite the timestamp truncation from nanoseconds to milliseconds, the two datasets show nearly identical record counts. In contrast, \texttt{tick} excludes odd lots in addition to truncating timestamps to the second, resulting in substantially fewer records.} For quotes data, Kibot analogously provides \texttt{tickbidaskms} with 
millisecond timestamps and \texttt{tickbidask} with second-level timestamps.

Before estimating daily realized variance, we conducted an exploratory analysis of the Kibot data, with a particular focus on listed stocks at the tick level, available with millisecond precision since 2015. To assess data quality, we ran a preliminary side-by-side comparison with a small snapshot of intra-daily stock data available on the NYSE Trade and Quote database\footnote{TAQ, accessed through the Wharton Research Data Services -- WRDS \url{https://www.wrds.wharton.upenn.edu}}, revealing a strong match. 

A further, more detailed comparison was conducted between daily prices and volumes (as they result from more granular Kibot data processing) and daily adjusted data from Yahoo Finance\footnote{\url{https://www.yahoo.com}} and CRSP\footnote{Center for Research in Security Prices (CRSP), available within WRDS.}. To that end, we focus on one year of Kibot tick data for Microsoft (MSFT), from 2023-01-03 to 2023-12-29. Transactions were first aggregated by timestamp to remove duplicates, ensuring that millisecond-level prices were consolidated before any further analysis. We then examined how different pre-processing choices, such as excluding trades after 16:00 {($\text{MSFT}_{1600}$)} or extending the trading window to 16:05\footnote{\citet{brownlees2006financial} had found that considering an extra five minutes - typically populated by trades with large volumes - could help capture the closing price, possibly recorded with a delay.} ($\text{MSFT}_{1605}$), affect daily price and volume measures. Daily prices were computed as the last available trade of each day (i.e., 16:00 or 16:05), while daily volumes were obtained by summing all intraday transactions.


\subsection{Price Analysis}
\autoref{tab:price_mainstats} shows that the daily-aggregated tick-level series (MSFT$_{1600}$ and MSFT$_{1605}$) reveals that both aggregation schemes produce virtually identical price levels. The high-frequency series are fully coherent with the daily reference data from Yahoo Finance (MSFT$_Y$) and Kibot (MSFT$_K$), showing no material discrepancies.

Examining the absolute percentage error\footnote{{The absolute percentage error between $x_1$ and $x_2$  is computed as $\text{PE} = \left|\frac{x_1 - x_2}{x_2}\right| \times 100$.}} statistics (\autoref{tab:price_error}) and the corresponding box plots (\autoref{fig:MSFTprcboxplot}), most deviations are extremely small,  typically below 0.1\% on average, with a few isolated outliers, further confirming the coherence of the price information across datasets and preprocessing choices.

\begin{table}[h!]
\centering
\resizebox{9cm}{!}{
\begin{tabular}{lrrrr}
\toprule
\multicolumn{5}{c}{\textbf{Price comparison}} \\
\midrule
& \multicolumn{2}{c}{\textbf{High-Frequency (HF) }} & \multicolumn{2}{c}{\textbf{Daily Reference (EOD)}} \\
& \textbf{MSFT$_{1600}$} & \textbf{$\text{MSFT}_{1605}$} & \textbf{$\text{MSFT}_Y$} & \textbf{$\text{MSFT}_K$} \\
\midrule
Count & 250 & 250 & 250 & 250 \\
Min & 219.10 & 219.10 & 219.16 & 219.10 \\
25\% & 279.75 & 280.07 & 279.68 & 279.74 \\
Median & 321.71 & 321.71 & 321.72 & 321.73 \\
75\% & 334.81 & 334.82 & 334.79 & 334.80 \\
Max & 380.62 & 380.57 & 380.62 & 380.62 \\
Mean & 311.04 & 311.16 & 311.03 & 311.04 \\
Std & 41.28 & 41.19 & 41.29 & 41.28 \\
\bottomrule
\end{tabular}}
\caption{Descriptive statistics for one year (2023-01-03 to 2023-12-29) daily MSFT prices across high-frequency aggregated series (MSFT$_{1600}$, MSFT$_{1605}$) and daily reference series (Yahoo Finance -- MSFT$_{Y}$, Kibot -- MSFT$_{K}$).}
\label{tab:price_mainstats}
\end{table}

\begin{minipage}[T]{0.47\textwidth}
\centering
    \resizebox{6cm}{!}{
    \begin{tabular}{lrrrr}
    \toprule
    \multicolumn{5}{c}{\textbf{Price Absolute Percentage error}} \\
    \midrule
     & \textbf{Mean} & \textbf{Std} & \textbf{Min} & \textbf{Max} \\
    \midrule
    MSFT$_{1600}$ vs $\text{MSFT}_{1605}$ & 0.12 & 0.52 & 0.00 & 5.03 \\
    MSFT$_{1600}$ vs $\text{MSFT}_Y$ & 0.01 & 0.04 & 0.00 & 0.53 \\
    MSFT$_{1600}$ vs $\text{MSFT}_K$ & 0.00 & 0.01 & 0.00 & 0.09 \\
    $\text{MSFT}_{1605}$ vs $\text{MSFT}_Y$ & 0.12 & 0.55 & 0.00 & 5.35 \\
    $\text{MSFT}_{1605}$ vs $\text{MSFT}_K$ & 0.12 & 0.54 & 0.00 & 5.30 \\
    $\text{MSFT}_Y$ vs $\text{MSFT}_K$ & 0.02 & 0.04 & 0.00 & 0.53 \\
    \bottomrule
    \end{tabular}}
    \captionof{table}{Absolute percentage errors analysis between high-frequency aggregated series (MSFT, $\text{MSFT}_{1605}$) and  daily price series ($\text{MSFT}_Y$, $\text{MSFT}_K$). One year of data (2023-01-03 to 2023-12-29).}
    \label{tab:price_error}
\end{minipage}%
\hspace{0.01\textwidth} 
\begin{minipage}[T]{0.49\textwidth}
\vspace{0pt}
\centering
    \includegraphics[width=1\linewidth]    {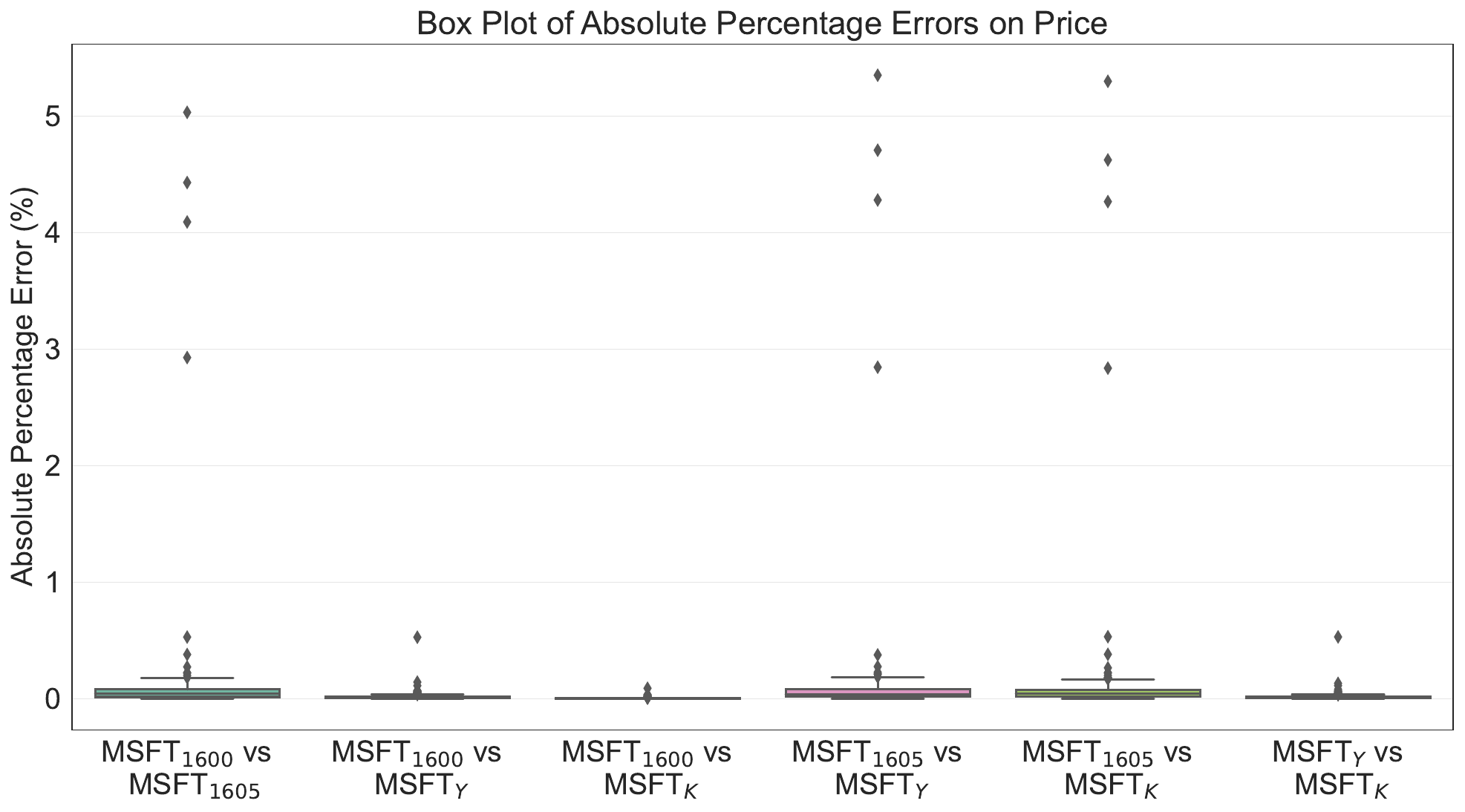}
    \captionof{figure}{Boxplot of absolute percentage errors across MSFT price series. One year of data (2023-01-03 to 2023-12-29).}
    \label{fig:MSFTprcboxplot}
\end{minipage}

\subsection{Volume Analysis}
In contrast to prices, trading volumes exhibit larger discrepancies across datasets.
High-frequency aggregated series (MSFT$_{1600}$, MSFT$_{1605}$), as well as daily aggregated Kibot volume (MSFT$_K$), report systematically lower volumes than what is available in the daily references (MSFT$_Y$ from Yahoo Finance, and MSFT$_C$ from CRSP): for individual trading days we show some examples in \autoref{tab:voldays} and we present some descriptive statistics over the one year full sample (2023-01-03 to 2023-12-29) in \autoref{tab:vol_mainstats}.
{To investigate whether extending the aggregation window can reconcile these differences, in the same table, we include also MSFT$_{all}$, which aggregates tickms-level volumes over the entire 24-hour trading period, including both regular hours and overnight sessions.}

According to the Kibot's documentation, these issues in part arise because odd-lot transactions (i.e., trades smaller than 100 shares, or 1,000 for low-priced securities; see below for a more detailed discussion) are excluded from the tick-level dataset, whereas they have been included in official exchange-reported volumes since October 2013, when U.S. exchanges began reporting odd-lot trades to the Consolidated Tape\footnote{The SEC approved including odd-lot transactions in the Consolidated Tape in October 2013 (Securities Exchange Act Release No. 70794), and the amendment became operative on December 9, 2013. \url{https://www.sec.gov/files/rules/sro/nms/2013/34-70794.pdf}}. As a result, Kibot’s end-of-day volumes tend to underestimate total trading activity.

Extending the aggregation window to 16:05 (MSFT$_{1605}$) increases the recorded volume and brings it closer to the Yahoo and CRSP benchmarks, suggesting that a small fraction of after-close trades contributes meaningfully to total reported volume.
{Further extending the aggregation to the full 24-hour period (MSFT$_{all}$) yields volumes nearly identical to the daily reference datasets, confirming that overnight trading activity, while typically representing a small fraction of total volume, fully accounts for the remaining discrepancy between high-frequency aggregations and the benchmark figures.}

Finally, \autoref{tab:volerror} and \autoref{fig:MSFTvolboxplot} summarize percentage differences across series. Mean errors exceed 20\% when comparing Kibot with Yahoo or CRSP, while discrepancies drop to around 4\% for the extended 16:05 window and to 0\% if we consider volume in the whole day period (MSFT$_{all}$). These results show that minor timing and filtering differences, particularly related to odd-lot exclusion, can substantially affect volume-based analyses.

\begin{table}[h!]
\centering
\begin{tabular}{lrrrrrr}
\toprule
\multicolumn{7}{c}{\textbf{Volume comparison}} \\
\midrule
& \multicolumn{3}{c}{\textbf{High-Frequency (HF) }} & \multicolumn{3}{c}{\textbf{Daily Reference (EOD)}} \\
\textbf{Date} & \textbf{MSFT$_{1600}$} & \textbf{MSFT$_{1605}$} & \textbf{MSFT$_{all}$ }& \textbf{MSFT$_Y$} & \textbf{MSFT$_C$} & \textbf{MSFT$_K$} \\
\midrule
2023-01-03 & 21,118,749 & 25,547,937 & 25,740,036& 25,740,000 & 25,723,760 & 20,138,969 \\
2023-01-04 & 44,396,551 & 49,392,966 & 50,623,394& 50,623,400 & 50,564,994 & 40,103,297 \\
2023-01-05 & 34,614,037 & 39,007,843 & 39,585,623& 39,585,600 & 39,541,865 & 32,111,249 \\
2023-01-06 & 36,697,134 & 42,411,714 & 43,613,574& 43,613,600 & 43,569,634 & 34,702,600 \\
2023-01-09 & 22,138,100 & 26,330,882 & 27,369,784& 27,369,800 & 27,334,038 & 20,896,326 \\
2023-01-10 & 21,648,976 & 26,549,791 &27,033,881  & 27,033,900 & 27,007,110 & 20,580,790 \\
2023-01-11 & 22,458,390 & 27,854,506 &28,669,331& 28,669,300 & 28,626,526 & 21,781,322 \\
\bottomrule
\end{tabular}
\caption{Example of daily trading volumes for selected days in January 2023 across high-frequency and daily reference MSFT datasets.}
\label{tab:voldays}
\end{table}

\begin{table}[h!]
\centering
\resizebox{13cm}{!}{
\begin{tabular}{lrrrrrr}
\toprule
\multicolumn{7}{c}{\textbf{Volume comparison}} \\
\midrule
& \multicolumn{3}{c}{\textbf{High-Frequency (HF) }} & \multicolumn{3}{c}{\textbf{Daily Reference (EOD)}} \\
& \textbf{MSFT$_{1600}$} & \textbf{MSFT$_{1605}$}  &  \textbf{MSFT$_{all}$}  & \textbf{MSFT$_{Y}$}  & \textbf{MSFT$_{C}$}  & \textbf{MSFT$_{K}$}  \\
\midrule
Count & 250 & 250 & 250 & 250 & 250 & 250 \\
Min & 9,220,482 & 9,220,792 &10,176,649& 10,176,600 & 10,114,326& 6,808,578 \\
25\% & 16,283,579 & 20,133,018 & 21,196,236& 21,186,475 & 21,164,668 & 15,049,934 \\
Median & 19,400,878 & 24,139,587 & 25,051,003 & 25,052,750 & 25,021,355 & 18,439,643 \\
75\% & 24,006,690 & 28,610,068 & 29,960,577& 29,960,600 & 29,922,746 & 22,178,547 \\
Max & 62,270,435 & 67,060,589 & 78,502,324 & 78,478,200 & 77,755,324 & 55,725,298 \\
Mean & 21,791,418 & 26,389,321 & 27,680,595 & 27,675,559 & 27,616,578 & 20,358,245 \\
Std & 8,662,096 & 9,899,126 & 10,642,406 & 10,640,234 & 10,614,316 & 8,321,219 \\
\bottomrule
\end{tabular}}
\caption{Descriptive statistics for one year of data (2023-01-03 to 2023-12-29): MSFT daily trading volumes based on high-frequency aggregation (MSFT$_{1600}$, MSFT$_{1605}$, , MSFT$_{all}$) and daily reference datasets (Yahoo Finance, CRSP, Kibot).}
\label{tab:vol_mainstats}
\end{table}

\begin{center}
\begin{minipage}[t]{0.47\textwidth}
\vspace{0pt}
\centering
    \resizebox{6cm}{!}{
    \begin{tabular}{lrrrr}
    \toprule
    \multicolumn{5}{c}{\textbf{Volume Percentage error}} \\
    \midrule
     & \textbf{Mean} & \textbf{Std} & \textbf{Min} & \textbf{Max} \\
    \midrule
    MSFT$_{1600}$ vs MSFT$_{1605}$ & 17.64 & 6.09 & 0.00 & 48.25 \\
    MSFT$_{1600}$ vs MSFT$_{all}$ & 21.18 & 7.26 & 2.79 & 57.34 \\
    MSFT$_{1600}$ vs MSFT$_Y$ & 21.16 & 7.27 & 2.79 & 57.32 \\
    MSFT$_{1600}$ vs MSFT$_C$ & 20.99 & 7.27 & 2.69 & 56.93 \\
    MSFT$_{1600}$ vs MSFT$_K$ & 10.44 & 6.83 & 0.04 & 38.15 \\
    MSFT$_{1605}$ vs MSFT$_{all}$ & 4.37 & 4.01 & 0.10 & 22.25 \\
    MSFT$_{1605}$ vs MSFT$_Y$ & 4.35 & 4.01 & 0.02 & 22.25 \\
    MSFT$_{1605}$ vs MSFT$_C$ & 4.15 & 4.03 & 0.04 & 22.05 \\
    MSFT$_{1605}$ vs MSFT$_K$ & 31.14 & 7.30 & 2.86 & 61.37 \\
    MSFT$_{all}$ vs MSFT$_Y$ & 0.02 & 0.06 & 0.00 & 0.58 \\
    MSFT$_{all}$ vs MSFT$_C$ & 0.28 & 0.32 & 0.00 & 3.64 \\
    MSFT$_{all}$ vs MSFT$_K$ & 37.32 & 8.69 & 20.84 & 76.40 \\
    MSFT$_Y$ vs MSFT$_C$ & 0.27 & 0.32 & 0.00 & 3.63 \\
    MSFT$_Y$ vs MSFT$_K$ & 37.30 & 8.69 & 20.84 & 76.40 \\
    MSFT$_C$ vs MSFT$_K$ & 37.00 & 8.66 & 20.57 & 76.11 \\
    \bottomrule
    \end{tabular}}
    \captionof{table}{Summary statistics of percentage differences in daily trading volumes across one year MSFT datasets.  One year of data (2023-01-03 to 2023-12-29).}
    \label{tab:volerror}
\end{minipage}
\hspace{0.005\textwidth} 
\begin{minipage}[t]{0.48\textwidth}
\vspace{0pt}
\centering
    \includegraphics[width=1\linewidth]{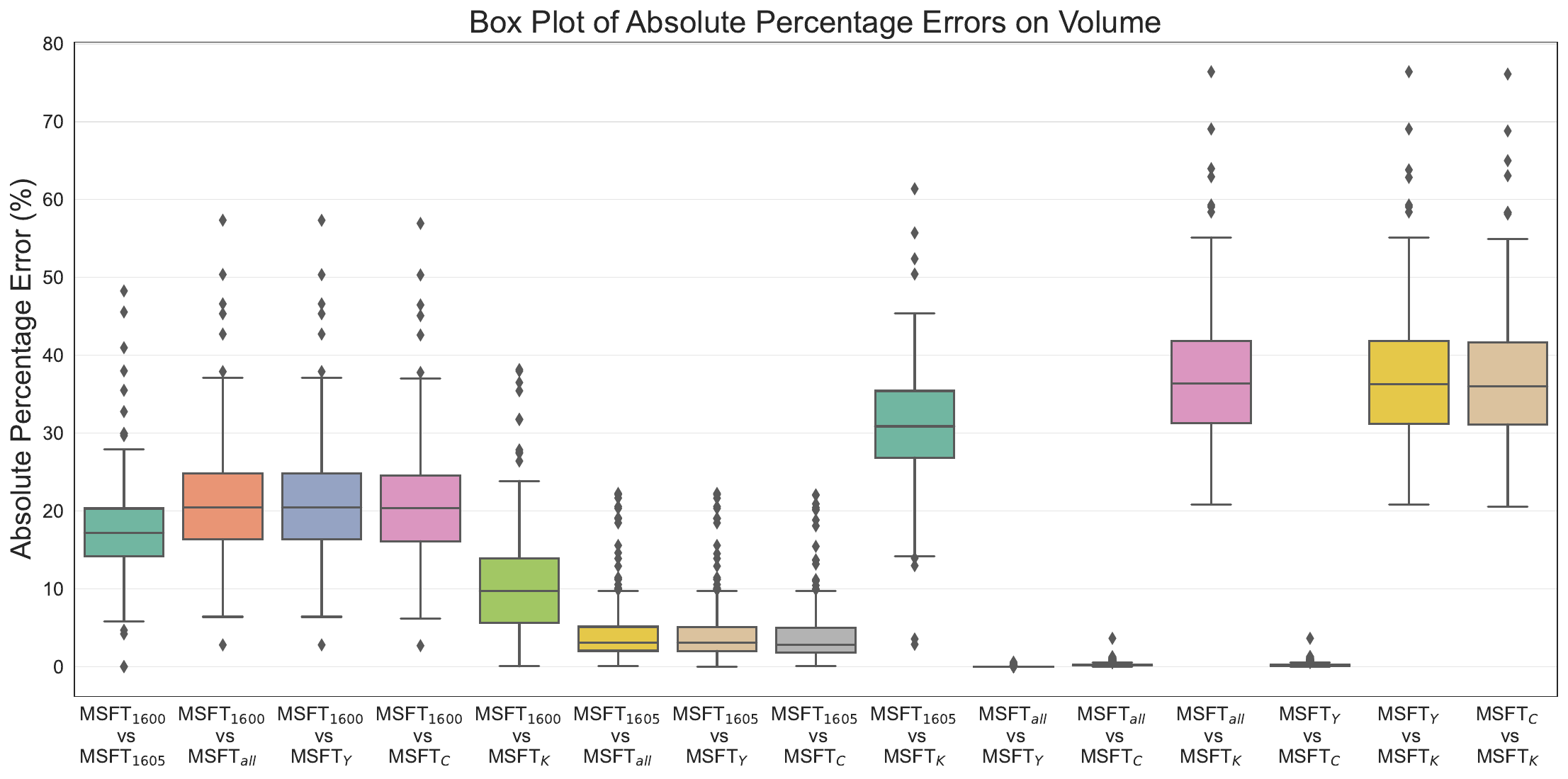}
    \captionof{figure}{Boxplot of percentage errors across MSFT volume series. One year of data (2023-01-03 to 2023-12-29).}
    \label{fig:MSFTvolboxplot}
\end{minipage}
\end{center}


\section{Some Insights on Odd-Lot Trades} \label{sec:oddlots}
Odd-lot transactions, defined as trades involving fewer than 100 shares (or fewer than 1,000 shares for low-priced securities), represent a distinct category of equity market activity. Historically, such trades were primarily associated with small investors and considered peripheral to price formation. However, in modern electronic markets, odd-lot activity has expanded dramatically and now accounts for a substantial fraction of total trades, particularly in high-frequency and algorithmic environments. \citep{ohara2014whats}

The treatment of odd-lot transactions varies considerably across data providers. In the Kibot dataset, they are excluded from the \texttt{tick} and fixed-interval intraday data (e.g., minute bars), but they are included in the higher-frequency \texttt{tickms} raw data.

The academic investigation of odd-lot trading can be traced to early contributions such as \citet{wu1972oddlots} who provides one of the earliest systematic analyses, documenting the persistence of odd-lot activity in U.S. equity markets and its relationship with investor behavior and market microstructure. In the post-decimalization period, \citet{johnson2014oddlot} show that the informational role of odd-lot trades increased markedly, as small-volume transactions became more closely tied to informed trading and algorithmic execution. \citet{ohara2014whats} demonstrate that excluding odd-lot trades from consolidated datasets such as TAQ can lead to systematic biases in measures of order imbalance, liquidity, and informational efficiency. More recently, \citet{bartlett2023market} reveal that odd-lot quotes often constitute an ''inside market'' offering better prices than the National Best Bid and Offer (NBBO), and therefore contribute materially to liquidity provision and price discovery.

As announced in October 2013, following U.S. Securities and Exchange Commission (SEC)
orders approving amendments to the UTP Plan and the CTA Plan,\footnote{See SEC Release
No.34--70793 (UTP Plan) and SEC Release No.34--70794 (CTA Plan), October 31, 2013, available at \url{https://www.sec.gov/files/rules/sro/nms/2013/34-70793.pdf} and
\url{https://www.sec.gov/files/rules/sro/nms/2013/34-70794.pdf}.}
Nasdaq and NYSE have officially incorporated odd-lot volume into their reported daily trading statistics, effective December 9, 2013.
Nevertheless, exchange rules stipulate that odd-lot trades cannot determine the official last price and therefore do not directly influence the daily open, high, low, or close (OHLC) values. The persistence of odd-lot activity, combined with heterogeneous vendor practices, introduces potential discrepancies in empirical datasets and may significantly affect intraday analyses that rely on unfiltered trade-level data. \citep{ohara2014whats}

\begin{table}[h!]
\centering
\resizebox{10cm}{!}{
\begin{tabular}{lrrrr}
\toprule
& \multicolumn{2}{c}{\textbf{Yearly Trading Activity}} & \multicolumn{2}{c}{\textbf{Odd Lots (Vol $<$ 100)}} \\
\cmidrule(lr){2-3} \cmidrule(lr){4-5}
{Symbol} & {Observations} & {Volume} & {Obs (\%)} & {Volume (\%)}  \\
\midrule
MSFT & 49,949,252 & 3,601,591,734 & 79.44 & 13.48 \\
JNJ & 13,859,636 & 1,230,836,830 & 71.93 & 11.54 \\
GE & 9,471,261 & 1,134,489,532 & 63.86 & 8.85 \\
MAT & 3,189,595 & 643,844,407 & 52.44 & 4.65 \\
LOCO & 543,523 & 56,608,498 & 69.14 & 9.61 \\
\bottomrule
\end{tabular} 
}
\caption{One year of high-frequency data from 2024-01-02 to 2024-12-31. Percentage of observations with a volume of less than 100 shares and their corresponding total volume percentage.}
\label{tab:oddlots} 
\end{table}

{After aggregating tick-level data recorded at the same millisecond, in \autoref{tab:oddlots}} we report some summary statistics about the number 
of high-frequency observations (and the corresponding overall volume) across one year (2024-01-02 to 2024-12-31) for five representative stocks. Interestingly, the majority of intraday trades are odd-lot transactions, accounting for 52-79\% of the observations, depending on the stock. However, their contribution to total trading volume is substantially smaller (5-13\%), highlighting a large discrepancy between transaction count and volume impact.

\begin{table}[h!]
\centering
\resizebox{14cm}{!}{
\begin{tabular}{lrrrrrrrrrrrrrrrrrrrr}
\toprule
& \multicolumn{10}{c}{\textbf{Daily Trading Activity}} \\
\cmidrule(lr){2-11} 
& \multicolumn{5}{c}{Observations} & \multicolumn{5}{c}{Volume} \\
\cmidrule(lr){2-6} \cmidrule(lr){7-11} 
Symbol & Min&  Q1 & Med & Q3 &Max &Min & Q1 & Med & Q3 & Max\\
\midrule
\midrule
MSFT & 79,340 & 167,946 & 186,966 & 215,314 & 515,694 & 6.0M & 11.0M & 13.1M & 16.0M & 40.9M  \\
JNJ & 28,992 & 48,046 & 54,082 & 60,071 & 93,759 & 2.5M & 3.8M & 4.6M & 5.6M & 12.2M  \\
GE & 17,506 & 30,034 & 34,804 & 41,146 & 111,160 & 1.8M & 3.0M & 4.0M & 5.1M & 17.6M  \\
MAT & 6,069 & 10,155 & 12,164 & 13,846 & 74,792 & 666K & 1.6M & 2.1M & 2.8M & 33.7M \\
LOCO & 992 & 1,616 & 1,962 & 2,567 & 11,582 & 62K & 135K & 187K & 250K & 3.1M  \\
\midrule
& \multicolumn{10}{c}{\textbf{Odd Lots (Vol $<$ 100)}} \\
\cmidrule(lr){2-11} 
& \multicolumn{5}{c}{Observations (\%)} & \multicolumn{5}{c}{Volume (\%)} \\
\cmidrule(lr){2-6} \cmidrule(lr){7-11} Symbol & Min&  Q1 & Med & Q3 &Max &Min & Q1 & Med & Q3 & Max \\
\midrule
\midrule
MSFT & 69.61 & 77.60 & 79.80 & 81.50 & 87.08 & 7.42 & 12.56 & 14.05 & 15.36 & 18.59 \\
JNJ & 59.36 & 69.56 & 72.40 & 75.01 & 82.00 & 6.90 & 10.32 & 11.98 & 13.48 & 20.82 \\
GE &  47.70 & 60.87 & 64.85 & 68.65 & 78.95 & 4.04 & 8.09 & 9.58 & 11.07 & 16.22 \\
MAT & 25.36 & 48.57 & 55.03 & 59.13 & 75.08 & 1.34 & 4.30 & 5.50 & 6.62 & 12.72 \\
LOCO & 39.33 & 62.17 & 70.55 & 76.88 & 88.77 & 2.20 & 8.28 & 10.44 & 14.29 & 28.15 \\
\bottomrule
\end{tabular}
}
\caption{Daily distribution of trading activity over 252 trading days in 2024. Top panel: total observations and volume. Bottom panel: odd lots (volume $<$ 100 shares) as percentage of total.}
\label{tab:oddlots_daily}
\end{table}

{Daily statistics (\autoref{tab:oddlots_daily}) confirm that odd-lot prevalence is coherent across the sample period. For each stock, we compute the daily incidence of odd-lot trades and their volume contribution, then report their distribution. The median daily percentages show that odd-lots typically represent 55\% (MAT) to 79\% (MSFT) of all trades, while contributing only 5-14\% of total volume. The narrow interquartile ranges indicate that these patterns are stable from day to day rather than driven by specific periods.}

\begin{figure}[h!]
    \centering
    \includegraphics[width=1\linewidth]{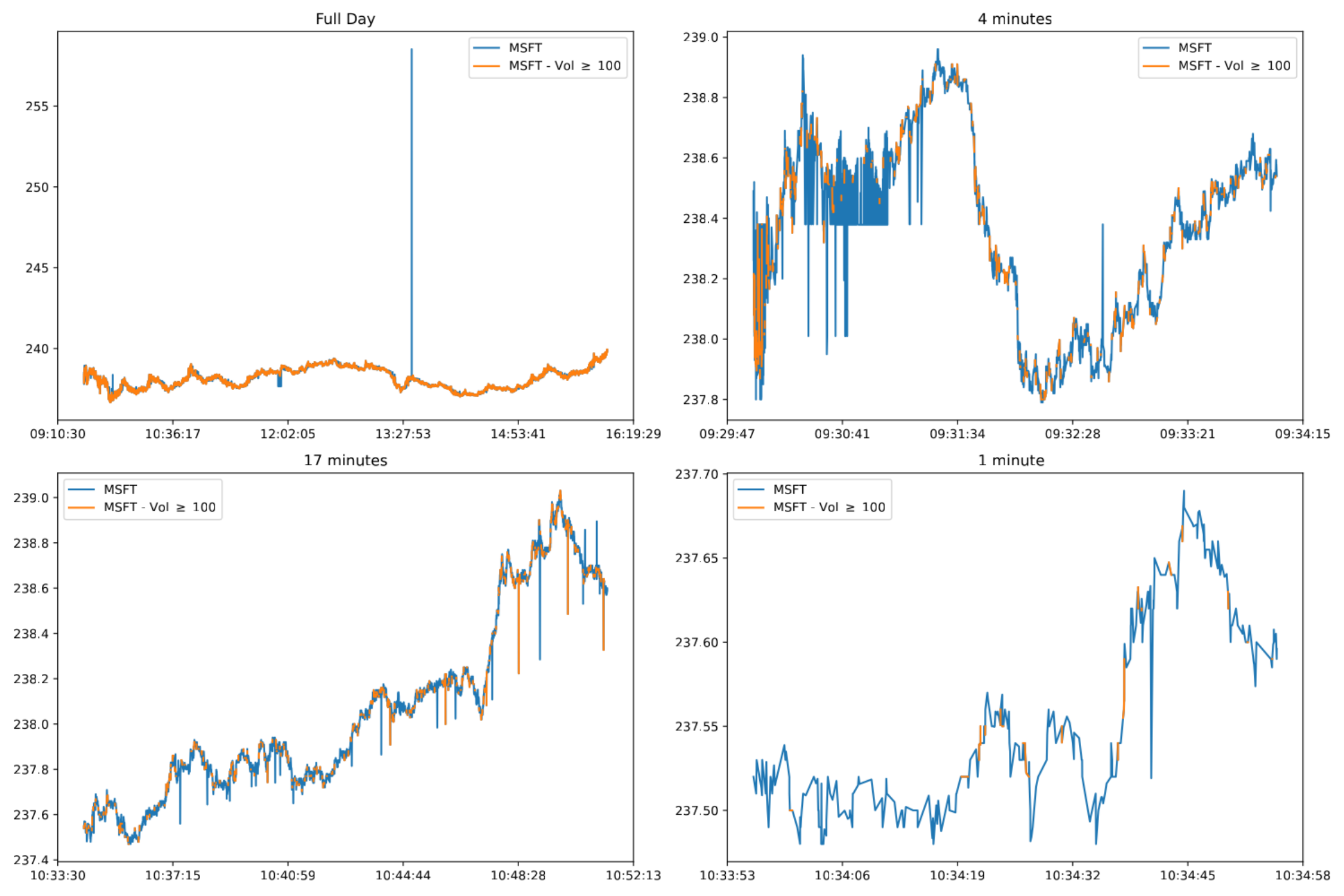}
    \caption{{Intraday price dynamics for MSFT, with and without odd-lot transactions (volume $<$ 100), displayed for different time windows at various trading times on December 30, 2022.}}
    \label{fig:odd_MSFT}
\end{figure}

\begin{figure}[h!]
    \centering
    \includegraphics[width=1\linewidth]{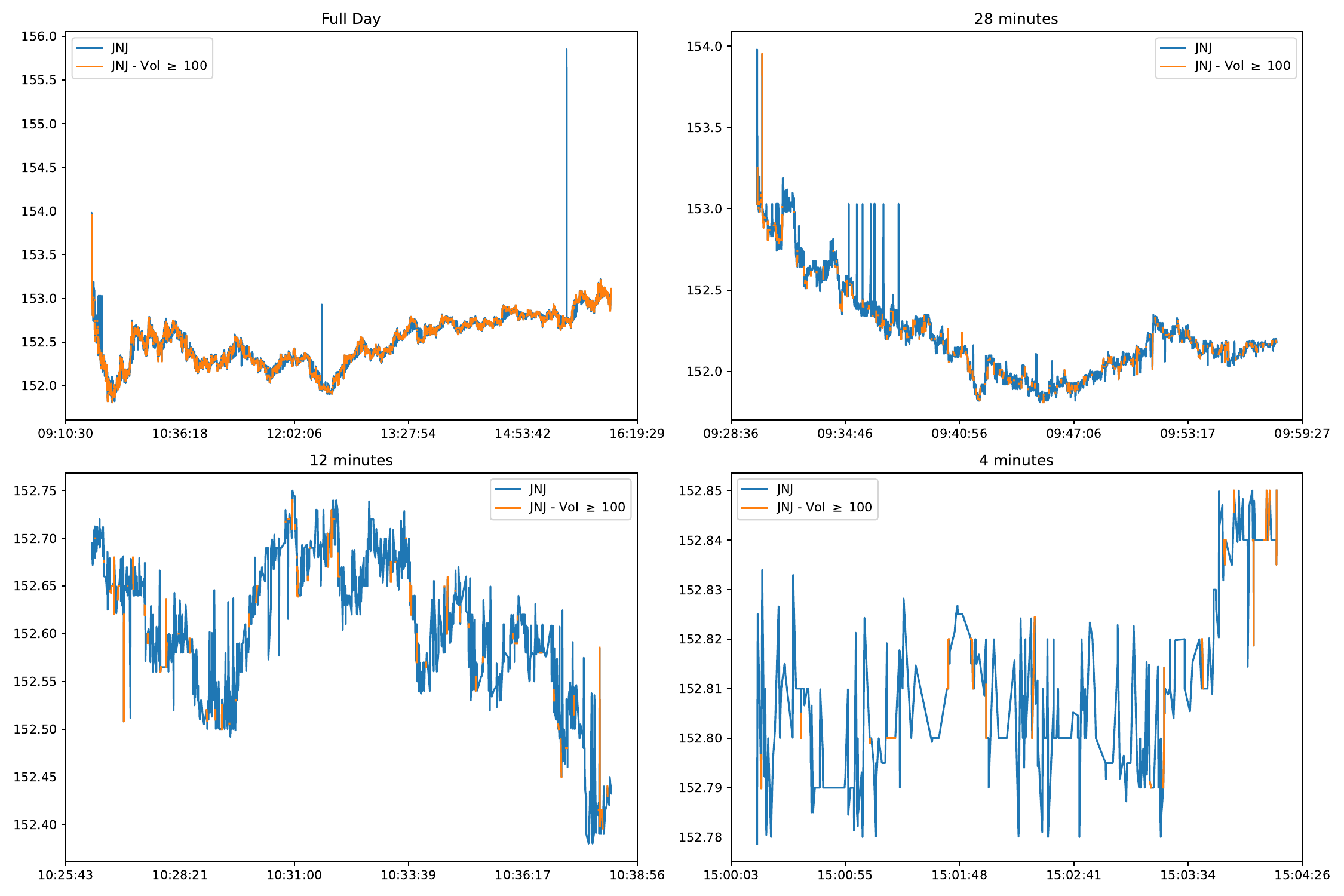}
    \caption{Intraday price dynamics for JNJ, with and without odd-lot transactions (volume $<$ 100), displayed for different time windows at various trading times on November 20, 2024.}
    \label{fig:odd_JNJ}
\end{figure}

\Cref{fig:odd_MSFT,fig:odd_JNJ} display both the original and the adjusted  (i.e., removing odd-lots) series for MSFT and JNJ, respectively. The four panels correspond to intervals with a different detail.

\begin{figure}[htbp]
    \centering
    \begin{minipage}[b]{0.48\textwidth}
        \centering
        \includegraphics[width=\textwidth]{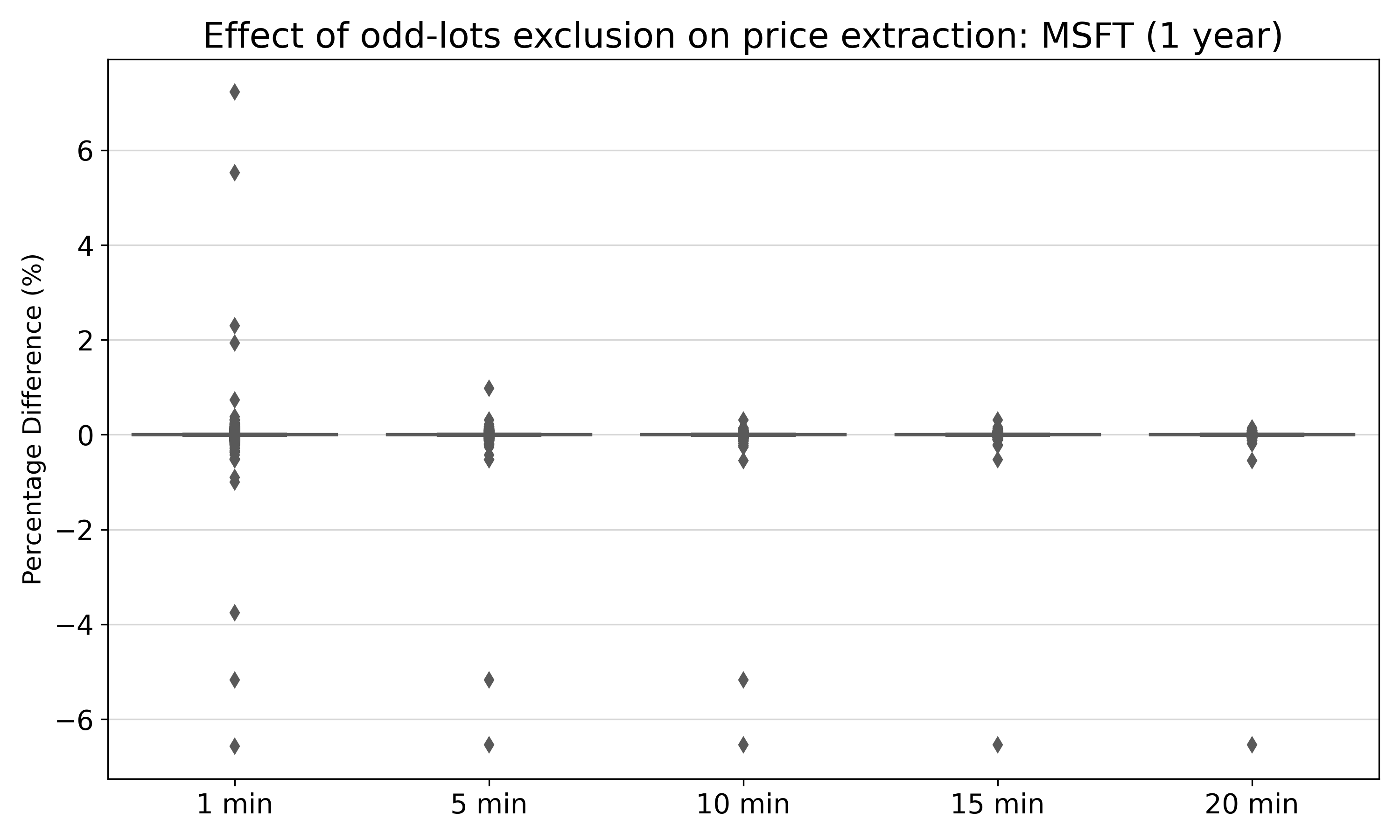}
    \end{minipage}
    \begin{minipage}[b]{0.48\textwidth}
        \centering
        \includegraphics[width=\textwidth]{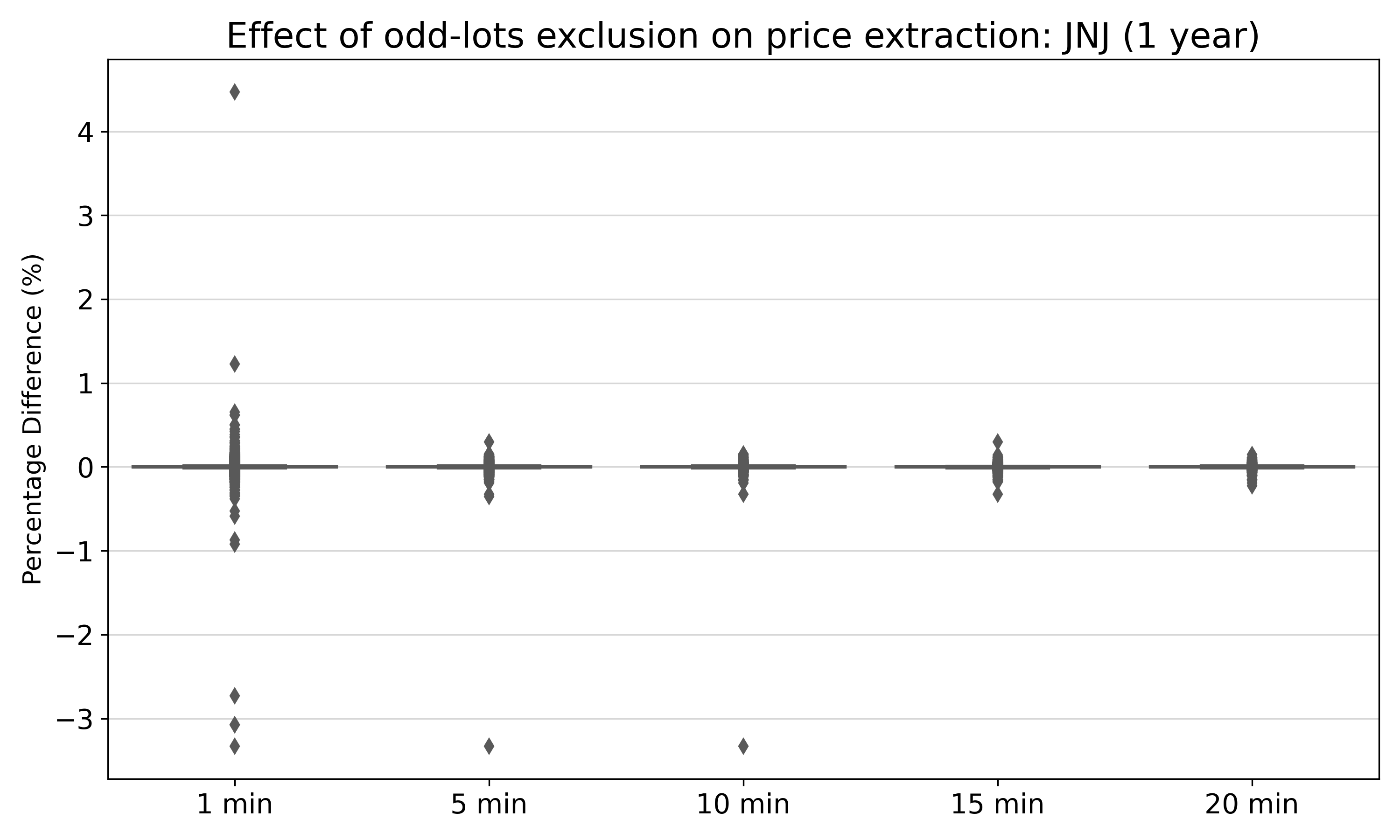}
    \end{minipage}
    
    
    \begin{minipage}[b]{0.48\textwidth}
        \centering
        \includegraphics[width=\textwidth]{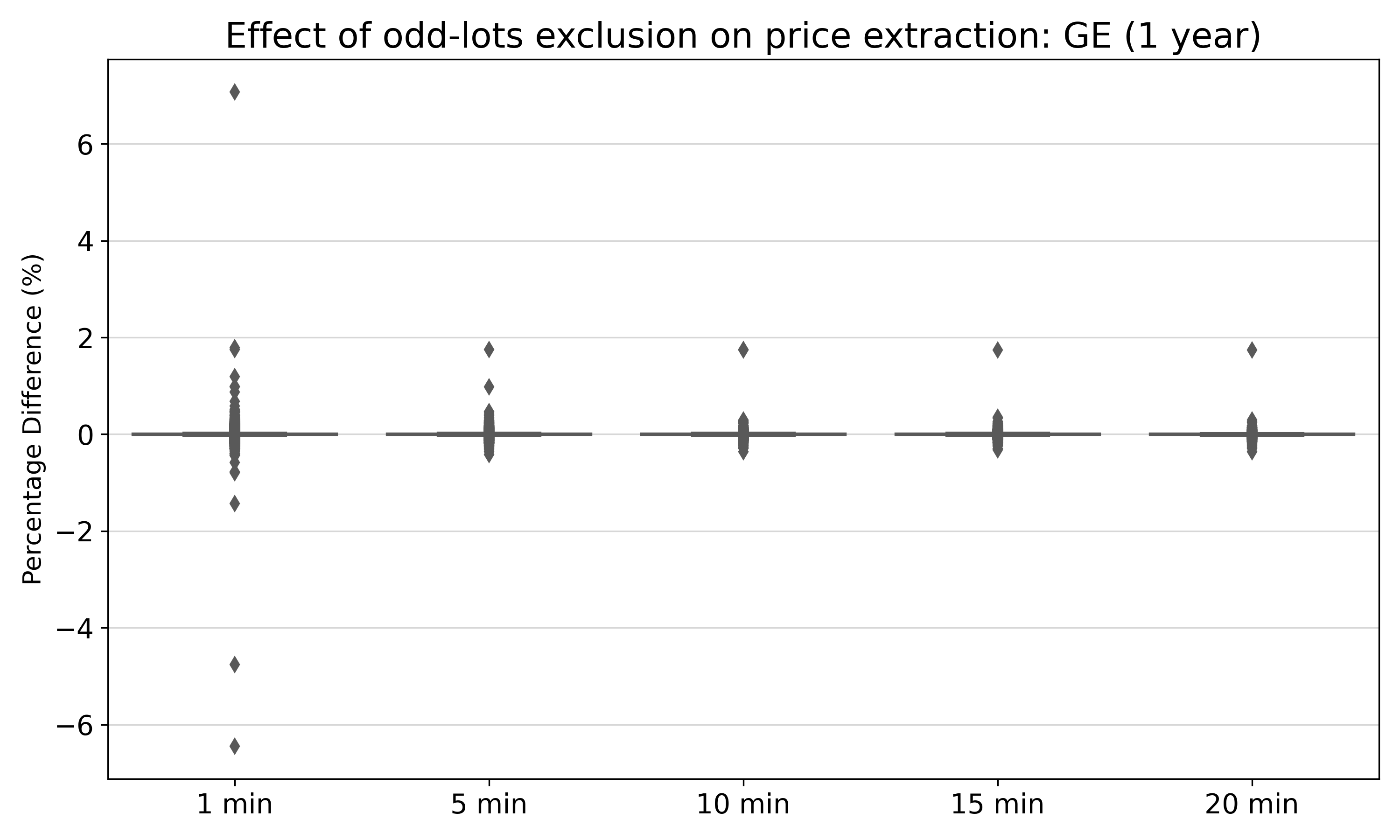}
    \end{minipage}
    \begin{minipage}[b]{0.48\textwidth}
        \centering
        \includegraphics[width=\textwidth]{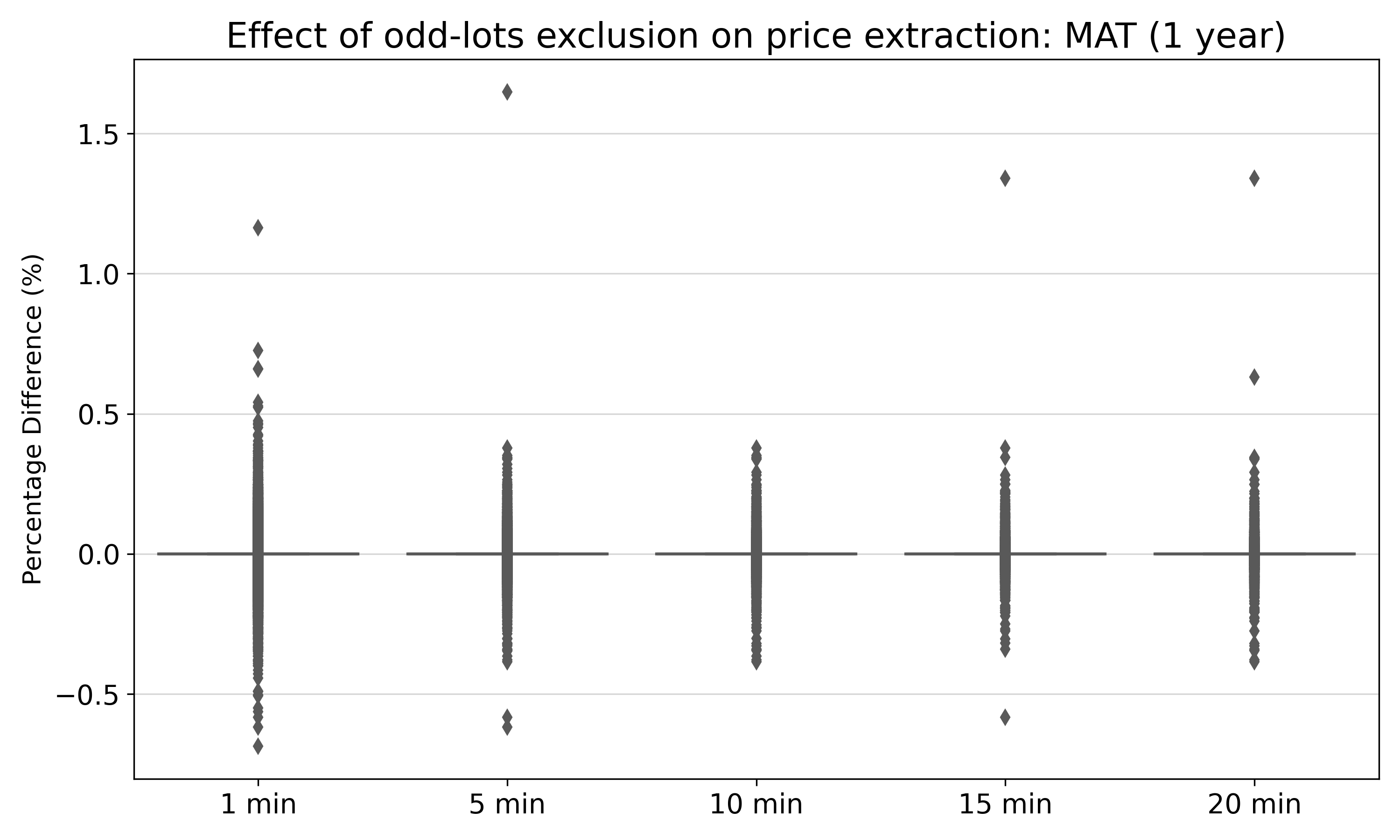}
    \end{minipage}
    
    
    \begin{minipage}[b]{0.48\textwidth}
        \centering
        \includegraphics[width=\textwidth]{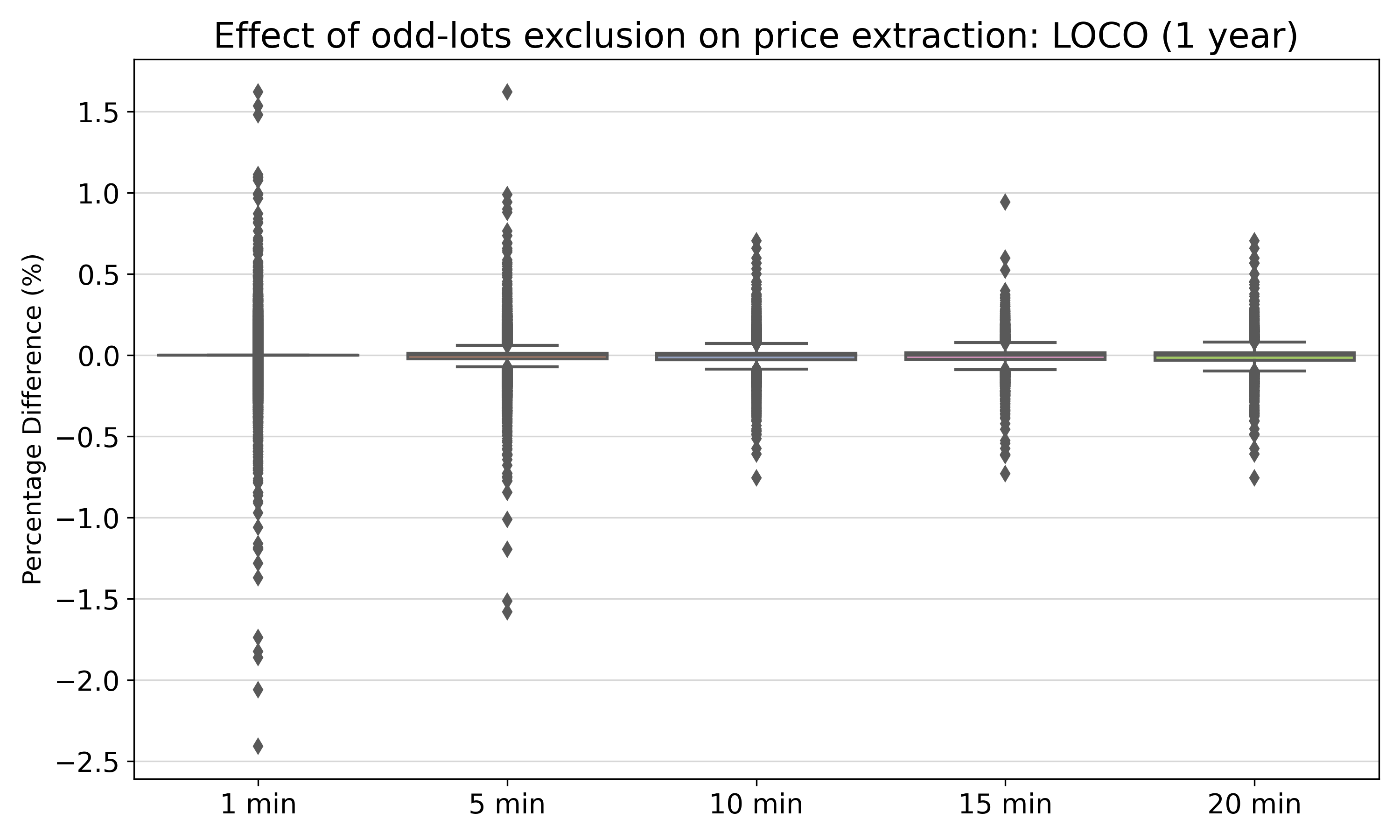}
    \end{minipage}
    
    \caption{{Percentage difference between original and odd-lot-adjusted price series for MSFT, JNJ, GE, MAT and LOCO over the period 2024-01-02 to 2024-12-31. Both series were sampled at 1, 5, 10, 15, and 20-minute intervals using the last available price.}}

    \label{fig:error_price_1y_noodds}
\end{figure}

\begin{figure}[htbp]
    \centering
    \begin{minipage}[b]{0.48\textwidth}
        \centering
        \includegraphics[width=\textwidth]{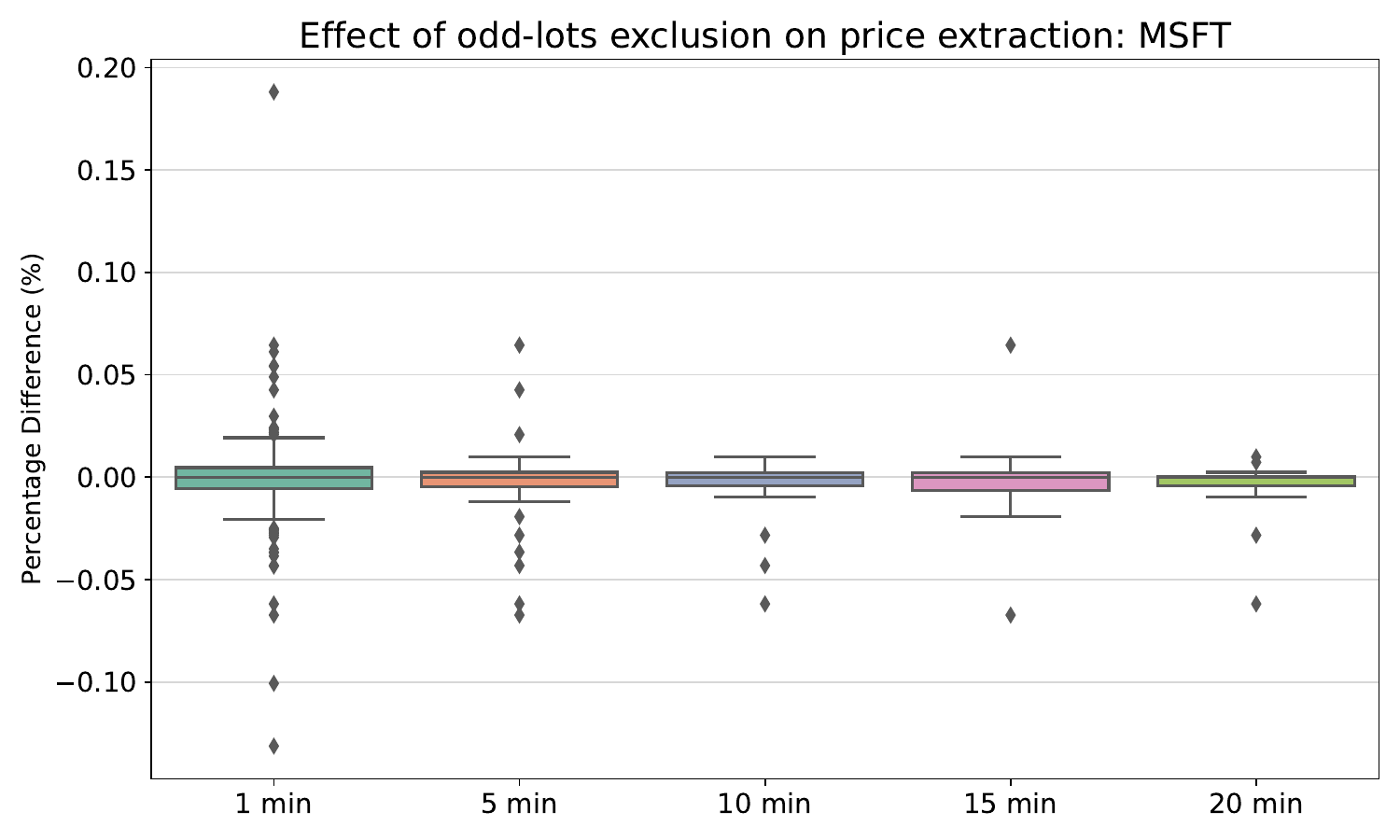}
    \end{minipage}
    \begin{minipage}[b]{0.48\textwidth}
        \centering
        \includegraphics[width=\textwidth]{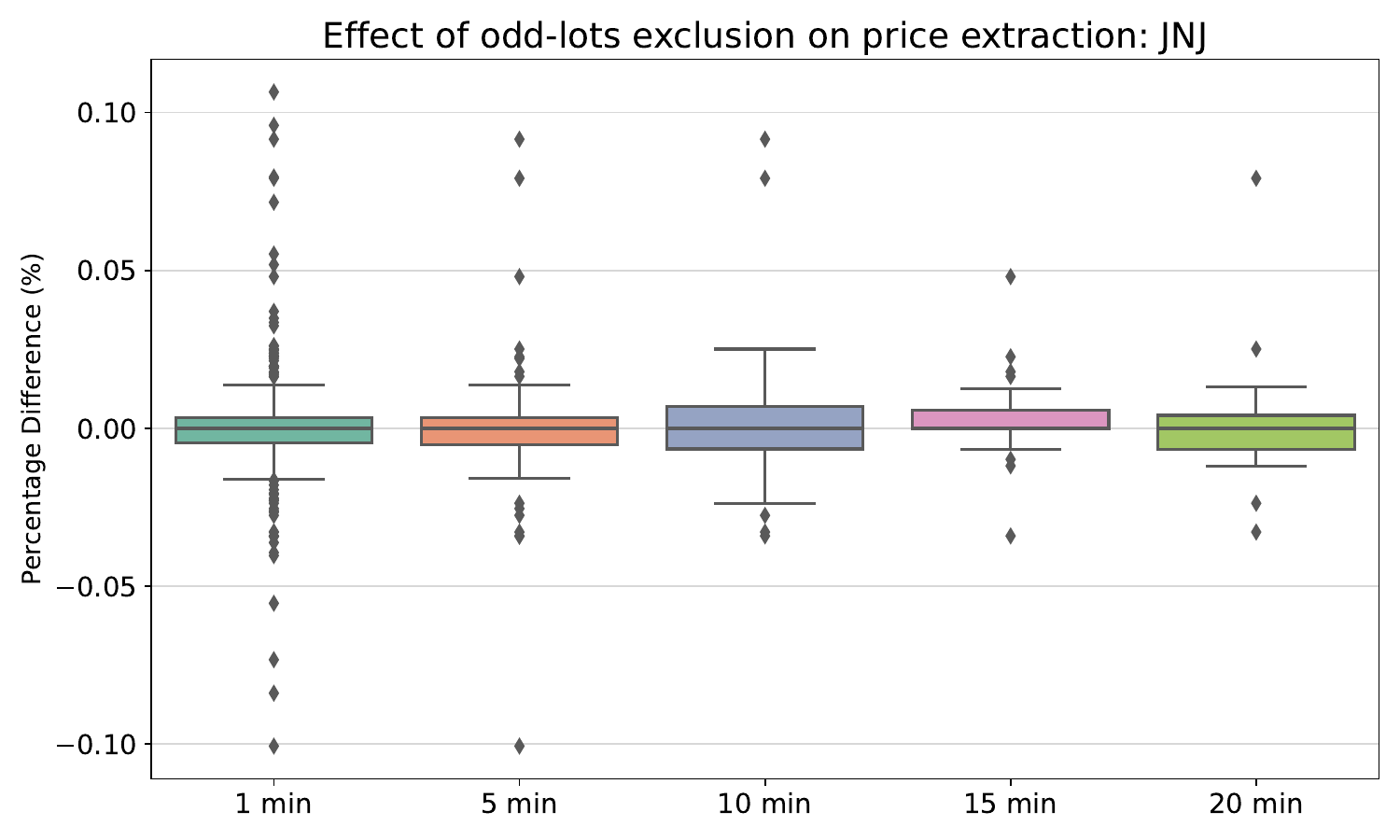}
    \end{minipage}
    
    
    \begin{minipage}[b]{0.48\textwidth}
        \centering
        \includegraphics[width=\textwidth]{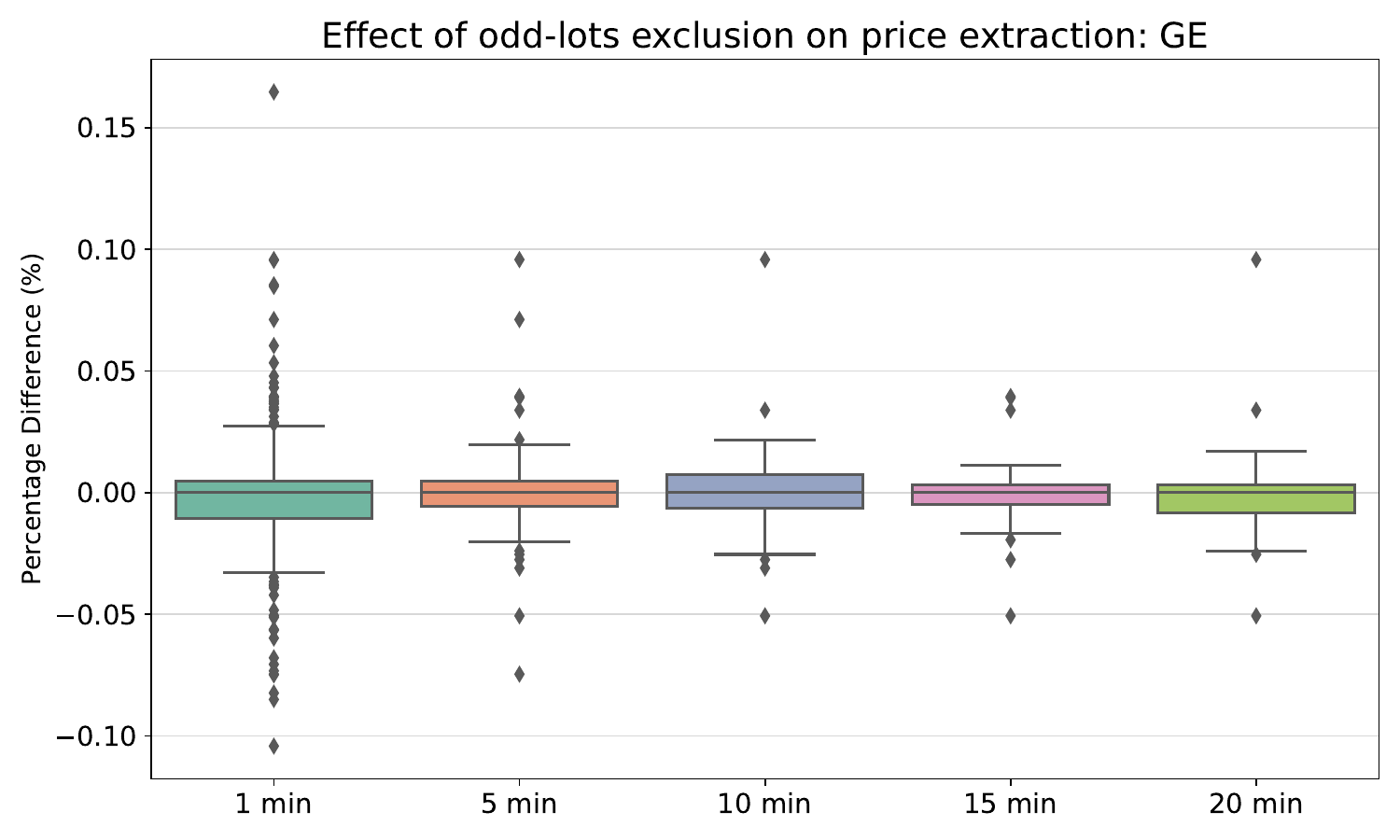}
    \end{minipage}
    \begin{minipage}[b]{0.48\textwidth}
        \centering
        \includegraphics[width=\textwidth]{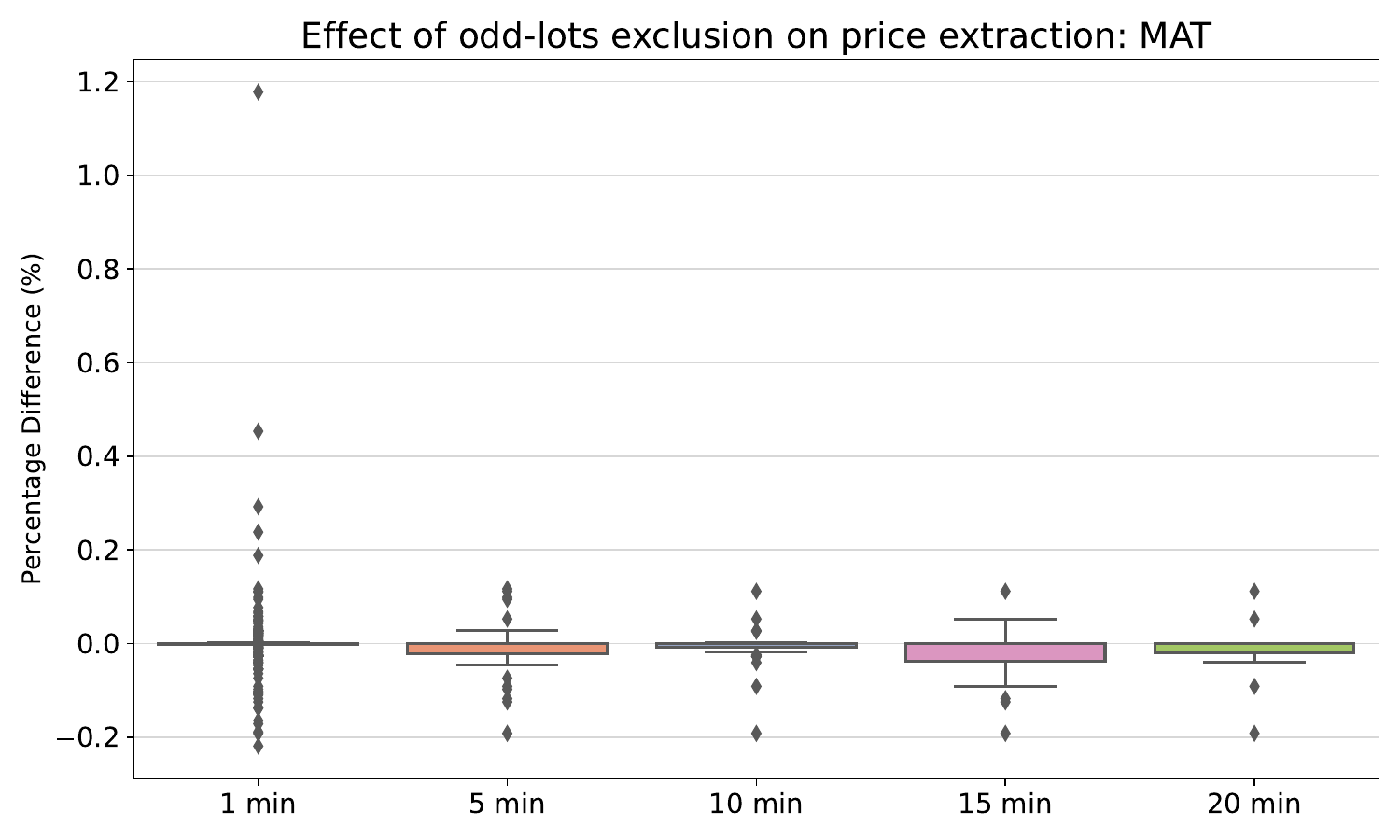}
    \end{minipage}
    
    
    \begin{minipage}[b]{0.48\textwidth}
        \centering
        \includegraphics[width=\textwidth]{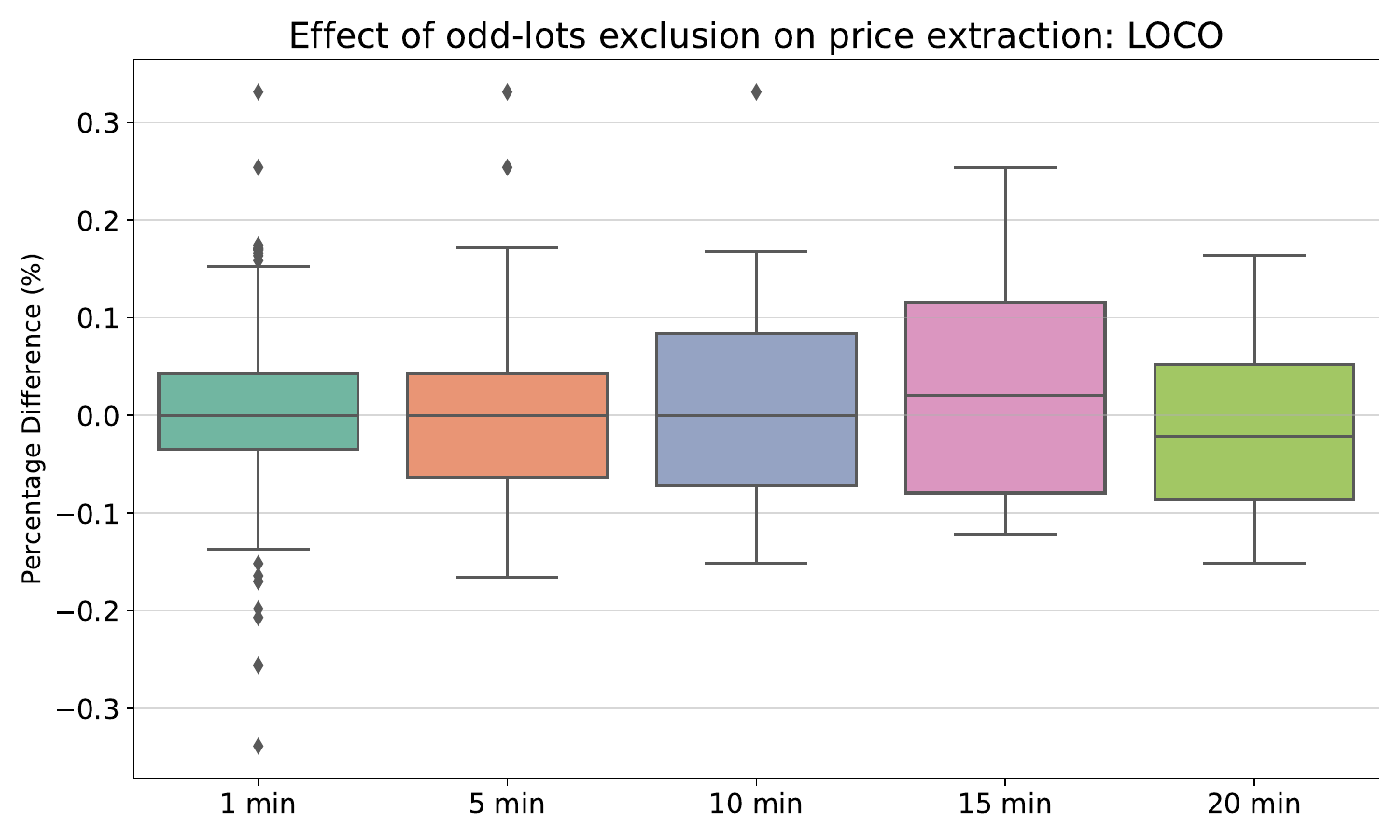}
    \end{minipage}
    
    \caption{{Percentage difference between original and odd-lot-adjusted price series for MSFT, JNJ, GE, MAT and LOCO on 2024-11-20. Both series were sampled at 1, 5, 10, 15, and 20-minute intervals using the last available price.}}
    \label{fig:error_price_aggr}
\end{figure}

We then constructed two parallel datasets (original and adjusted) for each stock, regularly-sampled at multiple time intervals (1, 5, 10, 15, and 20 minutes), using the last available price within each interval as representative for that period. {The percentage difference between the two series is computed as ${(\text{Price}_{\text{adjusted}} - \text{Price}_{\text{original}})}/{\text{Price}_{\text{original}}} \times 100$.} The resulting box plots (\Cref{fig:error_price_aggr,fig:error_price_1y_noodds}) illustrate the magnitude of the aggregation error attributable to the exclusion of odd-lot transactions. 
The median error is generally close to zero, indicating that excluding odd-lot trades has a minimal impact on the aggregated prices for most stocks. However, larger discrepancies are more pronounced for less liquid securities and shorter intra-daily sampling intervals, demonstrating that odd-lot trades can occasionally generate significant deviations in what is taken as the intraday prices. Overall, these plots confirm that while excluding odd-lot transactions has limited effects on highly liquid equities, it should be carefully considered when analyzing low-liquidity stocks or constructing high-frequency time series.

\clearpage

\section{The VOLARE Back-end: Data Preparation} \label{sec:constructingVolare}

As mentioned before, the high-frequency financial datasets were sourced from Kibot, which provides data for a wide range of financial instruments; we focus on three major asset classes: stocks, exchange rates, and futures.These datasets are stored on a dedicated server, providing the necessary components for computing realized variance and covariance measures over the sample period of interest, as well as additional market variables such as open, high, low, and close prices, trading volume, and the number of trades. The complete list of securities included in the library is reported in \Cref{tab:stocks_availability,tab:forex_availability,tab:futures_availability} in \autoref{sec:appendix}.

Each asset class presents distinct characteristics, mainly because of the market features (e.g., trading hours), but also for the format of the original data and the presence of outliers. 
In what follows, we describe the specific feature of each asset class and the preprocessing procedures implemented to ensure the coherence and comparability. 

\subsection{Asset Classes}

\paragraph{Stocks} 
The stock component of VOLARE is fed by unadjusted \texttt{tickms}. {Regular trading hours for U.S. equities officially run from 9:30 to 16:00 Eastern Time (ET), but we extend data collection to 16:05 following \citet{brownlees2006financial}, who found that including an additional five minutes beyond the official market close helps capture the closing price and associated trading volume, which are often recorded with a delay, as we show in our empirical analysis in \autoref{sec:Kibot}}. On certain pre-holiday sessions, such as the days preceding Christmas, Independence Day\footnote{\label{fn:closure}In certain years, the U.S. stock market closes at 13:00 on July 3rd ahead of Independence Day (July 4th). However, when July 4th falls on a weekend (Saturday or Sunday), the holiday is observed on the previous Friday or the following Monday, respectively, and no early market closure occurs. Likewise, if July 4th falls on a Monday, that day is observed as a full holiday with no early closure on the preceding Friday.}, and the day following Thanksgiving, the market closes early at 13:00. For these sessions, any observation after 13:00 is treated as pertaining to overnight activity and is excluded from the intraday sample.
Before computing realized measures, the raw tick data undergo a cleaning procedure to identify and remove outliers. This operation follows the methodology of \citet{brownlees2006financial}, which provides a robust framework for filtering anomalous price movements in high-frequency financial data (c.f \autoref{sec:outlier_detection} for details on the cleaning procedure).

\paragraph{Exchange Rates}
The exchange rates dataset used in VOLARE is constructed from \texttt{tickbidask}. We deliberately refrain from using millisecond-level observations, as the timestamps provided by Kibot correspond to the moments when trades were received by their system, rather than the precise execution times. This asynchrony between trade execution and data ingestion could otherwise distort the fine-scale temporal structure of the series.

Due to the overlap of global trading centers across different time zones, exchange rate markets are, in principle, continuously active five days a week, from Sunday at 17:00 ET to Friday at 17:00 ET; however, we excluded periods of limited or irregular trading activity from the sample. In fact, weekend trading is sporadic and limited to a few currency pairs, resulting in incomplete and incoherent data coverage. Similarly, during major global holidays (e.g., Christmas, New Year’s Day), trading activity is markedly reduced or absent for several currency pairs.
To ensure temporal coherence and liquidity comparability, all periods with abnormally low activity are omitted.

Since \texttt{tickbidask} data is not based on transaction prices, no outlier-filtering procedure is applied. In fact, using mid-quote prices (the average of bid and ask) for subsequent processing naturally mitigates the impact of transient anomalies or recording errors, and the resulting series are sufficiently regular and robust without the need for additional cleaning.

\paragraph{Futures} 
The futures dataset is derived from \texttt{tickbidask} as well. Although millisecond timestamps are technically available, CME Group’s official documentation indicates that the “Trade Time” field uses the HHMMSS format without milliseconds. Consequently, millisecond-level data offer no additional precision compared to second-resolution data, and their use could create misleading assumptions about timing accuracy. For this reason, all futures contracts in the VOLARE dataset are based on timestamps at the one-second resolution.

Futures contracts are traded on various exchanges, including NYMEX, COMEX, and CME. These markets operate nearly continuously from Sunday at 18:00 ET to Friday at 17:00 ET, with a daily maintenance break from 17:00 to 18:00 ET. Trading hours and holiday schedules vary across contract types and exchanges. To ensure temporal consistency and comparability, periods with no trading activity, such as weekends and major global holidays, are excluded from the sample.

As with the exchange rates, the subsequent use of mid-quote prices reduces the impact of transient anomalies; therefore, no outlier-filtering procedure is applied.

\medskip 

A summary of some relevant aspects of the different asset classes is reported in \autoref{tab:asset_type_info}.

\begin{table}[h!] 
\centering
\begin{tabular}{@{}lccc@{}}
\toprule
\textbf{Asset Type} & \textbf{Starting Date}  & \multicolumn{1}{c}{\textbf{Frequency}}  & \textbf{Outlier Detection} \\
\midrule
Stocks  & 2015-01-02 & 
\texttt{tickms}
&   yes\\
Exchange Rates  &  2009-09-25 & 
\texttt{tickbidask}
&  no \\
Futures & 2009-09-28 &  
\texttt{tickbidask}
& no\\
\bottomrule
\end{tabular}
\caption{Summary of some relevant aspects by asset class: starting date, data frequency, and outlier detection procedure.}
\label{tab:asset_type_info}
\end{table}

\subsection{Download Raw Data}
The raw high-frequency datasets used to build the VOLARE library are obtained via a custom download script interfacing with the Kibot API. The script accepts input parameters such as start and end dates, trading hours, and desired frequency (e.g., \texttt{tickms} for stocks and \texttt{tickbidask} for exchange rates and futures).

When multiple records within the same trading day share the same timestamp, the data are aggregated while preserving trade information by adding an additional column to the dataset. For stocks, aggregation computes the volume-weighted average price (VWAP), sums the total traded volume, and counts the number of trades\footnote{For stocks, the trade count accurately reflects the number of individual trades. However, for exchange rates and futures, trade and volume-related fields provided by the data source do not correspond to actual executed transactions. In particular, foreign exchange markets are decentralized and do not provide transaction-level volume information; any such fields typically reflect quote updates rather than traded quantities. For this reason, volume and trade counts are not included in the final datasets for exchange rates and futures, although they are retained at the download stage for traceability.}. For exchange rates and futures, aggregation is performed using the median of price-related variables, namely price, bid, and ask. In this context, the price variable does not play a distinctive role beyond bid and ask quotes and is included solely for coherence across asset classes. Volume and number of trades are excluded from the final datasets for these asset classes.
For each asset type, a dedicated folder is created; within it, each symbol has its own subfolder containing daily \texttt{parquet} files named according to the corresponding trading date. The organization of the raw data directories is illustrated in \autoref{fig:data_structure}. For stocks, each file contains at least \texttt{Time}, \texttt{Price}, \texttt{Volume}, and \texttt{Trades}; for exchange rates and futures, files include at least \texttt{Time}, \texttt{Price}, \texttt{Bid}, \texttt{Ask}, \texttt{Volume}, and \texttt{Trades}.

For housekeeping purposes, adjustment information (e.g., dividends or splits) is stored in an \texttt{adjustments.txt} file within each symbol folder and is updated incrementally upon download to preserve a complete record; to be clear, all realized measures computations are conducted using non-adjusted data. Additionally, a log file records metadata about the download process, allowing us to track any issues encountered.

The \texttt{parquet} format was chosen because it provides substantial storage savings compared to plain text files. For example, storing tick-level stock data in .parquet rather than .txt reduces storage requirements by over 60\% (\autoref{tab:storage}) while preserving full data fidelity. In addition, \texttt{parquet} files, being columnar, allow efficient access to individual columns and faster analytical processing, which is particularly important when working with large high-frequency datasets.

\bigskip

\noindent
\begin{minipage}{0.45\textwidth}
\centering
\begin{tabular}{@{}l@{}}
\texttt{/data/} \\
\texttt{\textbar-- stocks/} \\
\texttt{\textbar\phantom{--}\textbar-- AAPL/} \\
\texttt{\textbar\phantom{--}\textbar\phantom{--}\textbar-- 2025\_03\_10.parquet} \\
\texttt{\textbar\phantom{--}\textbar\phantom{--}\textbar-- 2025\_03\_11.parquet} \\
\texttt{\textbar\phantom{--}\textbar\phantom{--}\textbar-- ...} \\
\texttt{\textbar\phantom{--}\textbar\phantom{--}\textbar-- adjustments.txt} \\
\texttt{\textbar\phantom{--}\textbar-- MSFT/} \\
\texttt{\textbar\phantom{--}|-- TSLA/} \\
\texttt{\textbar-- exchange rates/} \\
\texttt{\textbar-- futures/} \\
\end{tabular}
\captionof{figure}{Structure of the raw data directory of the VOLARE library, organized by asset class and security.}
\label{fig:data_structure}
\end{minipage}
\hfill
\begin{minipage}{0.50\textwidth}
\centering
\begin{tabular}{lrr}
\toprule
\multicolumn{3}{c}{\textbf{Storage}} \\
\midrule
&\textbf{.txt } & \textbf{.parquet}\\
\midrule
MSFT&  2.11 GB& 594.3 MB  \\
JNJ&  575.2 MB & 181.2 MB\\
GE&  402.5 MB & 133.8 MB\\
XOM& 790.6 MB & 237.1 MB \\
KHC& 295 MB & 97.3 MB \\
\midrule
Total & 4.17 GB  & 1.24 GB \\
\bottomrule
\end{tabular}
\captionof{table}{File size comparison of .txt and .parquet formats for one year of raw data.}
\label{tab:storage}
\end{minipage}

\subsection{Data Cleaning Procedure for High-Frequency Prices} \label{sec:outlier_detection}

Each trading day can generate millions of observations at uneven intervals, often influenced by noise from human input errors, system glitches, delayed reporting of block trades, or genuine market anomalies. These irregularities can lead to outliers that, if left unaddressed, may distort realized variance measures.

Effective cleaning requires balancing data preservation with correction: excessive filtering may remove small but informative price movements (“over-cleaning”), while insufficient cleaning may retain spurious observations that inflate noise (“under-cleaning”). This balance is particularly challenging without exchange-provided trade condition flags, as some sources like ours do not offer metadata to identify cancelled or out-of-sequence trades, making data-driven cleaning procedures essential.

Widely adopted methods were proposed by \citet{brownlees2006financial} and \citet{barndorff2009realized}. The \citet{brownlees2006financial} algorithm identifies anomalous observations based on their deviation from neighboring prices using a trimmed mean and standard deviation within a local time window. The \citet{barndorff2009realized} method employs a rolling median filter, discarding prices deviating by more than ten mean absolute deviations (MAD) from a centered window of fifty observations. By using the median and MAD instead of the mean and standard deviation, this approach provides a non-parametric, highly robust filter that 
minimizes the influence of extreme values.

The VOLARE library employs the robust, data-driven cleaning procedure on the UHF stock price data based on the \citet{brownlees2006financial} proposal.

The method relies on local information around each observation to identify and remove values that deviate excessively from their surrounding neighborhood. Specifically, a price observation $p_i$ is flagged as an outlier when
        $$|p_i - \bar{p}_i(k)| \geq 3 \, s_i (k)+ \gamma $$
where $\bar{p}_i(k)$ denotes the $\delta$-trimmed mean and $s_i(k)$ the corresponding trimmed standard deviation computed over a symmetric neighborhood of $k$ observations around $i$. The parameter $\gamma$ is a granularity constant that prevents false detections when prices remain constant for several ticks (e.g., during periods of low liquidity).\\

The procedure is applied as follows.
\begin{enumerate}
     \item \textbf{Neighborhood definition.} For each price $p_i$, a neighborhood $N_i(k)$ of $k$ observations around $i$ is extracted, excluding the observation at index $i$. Specifically:
    \begin{itemize}
        \item If $i < k/2$, $N_i(k)$ includes the first $k+1$ prices of the series.  
        \item If $i \ge n - k/2$, $N_i(k)$ includes the last $k+1$ prices.  
        \item Otherwise, $N_i(k)$ is symmetric around $i$, from $i - k/2$ to $i + k/2 + 1$.  
    \end{itemize}
     \item \textbf{Trimmed mean and standard deviation.} The $\delta$-trimmed mean $\overline{p}_i(k)$ and standard deviation $s_i(k)$ are computed on the observations belonging to $N_i(k)$. The trim count (used to exclude the lowest and largest observations of $N_i(k)$ from the mean and standard deviation calculation) is determined as the integer value of $(\text{length of} N_i(k)) \times \delta / 2$ as $\lfloor (\text{length of } N_i(k)) \times \delta / 2 \rfloor$, where $\lfloor \cdot \rfloor$ denotes the floor function.
     \item \textbf{Outlier detection rule.} $p_i$ is flagged as an outlier if it deviates from $\overline{p}_i(k)$ by more than $3 \, s_i(k) + \gamma$. All detected outliers are recorded, and a cleaned series is obtained by marking these values.
     \item \textbf{Outlier treatment.} Prior to the outlier detection procedure, a new column is added to the original dataframe to store a flag for each observation. This column captures the status of each price as follows:
        \begin{itemize}
        \item Prices outside regular trading hours are assigned a value of 0.
        \item Observations identified as outliers are set to \texttt{NaN}.
        \item Valid prices are set to 1.
        \end{itemize}
During subsequent preprocessing for the computation of realized variance and covariance, each outlier is replaced by the mean of the two preceding and two following valid prices.
\end{enumerate}

The parameters of the \citet{brownlees2006financial} procedure are $k$ (neighborhood size), $\delta$ (trimming proportion), and $\gamma$ (granularity threshold):
$k$ controls local smoothness, with small windows being more sensitive to short-term fluctuations;
$\delta$ regulates the proportion of extreme values trimmed from each neighborhood;
$\gamma$, which turns out to be the most influential, determines the size of the filtering window, with higher values resulting in fewer observations removed.  

\begin{figure}[ht]
\centering
  \includegraphics[width=\linewidth]{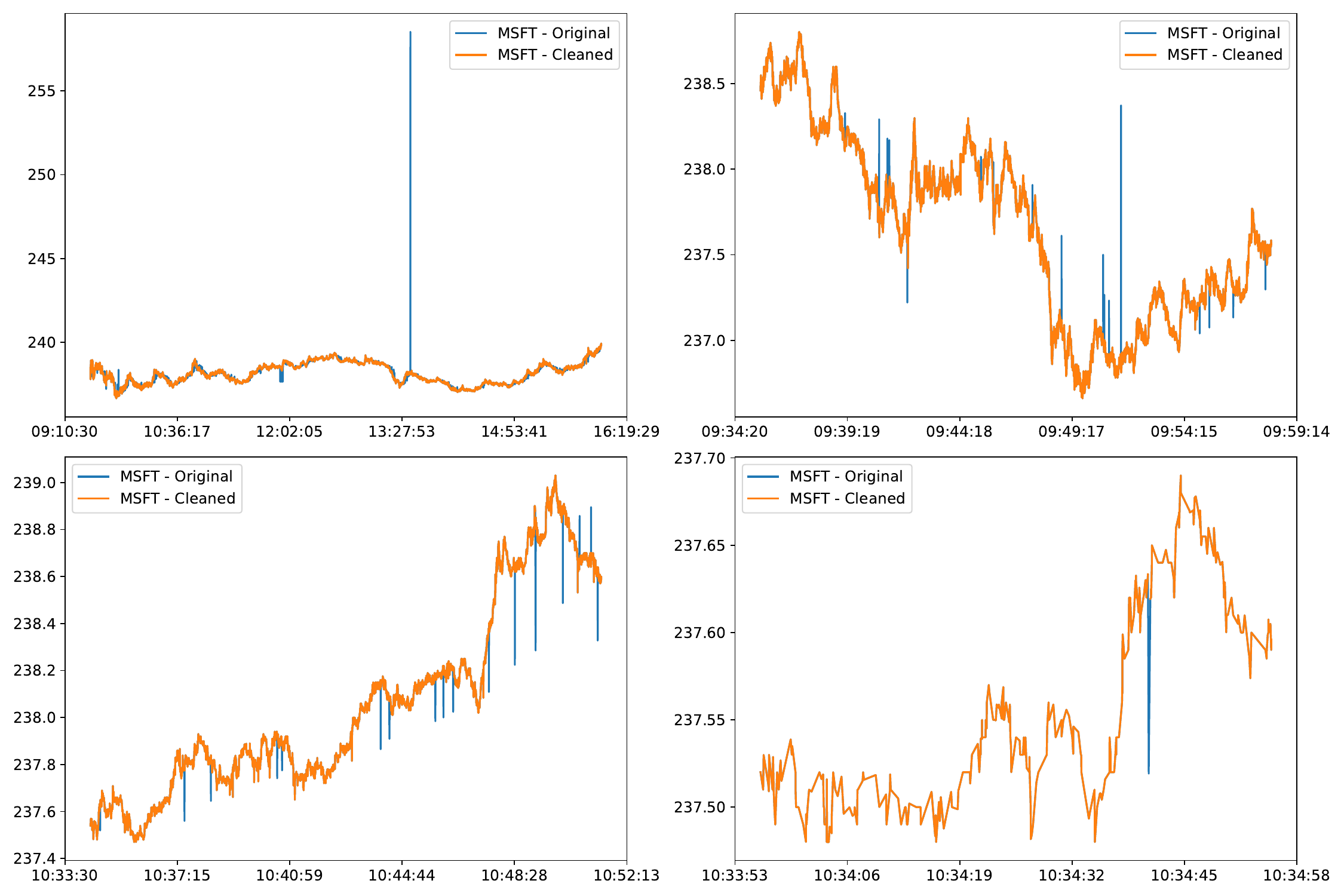}
  \caption{Original and cleaned daily price series for MSFT on December 30, 2022, with zoomed-in views of selected sub-periods to highlight the impact of outliers.}
\label{fig:MSFT_outliers}
\end{figure}

\begin{figure}[ht]
\centering
  \includegraphics[width=\linewidth]{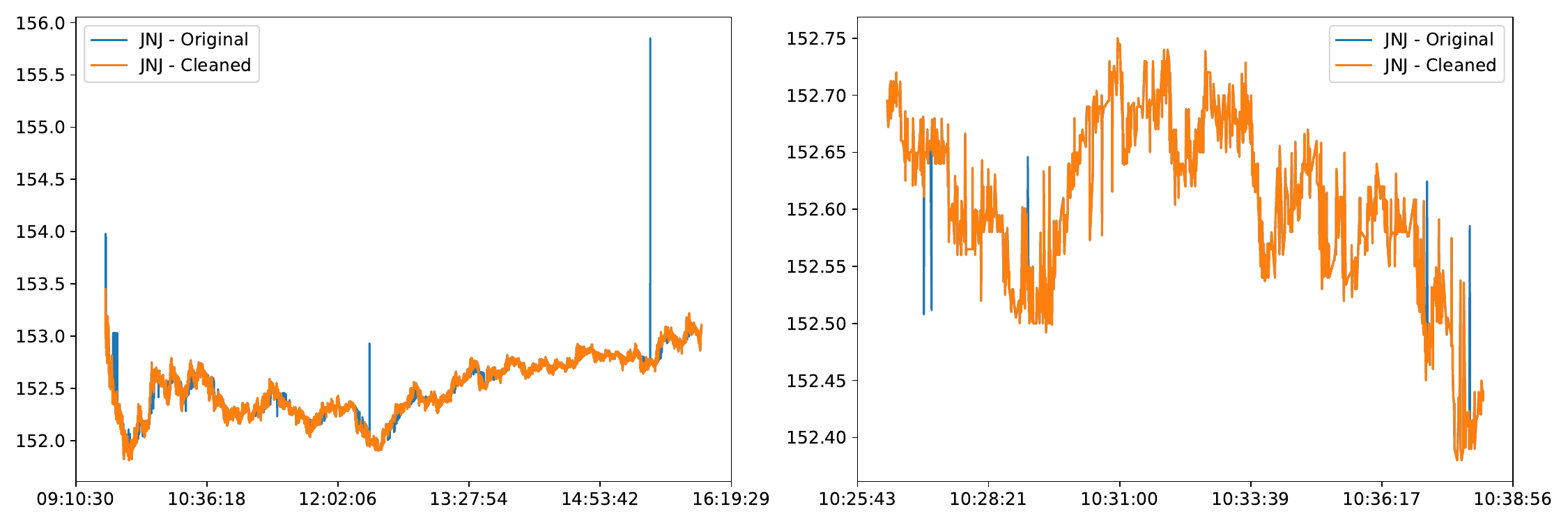}
  \caption{Original and cleaned daily price series for JNJ on November 20, 2024, with zoomed-in views of selected sub-periods to highlight the impact of outliers.}
\label{fig:JNJ_outliers}
\end{figure}

\begin{figure}[ht]
\centering
  \includegraphics[width=\linewidth]{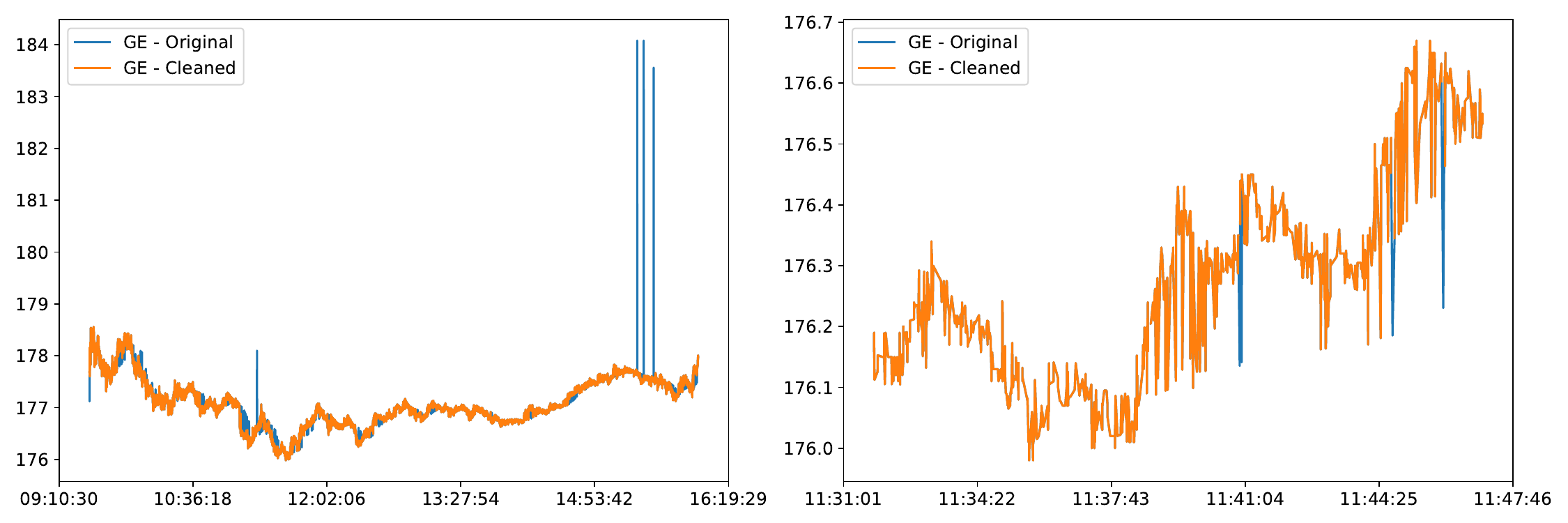}
  \caption{Original and cleaned daily price series for GE on November 20, 2024, with zoomed-in views of selected sub-periods to highlight the impact of outliers.}
\label{fig:GE_outliers}
\end{figure}

\begin{figure}[ht]
\centering
  \includegraphics[width=\linewidth]{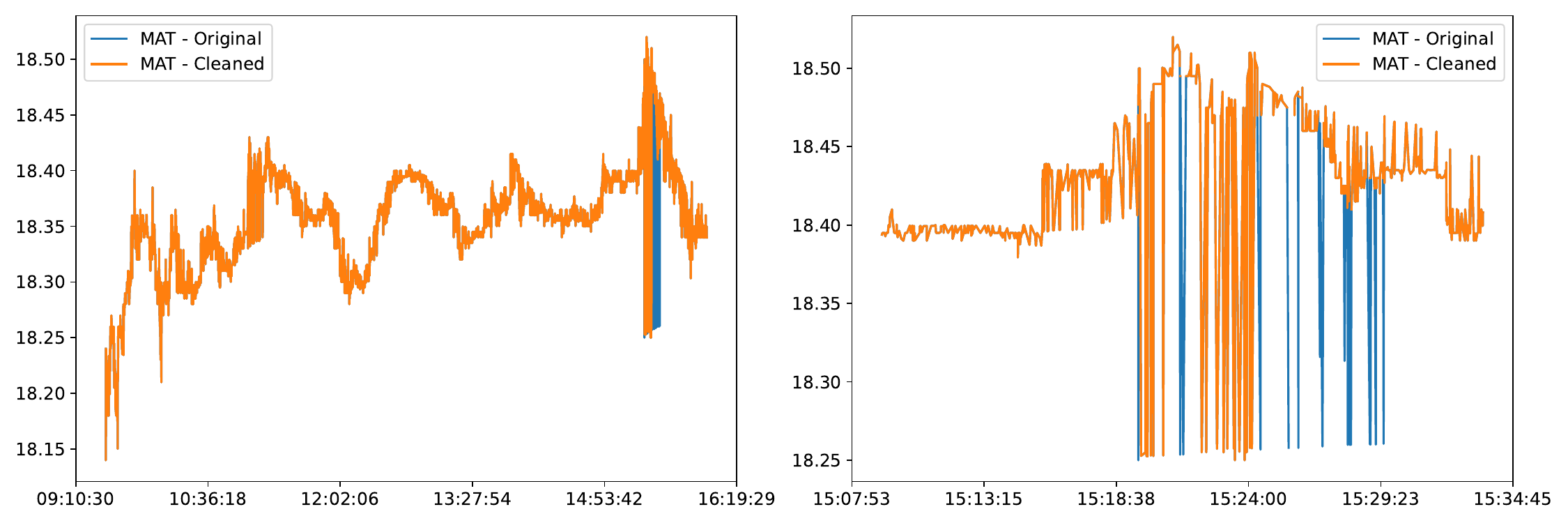}
  \caption{Original and cleaned daily price series for MAT on November 20, 2024, with zoomed-in views of selected sub-periods to highlight the impact of outliers.}
\label{fig:MAT_outliers}
\end{figure}

\begin{figure}[ht]
\centering
  \includegraphics[width=\linewidth]{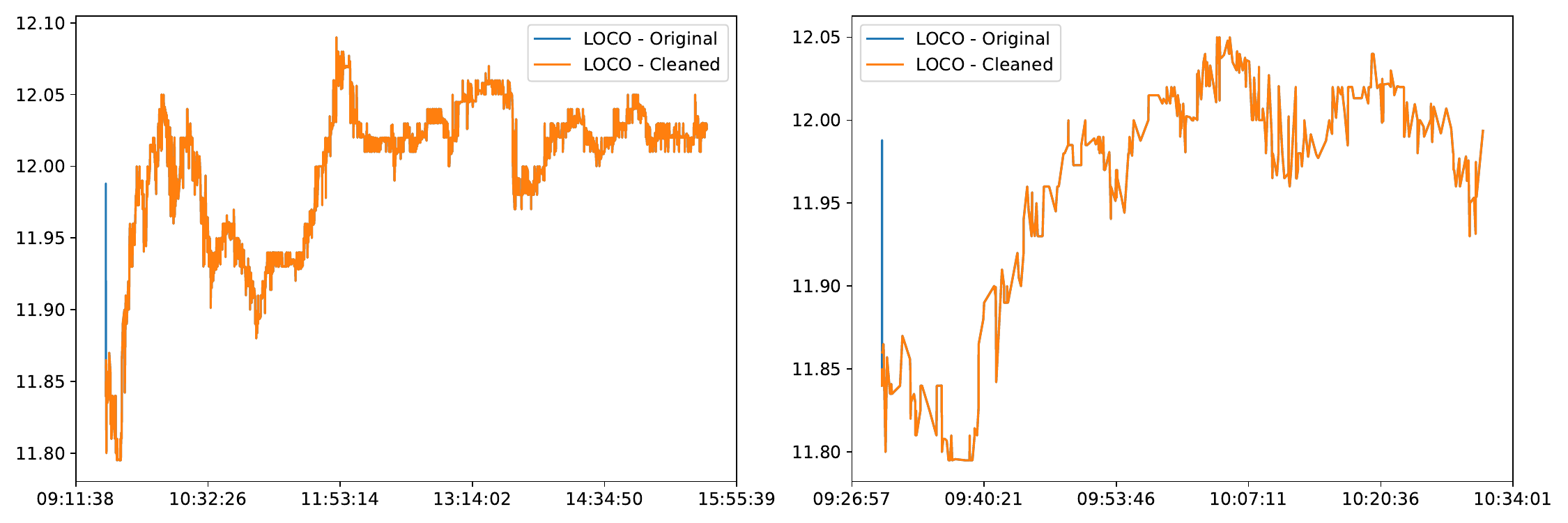}
  \caption{Original and cleaned daily price series for LOCO on November 05, 2024, with zoomed-in views of selected sub-periods to highlight the impact of outliers.}
\label{fig:LOCO_outliers}
\end{figure}

To illustrate the effect of the procedure, we applied it to five representative stocks: Microsoft (MSFT), Johnson \& Johnson (JNJ), General Electric (GE), Mattel (MAT), and El Pollo Loco (LOCO), using the fixed configuration $k=60$, $\delta=0.1$, and $\gamma=0.06$. The number of detected outliers for these assets is reported in \autoref{tab:num_outliers}, while \Crefrange{fig:MSFT_outliers}{fig:LOCO_outliers} visually compare the original and cleaned price series, showing how the filter effectively removes spurious spikes while preserving genuine market dynamics.

 \begin{table}[h!]
\centering
\begin{tabular}{lrrrr}
\toprule
\textbf{Symbol} & \textbf{Mkt. Cap. (USD)} & \textbf{Date} &\textbf{N. Obs.}&\textbf{N. Outliers} \\
\midrule
MSFT &  3.07 T& 2022-12-30 & 140,653&  434 \\
JNJ& 374.38 B& 2024-11-20 & 46,069& 191\\
GE &  193.40 B & 2024-11-20 & 29,269& 181\\
MAT&  6,18 B & 2024-11-20 &  13,823& 20 \\
LOCO&  0.36 B & 2024-11-05 & 3,400& 1\\
\bottomrule
\end{tabular}
\caption{Number of detected outliers for stocks of different market capitalizations, analyzed considering $k=60$, $\delta=0.1$, and $\gamma=0.06$.
}
\label{tab:num_outliers}
\end{table}

\begin{figure}[htbp]
    \centering
    \includegraphics[width=0.9\linewidth]{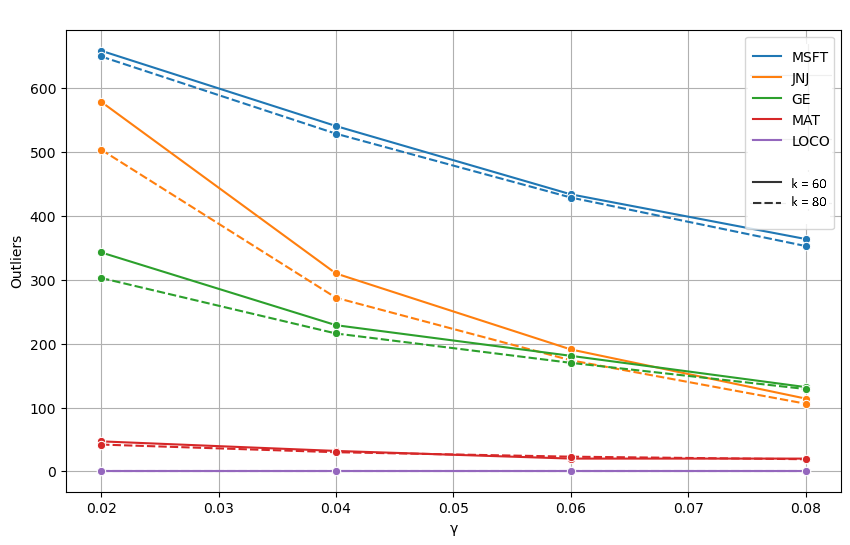}
    \caption{Number of outliers detected for different stocks as a function of the parameter $\gamma$, for two values of $k$ (60 and 80). The plot highlights how the number of outliers decreases as $\gamma$ increases, and compares stocks with different levels of market capitalization.}
    \label{fig:BG_sensitivity}
\end{figure}

On the same stocks and trading days, we conducted a sensitivity analysis by varying $k$ (i.e., 60 and 80) and $\gamma$ over a range of values to examine how the number of identified outliers changes under different parameter settings (\autoref{fig:BG_sensitivity}). The results show that the parameter $\gamma$ predominantly controls the aggressiveness of the filter, as reflected by the substantial variation in the number of detected outliers, while changes in $k$ have a comparatively limited impact on the overall removal rate.

Based on these findings, we applied across all stocks of the VOLARE library the parameter configuration $k=60$, $\delta=0.1$, and $\gamma=0.06$. The overall impact of the outlier detection on the complete dataset (2015-01-02 to 2026-01-30) is summarized in \autoref{tab:outliers_VOLARE_summary}. 
The procedure removes a modest proportion of observations while improving the quality and reliability of the high-frequency price series used for realized variance calculations.

\begin{table}[ht]
\centering
\resizebox{.8\textwidth}{!}{%
\begin{tabular}{l l r r r r r }
\toprule
\multicolumn{7}{c}{\textbf{Outlier Detection}} \\
\midrule
\textbf{Symbol} & \textbf{Days} & \textbf{Avg} & \textbf{Avg (\%)} & \textbf{Min} & \textbf{Max} & \textbf{Median} \\
\midrule
AAPL	&	2786	&	698.03	&	0.24	&	0	&	14247	&	289.5	\\
ADBE	&	2786	&	126.38	&	0.29	&	0	&	2020	&	49	\\
AMD	&	2786	&	440.89	&	0.18	&	0	&	10178	&	97	\\
AMGN	&	2786	&	53.75	&	0.21	&	0	&	1197	&	24	\\
AMZN	&	2786	&	791.85	&	0.48	&	4	&	10975	&	370	\\
AXP	&	2786	&	60.50	&	0.18	&	0	&	1283	&	16	\\
BA	&	2786	&	134.53	&	0.22	&	0	&	2575	&	52	\\
CAT	&	2786	&	75.83	&	0.23	&	0	&	1674	&	26	\\
CRM	&	2786	&	158.45	&	0.26	&	0	&	3416	&	49	\\
CSCO	&	2786	&	39.41	&	0.06	&	0	&	1915	&	16	\\
CVX	&	2786	&	113.18	&	0.18	&	0	&	2925	&	41	\\
DIS	&	2786	&	110.61	&	0.16	&	0	&	2574	&	54	\\
GE	&	2786	&	78.80	&	0.17	&	0	&	2692	&	15	\\
GOOGL	&	2786	&	525.10	&	0.49	&	3	&	11254	&	165.5	\\
GS	&	2786	&	83.43	&	0.25	&	0	&	1876	&	25	\\
HD	&	2786	&	112.86	&	0.27	&	0	&	2460	&	50	\\
HON	&	2786	&	41.84	&	0.15	&	0	&	1221	&	16	\\
IBM	&	2786	&	74.83	&	0.18	&	0	&	2316	&	11	\\
JNJ	&	2786	&	100.86	&	0.17	&	0	&	1835	&	43	\\
JPM	&	2786	&	162.75	&	0.21	&	0	&	3247	&	56	\\
KO	&	2786	&	37.34	&	0.05	&	0	&	1459	&	13	\\
MCD	&	2786	&	75.50	&	0.21	&	0	&	1715	&	29	\\
META	&	2786	&	545.47	&	0.33	&	0	&	17873	&	146.5	\\
MMM	&	2786	&	44.46	&	0.15	&	0	&	1187	&	18	\\
MRK	&	2786	&	71.21	&	0.13	&	0	&	1940	&	24	\\
MSFT	&	2786	&	645.50	&	0.31	&	0	&	15051	&	239	\\
NFLX	&	2786	&	271.17	&	0.34	&	0	&	8358	&	104.5	\\
NKE	&	2786	&	82.81	&	0.14	&	0	&	2826	&	33	\\
NVDA	&	2786	&	1023.77	&	0.21	&	0	&	18109	&	157	\\
ORCL	&	2786	&	220.48	&	0.20	&	0	&	9097	&	15	\\
PG	&	2786	&	80.65	&	0.16	&	0	&	1705	&	32	\\
PM	&	2786	&	39.86	&	0.12	&	0	&	1469	&	10	\\
SHW	&	2786	&	37.43	&	0.29	&	0	&	676	&	16	\\
TRV	&	2786	&	16.92	&	0.11	&	0	&	370	&	5	\\
TSLA	&	2786	&	1053.67	&	0.24	&	0	&	38548	&	311	\\
UNH	&	2786	&	186.01	&	0.29	&	0	&	8559	&	43	\\
V	&	2786	&	153.02	&	0.25	&	0	&	3329	&	57	\\
VZ	&	2786	&	26.63	&	0.04	&	0	&	1094	&	14	\\
WMT	&	2786	&	107.12	&	0.14	&	0	&	3310	&	32	\\
XOM	&	2786	&	111.00	&	0.14	&	0	&	2727	&	36.5	\\
\bottomrule
\end{tabular}%
}
\caption{Daily outlier detection summary by symbol (2015-01-02 to 2026-01-30). The table reports the daily average number of outliers and the average daily percentage of records identified as outliers during regular trading hours. Day-to-day variability is captured via minimum, maximum, median, and standard deviation. Outliers are detected using the \citet{brownlees2006financial} procedure with $k=60$, $\delta= 0.1$, $\gamma=0.06$.}
\label{tab:outliers_VOLARE_summary}
\end{table}

\clearpage

\subsection{From Tick Data to Regular Intervals}\label{sec:sampling}

To address the challenge of the unevenly spaced time of financial tick data, it is necessary to transform the series into a regular interval series. This facilitates analysis and helps in reducing the impact of market microstructure noise \citep{genccay2001introduction, hansen2006realized, bandi2008microstructure}. 
In the VOLARE library we consider a clock-based sampling approach, which organizes price data into fixed time intervals, such as one-minute or five-minute windows. Among various approaches proposed in the literature, we adopt the previous-tick approach, which selects the last observed price before or at the end of each fixed-time interval. If no price is recorded within a given interval, the price from the previous interval is carried forward.\citep{barndorff2009realized}. The resulting series spans from 9:30:00.000 (the first recorded price after the market opens) to 16:05:00.000 (by convention the last price), with a recorded price for each fixed-time interval. Each interval is labeled using the right endpoint, ensuring that the upper boundary of the interval is included in the aggregation.


\section{The VOLARE Back-end: Realized Measures} \label{sec:variance}

\subsection{Univariate Series}\label{sec:univariate}

The VOLARE library provides several indicators of market activity, such as well-established realized variance measures, widely popular in high-frequency financial econometrics, and realized quarticity (\autoref{tab:rv_measures}). In addition to these, the dataset provides the ticker symbol, date, high price ($H$), low price ($L$), open price ($O$), close price ($C$)\footnote{We select $H$ and $L$ values from all tick data of the day, including odd lots. For $O$ and $C$ prices, we use the first and last available prices of the day's series, excluding odd lots.}, and, for stocks, total number of trades ($N$), and traded volume (VOL). 

All realized measures are computed using returns on equally-spaced intervals, typically at 1-minute or 5-minute frequencies. For the realized kernel, we use returns sampled every second, as this estimator is designed to effectively mitigate market microstructure noise. 
Together, these estimators capture different dimensions of intraday return variability, such as overall variation, jump components, and asymmetric features of high-frequency data.

\begin{table}[h!]
\centering
\resizebox{\textwidth}{!}{
\begin{tabular}{llll}
\toprule
\textbf{Measure} & \textbf{Reference} & \textbf{Acronym} & \textbf{Subsampling} \\
\midrule
Parkinson Range & \citet{parkinson1980extreme} & pr & - \\ 
Garman-Klass Range & \citet{garman1980estimation} & gkr & - \\ 
Realized Range &  \citet{martens2007measuring} & rr5 & -  \\ 
  & \citet{christensen2007realized} \\ 
Realized Variance & \citet{andersen1998answering} &  rv1, rv5 & rv5\_ss \\ 
  & \citet{andersen2003modeling}  \\ 
Realized Quarticity & \citet{barndorff2002econometric} & rq1, rq5 &rq5\_ss\\
Bipower Variation & \citet{barndorff2004power} & bv1, bv5 & bv5\_ss\\
Realized Semivariances (pos e neg) & \citet{barndorff2010semivariance} &  rsp1, rsp5, rsn1, rsn5 & rsp5\_ss, rsn5\_ss \\ 
Median Realized Variance & \citet{andersen2012jump} &  medrv1, medrv5  & medrv5\_ss \\ 
Minimum Realized Variance &  \citet{andersen2012jump} &   minrv1, minrv5  & minrv5\_ss \\ 
Realized Kernel & \citet{barndorff2009realized} &  rk  & - \\ 
\hline
\end{tabular}} 
\caption{Realized measures available in the VOLARE library. The number following the acronym indicates the time interval used to sample the series at equally spaced times (e.g., rv1 refers to the realized variance sampling prices at 1-minute intervals). For each subsampled realized measure (suffix $ss$), we evaluate it using 5 subsample sets, each shifted by 1 minute.}
\label{tab:rv_measures}
\end{table}

{To ensure estimation reliability, realized measures are computed only for trading days with at least 40 intraday observations or with trading activity spanning less than two hours.
}

We also provide subsampled versions of several estimators \citep{zhou1996high, zhang2005tale}. The subsampling technique computes realized measures on multiple, slightly shifted sampling grids and averages the resulting estimates. In practice, we start from a fixed sampling frequency (every 5 minutes in our library) and we create several subsamples by shifting the grid by a fraction of that interval (in our case, one minute). Each realized measure is computed separately for each subsample, and the final estimator is obtained by averaging the subsample results. Compared to the corresponding estimator computed on a single, fixed sampling grid, this procedure improves estimator efficiency by reducing both variance and bias, while balancing the trade-off between high-frequency sampling (more but noisier information), and low-frequency sampling (less noise but it relies on fewer observations).

\subsubsection{Parkinson Range (pr)}
The Parkinson variance range \citep{parkinson1980extreme} 
is calculated as 
$$pr = \frac{1}{4 \ln(2)} \ln\left(\frac{H}{L}\right)^2.$$

\subsubsection{Garman-Klass Range (gkr)}
The Garman-Klass variance range \citep{garman1980estimation} 
is computed as 
$$gkr = \frac{1}{2} \ln\left(\frac{H}{L}\right)^2 - (2\ln(2) - 1) \ln\left(\frac{C}{O}\right)^2.$$

\subsubsection{Realized Range (rr)}
The Realized Range, introduced by \citet{martens2007measuring} and \citet{christensen2007realized}, is obtained as
$$rr = \frac{1}{4 \ln(2)} \sum_{i=1}^{m} \ln\left(\frac{H_{i}}{L_{i}}\right)^2,$$
where $H_i$ and $L_i$ are the highest and the lowest prices, respectively, of the $i$-th intraday interval, and $m$ is the number of such intervals. 

\subsubsection{Realized Variance (rv)} 
\label{rv}
The plain vanilla realized variance \citep{andersen1998answering, andersen2003modeling}
is defined as
$$rv = \sum_{i=1}^{m} r_{i}^2,$$
where $r_{i} = \ln(p_i / p_{i-1})$ denotes the return of the $i$-th interval. 
{Under the assumption of no jumps, it provides a consistent estimator of the integrated variance $IV = \int_0^T \sigma_u^2 du$ as the sampling frequency increases.}

\subsubsection{Realized Quarticity (rq)}\label{sec:quarticity}
Following \citet{barndorff2002econometric}, the realized quarticity is defined as
$$rq = \frac{m}{3}\sum_{i=1}^{m} r_{i}^4 \, .$$
The factor $m/3$ acts as a normalization constant that adjusts for the sampling frequency and ensures that $rq$ is an unbiased estimator of the integrated quarticity $IQ = \int_0^T \sigma_u^4 du$.

\subsubsection{Bipower Variation (bv)}
The Bipower Variation \citep{barndorff2004power} is a robust volatility estimator that captures the continuous component of price movements. It is defined as
$$bv = \frac{\pi}{2} \sum_{i=2}^{m} |r_{i}||r_{i-1}|.$$


\subsubsection{Realized Semivariances (rsp, rsn)}
Realized semivariances \citep{barndorff2010semivariance} decompose price volatility into its upside and downside components, providing critical insight into the asymmetric nature of financial market fluctuations. The downside (upside) realized semivariance $rs^-$ ($rs^+$)
captures the variance contribution from negative (positive) returns, as defined by
$$rs^- = \sum_{i=1}^{m} r_{i}^2 \; I_{[r_i <0]} 
\qquad
rs^+ = \sum_{i=1}^{m} r_{i}^2 \; I_{[r_i >0]}.$$

The indicators $I_{[r_i <0]}$ and $I_{[r_{i} >0]}$
ensure that only negative or positive returns, respectively, contribute to each measure. The two components sum to the total realized variance, i.e. 
$rv = rs^+ + rs^-$. 

\subsubsection{Median Realized Variance (medrv)}
Median realized variance \citep{andersen2012jump} provides a robust alternative to standard realized variance estimators by effectively filtering out jumps in the price process. It uses the median of three consecutive squared returns to minimize the impact of outliers and price discontinuities, making it particularly valuable for distinguishing between continuous volatility components and jump variations. Formally, the median realized variance is computed as
$$\text{medRV} = \frac{\pi}{6-4\sqrt{3}+\pi} \cdot \frac{m}{m-2} \sum_{i=2}^{m-1} \text{med}(|r_{i-1}|, |r_{i}|, |r_{i+1}|)^2,$$
where $\text{med}(\cdot)$ denotes the median operator. 
The scaling factor $\frac{\pi}{6-4\sqrt{3}+\pi}$ ensures consistency toward the integrated variance, while the $\frac{m}{m-2}$ term provides a small-sample bias adjustment.

\subsubsection{Minimum Realized Variance (minrv)}

Minimum realized variance \citep{andersen2012jump} is another robust estimator of realized variance designed to mitigate the impact of jumps. It is based on the minimum of two adjacent squared returns, effectively filtering out large price movements that might represent jumps, rather than continuous volatility. The estimator is defined as
$$\text{minRV} = \frac{\pi}{\pi-2} \cdot \frac{m}{m-1} \sum_{i=2}^{m} \min(|r_{i-1}|, |r_{i}|)^2,$$
where the scaling factor $\frac{\pi}{\pi-2}$ ensures consistency toward the integrated variance, while $\frac{m}{m-1}$ provides a finite-sample bias adjustment.

\subsubsection{Realized Kernel (rk)}
The realized kernel variance estimator, introduced by \cite{barndorff2008designing, barndorff2009realized}, provides a consistent and efficient way to estimate the integrated variance in presence of market microstructure noise. Denoting with $r_i$ a generic high-frequency return, the realized kernel estimator is
$$ rk = \gamma_0 + 2\sum_{h=1}^{H} k\left(\frac{h}{H+1}\right) \gamma_h,$$
where $\gamma_0$ is the {plain vanilla} realized variance, and $\gamma_h = \sum_{i=h+1}^{m} r_i r_{i-h}$ are the autocovariances of order $h, \ h\geq 0$,
$k(\cdot)$ is a kernel function, and $H$ is the bandwidth parameter.
The choice of kernel function $k(\cdot)$ is crucial for the estimator's properties. In our calculations, we employ the Parzen kernel due to its excellent balance between efficiency and robustness.\footnote{ \citet{barndorff2009realized} show that the Parzen kernel fulfills all necessary conditions to guarantee the positivity of the estimator and consistent estimation of integrated variance in the presence of microstructure noise.} Its expression is:
$$
k(x) = \begin{cases} 
1 - 6x^2 + 6x^3 & 0 \leq x \leq 1/2 \\
2(1 - x)^3 & 1/2 < x \leq 1 \\
0 & x > 1.
\end{cases}$$

We select the bandwidth as
$$H^* = c^* \xi^{4/5} m^{3/5},$$
where
$$c^* = \left\{ \frac{k''(0)^2}{k_{0,0}} \right\}^{1/5},$$ with $k''(0)^2$ representing the squared second derivative of the kernel function evaluated at zero, and $k_{0,0} = \|k\|_2^2 = \int_0^1 k(x)^2 dx$ is the squared $L^2$-norm of the kernel function over its support.\footnote{For our Parzen kernel, $c^* = ((12)^2 / 0.269)^{1/5} = 3.5134$.}  

The term $\xi^2$ is a function of the variance $\omega^2$ of the market microstructure noise  contaminating observed prices, and the integrated quarticity:
$$\xi^2 = \frac{{\omega}^2}{({ T \cdot IQ})^{1/2}} \approx \frac{{\omega}^2}{{{IV}}},$$
where we justify the approximation $({ T \cdot IQ})^{1/2} \approx IV$ on the grounds that with moderate volatility variation, $IQ \approx IV^2/T$.\footnote{\citet{barndorff2009realized} note that this simplification is reasonable in practical applications and makes bandwidth selection more stable, as estimating $IV$ is significantly simpler than estimating IQ.}

\medskip

We estimate the numerator and denominator of $\xi^2$ as follows.

\paragraph{Noise variance. }
Following \citet[p. C5]{barndorff2009realized}, building on \citet{zhang2005tale} and \cite{bandi2008microstructure}, we estimate the noise variance as realized variance computed using 2-minute returns, $RV_{\text{2'}}$, divided by the number of non-zero returns, $n$:
$$\hat{\omega}^2 = \frac{RV_{\text{2'}}}{2 n}$$

\paragraph{Integrated variance.} 
Following \cite[p. C4]{barndorff2009realized}, we compute sparse realized variances based on 20-minute returns, on $p$ windows, each adding one second and dropping one second (hence $p=20 \times 60=1200$ subsamples). More specifically, if $\text{RV}^{(j)}$ represents the realized variance computed from the $j$-th subsample, then the estimated $IV$ is the average of the $p$ sparse realized variances
$$\widehat{\text{IV}} = \text{RV}_{\text{sparse}} = \frac{1}{p} \sum_{j=1}^{p} \text{RV}^{(j)}.$$

\paragraph{End effect.} 
Boundary treatment plays a crucial role in realized kernel estimation, as bias tends to accumulate at the edges of the sampling period. \citet{barndorff2009realized} suggest using jittering techniques and endpoint corrections to address boundary effects. These methods involve adjusting the first and last observations before sampling prices, using the average of the two preceding and two succeeding prices to mitigate any edge-related bias.


\subsection{Multivariate Series} \label{sec:covariance}
A key challenge in estimating realized covariance from high-frequency financial data is the lack of synchronicity between price observations across assets. There are two popular approaches to address this issue: (i) the ``previous-tick'' method, which uses a regularly spaced time grid and set the price of each asset to the last observed before each grid time \citep{zhang2011estimating}; (ii) the refresh time scheme, which considers irregularly spaced data where each time is the instant at which all assets have been traded at least once from the previous time \citep{barndorff2011multivariate}.

We adopt the previous-tick method, using the most recent transaction price available before each time of a 1-minute grid. This procedure yields a synchronized, regularly spaced return series across all assets, allowing the use of standard realized covariance estimators.

{When applying the previous-tick synchronization to futures contracts, an additional complication arises from heterogeneous trading schedules across exchanges. Unlike equities or exchange rates that typically share common trading hours, futures contracts trade on different exchanges (NYMEX, COMEX, CBOT, CME) with varying session structures.\\
On days when trading hours differ substantially across contracts, for instance, when corn trades only during evening hours (19:00–23:59 ET) while other futures operate continuously, we assign zero returns to the asset with limited hours during non-overlapping periods. This may introduce a downward bias in the estimated covariances between contracts with misaligned trading schedules, particularly affecting pairs involving CBOT agricultural contracts.
These non-synchronous trading episodes occur on a limited number of days in our sample, primarily during exchange-specific holidays or restricted trading sessions.}

Considering $N$ assets observed at $m$ equally-spaced intervals during a single trading day, we denote as $r_{i,k}$ the return of the $i$-th asset during the $k$-th intraday interval ($i = 1, 2, \ldots, N$, $k = 1, 2, \ldots, m$), and as $\mathbf{r}_k = [r_{1,k}, r_{2,k}, \ldots, r_{N,k}]$ the vector of returns in the interval. 
We provide realized covariance matrices for groups of assets categorized by type (e.g., stocks, exchange rates, and futures). 

\autoref{tab:rcov_measures} summarizes the covariance measures calculated.

\begin{table}[h!]
\centering
\resizebox{\textwidth}{!}{
\begin{tabular}{lll}
\toprule
\textbf{Measure} & \textbf{Reference} & \textbf{Acronym}  \\
\midrule
Realized Covariance & \citet{barndorff2004covariation} &  rcov \\ 
Bipower Covariance & \citet{barndorff2004measuring} & rbpcov\\ 
Realized Semicovariances & \citet{bollerslev2020realized} &  rscov\_p, rscov\_n,  \\ 
(pos, neg, mixed) & &   rscov\_mp, rscov\_mn \\ 

\bottomrule
\end{tabular}} 
\caption{Covariance measures provided in the VOLARE library. For all estimates, we  to obtain equally spaced, 1-minute data using the ``previous-tick'' method.}
\label{tab:rcov_measures}
\end{table}

\subsubsection{Realized Covariance (rcov)}

Following \citet{barndorff2004covariation}, the realized covariance between two assets $i$ and $j$ over a single trading day is defined as the sum of the products of their synchronized intraday returns:

\begin{equation}
\text{rc}_{i,j} = \sum_{k=1}^{m} r_{i,k} \cdot r_{j,k}.
\end{equation}

For a set of $N$ assets, the realized covariance matrix aggregates all pairwise covariances into the $N \times N$ matrix
\begin{equation}
\mathbf{rc} = \sum_{k=1}^{m} \mathbf{r}_{k} \mathbf{r}_{k}'.
\end{equation}


\subsubsection{Realized Bipower Covariance (rbpcov)}
The realized bipower covariance is a robust estimator of integrated covariance that is less sensitive to jumps compared to the standard realized covariance. Following \citet{barndorff2004measuring}, the bipower covariance between assets $i$ and $j$ is defined as:


$$\text{bc}_{i,j} = \frac{\mu_1^{-2}}{4}  \sum_{k=2}^{m} \biggl( \left| r_{i,k-1} + r_{j,k-1} \right| \cdot \left| r_{i,k} + r_{j,k} \right| - \left| r_{i,k-1} - r_{j,k-1} \right| \cdot \left| r_{i,k} - r_{j,k} \right| \biggr).$$

with $\mu_1 = \sqrt{2/\pi} \approx 0.7979$ being the first moment of the absolute value of a standard normal random variable
When $i = j$, this formulation simplifies to the standard realized bipower variation of asset $i$.

\subsubsection{Realized Semicovariances (rscov)}
Realized semicovariances extend the traditional covariance framework by decomposing co-movements between assets based on the signs of their returns, thereby uncovering asymmetric dependence structures that vary across market conditions.

Following \citet{bollerslev2020realized}, we decompose the realized covariance matrix into four components based on the signs of the underlying high-frequency returns. We first define the signed return vectors as:
\begin{align}
\mathbf{r}_k^+ &= \mathbf{r}_k \odot \mathbf{1}\{\mathbf{r}_k > \mathbf{0}\}, \nonumber \\
\mathbf{r}_k^- &= \mathbf{r}_k \odot \mathbf{1}\{\mathbf{r}_k \leq \mathbf{0}\}, \nonumber
\end{align}
where $\odot$ denotes element-wise multiplication and $\mathbf{1}\{\cdot\}$ is the indicator function applied element-wise. The four daily realized semicovariance matrices are then defined as:
\begin{align}\nonumber
\mathbf{rs^{++}} &= \sum_{k=1}^{m} \mathbf{r}_k^+ (\mathbf{r}_k^+)', \quad \text{(concordant positive)} \nonumber\\
\mathbf{rs^{--}} &= \sum_{k=1}^{m} \mathbf{r}_k^- (\mathbf{r}_k^-)', \quad \text{(concordant negative)} \nonumber\\
\mathbf{rs^{+-}} &= \sum_{k=1}^{m} \mathbf{r}_k^+ (\mathbf{r}_k^-)', \quad \text{(mixed/discordant positive-negative)} \nonumber\\
\mathbf{rs^{-+}} &= \sum_{k=1}^{m} \mathbf{r}_k^- (\mathbf{r}_k^+)', \quad \text{(mixed/discordant negative-positive)} \nonumber
\end{align}

The positive semicovariance $\mathbf{rs^{++}}$ captures co-movement during periods  when assets experience positive returns, reflecting synchronized upward movements. The negative semicovariance matrix $\mathbf{rs^{--}}$ measures co-movement during joint downward movements, which is particularly relevant for risk management and portfolio diversification analysis. The mixed semicovariance matrices $\mathbf{rs^{+-}}$ and $\mathbf{rs^{-+}}$ capture the co-movement when assets move in opposite directions.

By construction, the standard realized covariance matrix can be decomposed as
\begin{equation}
\mathbf{rc} = \mathbf{rs^{++}} + \mathbf{rs^{--}} + \mathbf{rs^{+-}} + \mathbf{rs^{-+}}.
\end{equation}
Notice that while $\mathbf{rs^{++}}$ and $\mathbf{rs^{--}}$ are symmetric and positive semidefinite, the mixed semicovariance matrices $\mathbf{rs^{+-}}$ and $\mathbf{rs^{-+}}$ are generally asymmetric, with zero diagonal elements by construction.

\section{The VOLARE Front-end: Architecture and User Experience} \label{sec:frontend} 
The VOLARE platform integrates a structured back-end infrastructure, with a user-centered front-end interface.

The back-end forms the operational core of the system, automating the entire data lifecycle: from the initial download of high-frequency market data, via the Kibot API, to the documented operations leading to the computation of realized variance and covariance measures. All raw and processed data are stored in a MinIO-based data lake, while final statistical outputs are housed in a PostgreSQL relational database, ensuring full traceability, coherence, and streamlined data retrieval. Efficient storage and indexing enable fast query response times, while the infrastructure setup involves  automated monthly updates and regular monitored backups, to guarantee continuity and rapid recovery in case of failures. Future extensions across assets and classes of assets would maintain the same design.


The VOLARE front-end, developed in React, is designed to provide users with an intuitive interface for accessing and exploring high-frequency financial data. 

\subsection{Homepage}
The {Homepage} serves as the entry point to the platform, welcoming users with a presentation of VOLARE's purpose and capabilities. It provides an access to comprehensive realized volatility and covariance measures derived from high-frequency financial data, contextualizing the project within financial econometrics research.

From the Homepage, users can access the platform's two core functionalities through prominent action buttons: Download Data for direct dataset access and Visualize Data for the Interactive Dashboard. This dual-path design acknowledges different user workflows: some prefer downloading complete datasets for offline analysis, while others benefit from interactive web-based exploration tools.

Additional information pages are accessible through the navigation bar:

\paragraph{Documentation} The page provides comprehensive insight into the technical and methodological foundations of the platform, detailing the complete data processing pipeline from high-frequency tick data acquisition through the Kibot API to the computation of realized measures. Users can access detailed explanations of data sources and coverage, cleaning and preprocessing procedures, and statistical methodologies employed for computing realized variance and covariance measures.
\paragraph{About} The page presents VOLARE's foundational mission and motivations. VOLARE (VOLatility Archive for Realized Estimates) was created to address a fundamental challenge in financial research: making high-quality, research-grade realized volatility data accessible to financial researchers, quantitative analysts, and academics. The page articulates the platform's origin story, explaining how direct experience with the complexity of processing, cleaning, and extracting meaningful volatility estimates from ultra-high-frequency data drove the creation of VOLARE.
\paragraph{FAQ} The Frequently Asked Questions page addresses common questions regarding platform usage, data access, and methodology, offering concise answers about VOLARE's features.

\subsection{Download Data: Direct Dataset Access}
For users who prefer working by themselves outside the VOLARE platform, the Download section offers direct access to pre-organized data packages. This section provides a catalog of available datasets organized by asset type (stocks, exchange rates, futures), allowing users to download complete historical series in standardized formats. Each dataset comes with a README file that documents asset type and data type (variance or covariance), temporal coverage with date range and total number of records, the list of included assets with their symbols, and all available measures. This structured documentation ensures that users have complete metadata for proper integration into their analytical environments and suitable citation in research outputs.

\subsection{Visualize Data: The Core Analytical Interface}
Accessible via the \textbf{Visualize Data} button on the Homepage, the Interactive Dashboard is the heart of the VOLARE platform's exploratory analysis capabilities. The dashboard guides users through a sequential workflow with the following steps:
\begin{itemize}
\item \textbf{Asset Type Selection}: Users select the category of financial instruments to analyze (Stocks, Exchange Rates, or Futures)
\item \textbf{Symbol Selection}: Users can either search for a specific symbol or choose from the list of available assets within the selected category
\item \textbf{Date Range Selection}: Users can define the temporal scope of the analysis by selecting a custom date range. If no range is specified, the system will automatically set the date range to eleven months from the current date to ensure data availability (e.g., on November 13, 2025, the default range would be from October 13, 2024, to October 13, 2025)
\item \textbf{Measure Selection}: Users specify which realized volatility or covariance statistics to analyze from the available measures
\end{itemize}

Upon completing the workflow, the platform transitions to the Results page, where requested data is presented through an interactive visualization environment designed to facilitate both exploratory analysis and formal statistical investigation. The centerpiece is a dynamic graphing system with zoom and functionality for detailed temporal inspection and hover tooltips displaying values and dates.

Alongside the graphical display, a panel provides key statistical summary metrics computed in real-time: average volatility, volatility of volatility, average returns, average volume.

Moving beyond standard visualization, the platform provides advanced analytical features that enhance exploratory research. Users can add additional measures to the current view for multi-measure comparison without returning to the configuration workflow, enabling side-by-side evaluation of different volatility estimators. As presented in detail in \autoref{sec:volatilitymodels}, the platform offers integration with established volatility forecasting models including: HAR, HAR-Q, MEM(1,1), AMEM(1,1), and AMEM(2,1), allowing users to overlay model forecasts onto historical data and assess forecast accuracy through visual inspection and quantitative metrics.

Recognizing that users often need to continue analysis outside the web platform, VOLARE provides comprehensive export capabilities including high-resolution graph export of current visualizations and raw data downloads ready for import into statistical software.

\section{The VOLARE Front-end: Volatility Models} \label{sec:volatilitymodels}
Volatility modeling plays a central role in financial econometrics, with applications ranging from risk management and derivatives pricing to portfolio allocation. Over the years, a variety of models have been developed to capture the persistence and dynamics of financial market volatility (see, e.g., \citet{andersen1998answering, hansen2005forecast, bollerslev2016exploiting, cipollini2025multiplicative20y}). 

For realized volatility forecasting, two popular frameworks are the Heterogeneous Autoregressive (HAR) model \citep{corsi2009simple} and the Multiplicative Error Model (MEM) family \citep{engle2002new, engle2006multiple}. VOLARE currently supports HAR and HAR-Q specifications, as well as MEM, {Asymmetric MEM (AMEM)}, and AMEM(2,1) models. Users can estimate these models with a variety of realized variance measures, including Parkinson range, Garman–Klass, realized range, realized variance, bipower variation, median realized variance, minimum-variance, and realized kernel.

In practice, users may specify the sample period, asset ticker, and volatility measure of interest. The selected data is first visualized, after which a model can be chosen for estimation. Model estimation is performed only if the chosen range contains at least 750 observations (corresponding roughly to three years of daily data). Once this condition is met, VOLARE reports parameter estimates, together with standard diagnostic checks on the residuals.
Moreover, when enough data past the selected window for estimation are available,  one-step-ahead forecasts are generated for at least five and up to twenty-two out-of-sample periods. The graphical output overlays both the in-sample estimates and the out-of-sample forecasts on the volatility series. Confidence intervals are provided only for the forecasts, together with evaluation metrics, enabling users to assess predictive accuracy against observed outcomes.

Before delving into the details of model estimation, it is important to clarify some aspects regarding the scales adopted for estimation and forecasting. 
HAR models are estimated on daily variance measures, to preserve the linear structure of the model; in contrast, MEM family are estimated using annualized volatility, calculated as $\sqrt{252 \times \text{variance}} \times 100$. Accordingly, the reported parameter estimates refer to such scales.
In the graphical outputs, however, both model families display results on the annualized volatility scale to facilitate comparison.

The following sections introduce the models considered and describe the procedures employed for their estimation and forecasting. {An example of the analysis of the estimated volatility models across different realized variance measures for assets in different classes is provided in Appendix~\ref{app:vol_model_analysis}.}

\subsection{Heterogeneous Autoregressive Models}
\subsubsection{HAR}
The Heterogeneous Autoregressive (HAR) model, introduced by \citet{corsi2009simple}, makes the current variance measure to linearly depend on its own past, up to one month, by means of suitably restricted coefficients. 
Formally, the model can be specified as:
\begin{equation}
y_t = \omega + \alpha_d y_{t-1} + \alpha_w y_{t-1}^{(w)} + \alpha_m y_{t-1}^{(m)} + \varepsilon_t,
\end{equation}
where $y_t$ denotes the daily variance measure, $y_{t-1}^{(w)}$ and $y_{t-1}^{(m)}$ are the weekly and monthly components defined as
\begin{align}
y_{t-1} &  \quad \text{(daily component)} \nonumber \\
y^{(w)}_{t-1} &= \frac{1}{4}\sum_{i=2}^{5} y_{t-i} \quad \text{(weekly component)} \nonumber \\
y^{(m)}_{t-1} &= \frac{1}{16}\sum_{i=6}^{22} y_{t-i} \quad \text{(monthly component)} \nonumber
\end{align}
The weekly component starts from $i=2$ to avoid overlap with the daily term, while the monthly component is computed over days $t-6$ to $t-22$ so as to exclude the lags already used in the daily and weekly components. This construction ensures that each component (daily, weekly, and monthly) contributes distinct and non-redundant information to the model, thereby enhancing clarity and interpretability.

\subsubsection{HAR-Q}
The HAR-Q model  \citep{bollerslev2016exploiting} extends the HAR specification for realized variance measures, by incorporating realized quarticity (rq) as a proxy for the conditional variance of measurement errors. The specification is
\begin{equation}
y_t = \omega + \left(\alpha_d + \alpha_{Q} rq_{t-1}^{1/2}\right) y_{t-1} + \alpha_w y_{t-1}^{(w)} + \alpha_m y_{t-1}^{(m)} + \varepsilon_t,
\end{equation}
where $rq_{t-1}^{1/2}$ is the square root of the lagged realized quarticity (\ref{sec:quarticity}) 
demeaned for interpretability. The time-varying coefficient on the daily lag becomes $\alpha_{1,t} = \alpha_d + \alpha_{Q} rq_{t-1}^{1/2}$.

\subsubsection{Parameter Estimation Procedure}
Once the functional form of the model HAR or HAR-Q is specified, parameter estimation proceeds via Ordinary Least Squares (OLS).  

\begin{enumerate}
    \item \textbf{Estimation.} The parameters $\omega, \alpha_d, \alpha_w, \alpha_m$, and $\alpha_Q$ are obtained by regressing the variance measure on its daily, weekly, and monthly components, as well as the HAR-Q interaction term.  
    \item \textbf{Inference.} To ensure valid inference in the presence of possible heteroskedasticity and serial correlation, robust standard errors are computed using the Heteroskedasticity and Autocorrelation Consistent (HAC) covariance estimator of \citet{newey1987simple}. The automatic lag selection rule of \citet{newey1994automatic} is employed:
    $$\text{maxlags} = \left\lfloor  4 \cdot \left(\frac{T}{100}\right)^{2/9} \right\rfloor,$$
    where $T$ denotes the sample size.
    This procedure adjusts the covariance matrix in a data-driven manner, providing reliable standard errors without requiring explicit distributional assumptions.  
    \item \textbf{Residual diagnostic.} The adequacy of the model is further assessed by analyzing the residuals $\hat{\varepsilon}_t$. The Ljung–Box test \citep{ljung1978measure} is applied to residuals and squared residuals (5 lags) to detect potential autocorrelation and conditional heteroskedasticity. In addition, Engle’s ARCH test \citep{engle1982autoregressive} with 5 lags is used to formally test for remaining autoregressive conditional heteroskedasticity.
\end{enumerate}

\subsection{The Multiplicative Error Models (MEM)}

The Multiplicative Error Model (MEM), introduced by \citet{engle2002new} and extended by \citet{engle2006multiple}, is designed to model non-negative time series such as realized volatility, trading volume, or durations. The key idea is to decompose the observed variable into the product of a deterministic conditional mean and a non-negative innovation,
$$y_t = \mu_t \cdot \varepsilon_t, \quad \varepsilon_t \overset{\text{i.i.d}}{\sim} D^+(1, \sigma^2),$$
where $\mu_t$ denotes the conditional mean of $y_t$ given past information $\mathcal{I}_{t-1}$, and $\varepsilon_t$ is a non-negative innovation term. 
Common choices for the distribution $D^+$ include Gamma and Lognormal distributions.

\subsubsection{MEM(1,1)}
The conditional mean $\mu_t$ is typically modeled through a dynamic autoregressive structure analogous to a GARCH(1,1) process:
\begin{equation} \label{eq:mem}
\mu_t = \omega + \alpha_1 \, y_{t-1} + \beta_1 \, \mu_{t-1} .
\end{equation}
The parameters of the model must satisfy
\begin{itemize}
    \item $\omega > 0$, 
    \item $\alpha_1 \geq 0$, 
    \item $ 0 \leq \beta_1 < 1$, 
    \item  persistence $ = $ $\alpha_1 + \beta_1 < 1$
\end{itemize}
to guarantee non-negativity and covariance stationarity.

\subsubsection{Asymmetric MEM (AMEM)}

The Asymmetric Multiplicative Error Model (AMEM) \citep{engle2006multiple, cipollini2021realized} extends the MEM by allowing for asymmetric effects of past shocks on current volatility. Empirical evidence suggests that negative returns often trigger stronger volatility responses than positive returns of the same magnitude, a phenomenon referred to as the leverage effect.  

\paragraph{AMEM(1,1)} The specification of the conditional mean is:
\begin{equation} \label{eq:amem}
\mu_t = \omega + \alpha_1 \, y_{t-1} +\beta_1 \, \mu_{t-1}  + \gamma_1 \, y^{(-)}_{t-1},
\end{equation}
where $y^{(-)}_{t-1} = y_{t-1} \cdot \mathbb{I}(r_{t-1} < 0)$ captures the asymmetric component linked to negative returns, and $r_{t-1}$ denotes the return at time $t-1$.

To ensure positivity and covariance stationarity, the parameters are restricted as follows
\begin{itemize}
\item $\omega > 0$, 
\item $\alpha_1, \gamma_1 \geq 0$,
\item $ 0 \leq \beta_1 < 1$, 
 \item  persistence $ = $ $ \alpha_1 + \beta_1 +\frac{\gamma_1}{2} < 1$.  
\end{itemize}

\paragraph{AMEM(2,1)} This is an extension of the AMEM allowing for two lags of the dependent variable,
\begin{align} \label{eq:amem21}
\mu_t &= \omega + \alpha_1 \, y_{t-1} + \alpha_2 \, y_{t-2} + \beta_1 \, \mu_{t-1} + \gamma_1 \, y^{(-)}_{t-1}.
\end{align}
In this case, to ensure positivity and stationarity the following parameter restrictions are imposed
\begin{itemize}
\item $\omega > 0$,
\item $\alpha_1 \geq 0$, $\alpha_1 + \gamma_1 \geq 0$,
\item $0 \leq \beta_1 < 1$, 
\item $\alpha_2+ \alpha_1 \beta_1 \geq 0$, $\alpha_2+ ( \alpha_1 + \gamma_1) \beta_1 \geq 0$,  
\item persistence $ =  \alpha_1 + \alpha_2 +\beta_1 + \frac{\gamma_1}{2} < 1$. 
\end{itemize}

\subsubsection{Parameter Estimation Procedure}

Once the functional form of the conditional mean and the distributional assumption for the multiplicative error are specified, parameter estimation proceeds via quasi maximum likelihood (QML). The procedure is inherently recursive: for each candidate parameter vector $\theta$, the sequence of conditional means $\{\mu_t\}_{t=1}^T$ must be generated to evaluate the quasi log-likelihood. The steps are as follows.

\begin{enumerate}
    \item \textbf{Initialization.} 
     Initial values for the parameter vector $\theta$ are directly assigned so as to satisfy admissibility conditions (non-negativity and stationarity).
     The initial conditional mean is set to the unconditional mean implied by the model, e.g. $\mu_0 = \omega / (1 - \alpha_1 - \beta_1)$ in a MEM(1,1). Parameter values are initialized as follows: the intercept $\omega$ is chosen to match the sample mean $\bar{y}$ via $\omega = \bar{y} \cdot (1 - \text{persistence})$; the autoregressive coefficient is set to $\beta_1 \approx 0.6$; the innovation coefficient to $\alpha_1 \approx 0.1$; and, for extended models, additional parameters, such as $\alpha_2$ and $\gamma_1$, are initialized at zero.

    \item \textbf{Filtering step.} 
    For a given candidate $\theta$, the dynamic equation is used to recursively compute the path of conditional means. For example, in a MEM(1,1)
    $$\mu_t = \omega + \alpha_1 y_{t-1} + \beta_1 \mu_{t-1}.$$

    \item \textbf{Log-likelihood evaluation.} 
    Given $\mu_t$, the current multiplicative residual is $\varepsilon_t = y_t / \mu_t$, so that the quasi log-likelihood is proportional to
    $$\ell(\boldsymbol{\theta}) \propto  \sum_{t=1}^T \left[ \log\left(\varepsilon_t \right) - \varepsilon_t  +1\right] \,.$$
   
    \item \textbf{Likelihood maximization.} 
    Numerical optimization (implemented via a Sequential Least Squares Programming routine) iteratively updates $\theta$, recomputes $\{\mu_t\}$ and the quasi-log-likelihood, and adjusts parameters until convergence. Non-negativity constraints and stationarity restrictions are enforced throughout.

    \item \textbf{Estimation of the error variance.} 
    The estimator of $\sigma^2$ is
    $$\hat{\sigma}^2 = \frac{1}{T} \sum_{t=1}^T (\hat{\varepsilon}_t - 1)^2,$$
    where 
    $\widehat{\varepsilon}_t = y_t / \widehat{\mu}_t$, and $\widehat{\mu}_t$ is the conditional mean evaluated at the estimated $\widehat{\theta}$.  
    
    \item \textbf{Inference.}
    Following \citet{brownlees2012mem}, the asymptotic covariance matrix is estimated by
    $$\hat{\mathbf{V}}(\hat{\boldsymbol{\theta}}) = 
    \hat{\sigma}^2 \left[\sum_{t=1}^{T} {\mu_t^2}\frac{\partial \mu_t}{\partial \boldsymbol{\theta}}\frac{\partial\mu_t}{\partial\boldsymbol{\theta}'} \Big{|}_{\boldsymbol{\theta} = \widehat{\boldsymbol{\theta}}}\right]^{-1},$$
    so that standard errors are obtained as 
    $$\text{se}(\hat{\theta}_j) = \sqrt{[\hat{\mathbf{V}}(\hat{\boldsymbol{\theta}})]_{jj}} \, ,$$
    with corresponding $z$-statistics
    $$z_j = \hat{\theta}_j/\text{se}(\hat{\theta}_j).$$ 
    Under standard regularity conditions, $\widehat{\boldsymbol{\theta}}_j$ is asymptotically standard normal, so that two-sided $p$-values can be computed as
    $$p\text{-value}_j = 2[1 - \Phi(|z_j|)] \, ,$$ 
    where $\Phi(\cdot)$ denotes the standard normal cumulative distribution function.

    \item \textbf{Residual diagnostics.} 
    Model adequacy is evaluated using the zero-mean residuals $\widehat{u}_t = \hat{\varepsilon}_t - 1$. The Ljung–Box test \citep{ljung1978measure} is applied to both $\widehat{u}_t$ and $\widehat{u}_t^2$ (typically with 5 lags) to detect serial correlation and conditional heteroskedasticity. In addition, Engle’s ARCH test \citep{engle1982autoregressive}  with 5 lags provides another check for residual heteroskedasticity.
\end{enumerate}

\subsection{The Forecasting Setup}

{Following the standard approach in volatility forecasting, we perform out-of-sample evaluation by comparing one-step ahead forecasts to the "true" realized measure \citep{hansen2005forecast, patton2011volatility,  bollerslev2016exploiting, cipollini2021realized}}

The sample is split into two parts: an estimation (in-sample) period, from the beginning of the series up to a time $T$, and a forecasting (out-of-sample) period of length $H$, covering observations $t = T+1, \ldots, T+H$. Model parameters $\hat{\boldsymbol{\theta}}$ are estimated once using only the in-sample data and are kept fixed throughout the forecasting exercise. One-step-ahead forecasts are then generated sequentially, conditioning on the information set available at time $t-1$.

\paragraph{Forecast uncertainty.}
Prediction intervals are constructed using a non-parametric approach based on the empirical distribution of in-sample residuals.

For HAR models, residuals are defined as forecast errors 
$\hat{e}_t = y_t - \hat{y}_t$. 
The 95\% confidence interval is obtained by adding the empirical quantiles of $\hat{e}_t$ to the point forecast:
$$CI_{95\%}(y_{t|t-1}) = 
\left[\hat{y}_{t|t-1} + q_{0.025}(\hat{e}), \;
      \hat{y}_{t|t-1} + q_{0.975}(\hat{e}) \right],$$

where $y_{t|t-1}$ represents the forecasted value at time $t$, conditional on the information available at time $t-1$,
with the lower bound possibly truncated at zero to preserve non-negativity.

For MEM specifications, residuals are given by the multiplicative innovations
$\hat{\varepsilon}_t = y_t / \hat{\mu}_t$. 
The 95\% confidence interval is then obtained by scaling the point forecast with the empirical quantiles of $\hat{\varepsilon}_t$:

$$CI_{95\%}(y_{t|t-1}) = 
\left[\hat{\mu}_{t|t-1} \cdot q_{0.025}(\hat{\varepsilon}), \;
      \hat{\mu}_{t|t-1} \cdot q_{0.975}(\hat{\varepsilon}) \right].$$

\paragraph{Forecast evaluation.}
Predictive accuracy is assessed using two standard loss functions, the Mean Squared Error (MSE) and the Quasi-Likelihood (QLIKE), both consistent à la \citet{patton2011volatility}, computed over $H$ out-of-sample horizon
$$MSE = \frac{1}{H} \sum_{t=T+1}^{T+H} \left( y_t - \hat{y}_{t|t-1} \right)^2,$$
$$QLIKE =  \frac{1}{H} \sum_{t=T+1}^{T+H} 
    \left( \frac{y_t}{\hat{y}_{t|t-1}} - \log\left(\frac{y_t}{\hat{y}_{t|t-1}}\right) - 1 \right).$$

\section{Concluding Remarks}
VOLARE was conceived and implemented as a research infrastructure designed to restore and extend open access to high-quality realized volatility and covariance measures in financial econometrics. In a context where ultra-high-frequency (UHF) data are costly, computationally demanding, and methodologically delicate to process, VOLARE provides a transparent, reproducible, and academically rigorous pipeline transforming raw tick-level
observations into validated daily realized measures across multiple asset classes.

The fundamental contribution of the initiative lies in three interconnected dimensions. 

First, at the methodological level, VOLARE operationalizes state-of-the-art econometric procedures within a coherent and fully documented framework. The
implementation of the cleaning algorithm for stock data and the computation of a comprehensive set of realized measures ensure that the resulting series are not merely
convenient datasets, but carefully engineered research objects. The platform transforms millions of unevenly spaced tick observations into regularly sampled, microstructure-aware volatility measures suitable for econometric modelling and forecasting.

Second, at the infrastructure level, VOLARE integrates a robust back-end pipeline, covering data acquisition, cleaning, computation, and storage, with a web-based front-end designed for research usability. This dual access structure, full dataset extraction on the one hand, and real-time visualization and model estimation on the other, lowers the entry barrier to high-frequency volatility research while preserving methodological transparency.

Third, at the scientific ecosystem level, VOLARE was designed as a public good. By making daily realized variances and covariances available for a representative set of equities, exchange rates, and futures contracts, the platform facilitates replication,
comparative evaluation of models, and methodological innovation. In addition, by embedding real-time estimation of HAR- and MEM-type volatility models, VOLARE provides a unified environment where measurement and modelling coexist. This integration shortens the distance between raw data, econometric specification, diagnostic evaluation, and forecasting, encouraging experimentation and pedagogical use.

Beyond its current configuration, VOLARE is intrinsically expandable. The modular structure of the back-end allows for additional assets, and sampling frequencies; new realized measures can be incorporated with limited friction. Similarly, the modeling layer can be enriched with multivariate frameworks,
nonlinear specifications, or machine-learning-based volatility predictors.

\newpage 

\section*{Acknowledgments}

{This work was carried out as part of the ForVARD project (part of the  project GRINS - Grow\-ing Resilient, INclusive and Sustainable; grant GRINS PE00000018) funded by the European Union - Next Generation EU. However, the views and opinions expressed are solely those of the authors and do not necessarily reflect those of the European Union or the European Commission. Neither the European Union nor the European Commission can be held responsible for them. Without implicating, we are grateful to Demetrio Lacava and Luca Scaffidi Domianello, for useful suggestions on this paper. We also thank participants in the Conferences  \textit{2nd Workshop on Sustainable Finance – Spoke 4 GRINS} held in Venice, on December 2–3, 2024, \textit{Intermediate workshop of the ForVARD – Forecasting Volatility and Risk Dynamics project} held in Messina, on March 11, 2025, \textit{New Perspectives in Mathematical and Statistical Methods for Actuarial Sciences and Finance, Waiting for MAF} held in Salerno, on June 27-28, 2025, \textit{Methodological 
and Computational Challenges in Large-Scale Time Series Models for Economics and Finance}, held in Frascati on September 11–12, 2025, \textit{End of the PRIN 2022 Project: What’s Next? Insights, Ideas, and Future Collaborations}, held in Messina on January 12, 2026, and \textit{Volatility and Liquidity Workshop}, held in Pavia on January 22-23, 2026.}
\bibliographystyle{apecon} 
\bibliography{volatility}

@article{andersen1998answering,
 ISSN = {00206598, 14682354},
 author = {Torben G. Andersen and Tim Bollerslev},
 journal = {International Economic Review},
 number = {4},
 pages = {885--905},
 publisher = {[Economics Department of the University of Pennsylvania, Wiley, Institute of Social and Economic Research, Osaka University]},
 title = {Answering the Skeptics: Yes, Standard Volatility Models do Provide Accurate Forecasts},
 urldate = {2025-03-18},
 volume = {39},
 doi = {10.2307/2527343},
 year = {1998}
}

@article{andersen2003modeling,
author = {Andersen, Torben G. and Bollerslev, Tim and Diebold, Francis X. and Labys, Paul},
title = {Modeling and Forecasting Realized Volatility},
journal = {Econometrica},
volume = {71},
number = {2},
pages = {579-625},
doi = {10.1111/1468-0262.00418},
year = {2003}
}

@article{andersen2012jump,
title = {Jump-robust volatility estimation using nearest neighbor truncation},
journal = {Journal of Econometrics},
volume = {169},
number = {1},
pages = {75-93},
year = {2012},
note = {Recent Advances in Panel Data, Nonlinear and Nonparametric Models: A Festschrift in Honor of Peter C.B. Phillips},
issn = {0304-4076},
doi = {10.1016/j.jeconom.2012.01.011},
author = {Torben G. Andersen and Dobrislav Dobrev and Ernst Schaumburg},
}

@article{bandi2008microstructure,
  title={Microstructure noise, realized variance, and optimal sampling},
  author={Bandi, Federico M and Russell, Jeffrey R},
  journal={The Review of Economic Studies},
  volume={75},
  number={2},
  pages={339--369},
  year={2008},
  publisher={Wiley-Blackwell}
}

@article{barndorff2002econometric,
  title={Econometric analysis of realized volatility and its use in estimating stochastic volatility models},
  author={Barndorff-Nielsen, Ole E and Shephard, Neil},
  journal={Journal of the Royal Statistical Society Series B: Statistical Methodology},
  volume={64},
  number={2},
  pages={253--280},
  year={2002},
  publisher={Oxford University Press}
}

@article{barndorff2004covariation,
 ISSN = {00129682, 14680262},
 author = {Ole E. Barndorff-Nielsen and Neil Shephard},
 journal = {Econometrica},
 number = {3},
 pages = {885--925},
 publisher = {[Wiley, Econometric Society]},
 title = {Econometric Analysis of Realized Covariation: High Frequency Based Covariance, Regression, and Correlation in Financial Economics},
 urldate = {2025-05-21},
 volume = {72},
 year = {2004}
}

@article{barndorff2004power,
    author = {Barndorff-Nielsen, Ole E. and Shephard, Neil},
    title = {Power and Bipower Variation with Stochastic Volatility and Jumps},
    journal = {Journal of Financial Econometrics},
    volume = {2},
    number = {1},
    pages = {1-37},
    year = {2004},
    month = {01},
    issn = {1479-8409},
    doi = {10.1093/jjfinec/nbh001}
 }

@techreport{barndorff2004measuring,
  title={Measuring the impact of jumps in multivariate price processes using bipower covariation},
  author={Barndorff-Nielsen, Ole E and Shephard, Neil},
  year={2004},
  institution={Discussion paper, Nuffield College, Oxford University}
}

@article{barndorff2008designing,
author = {Barndorff-Nielsen, Ole E. and Hansen, Peter Reinhard and Lunde, Asger and Shephard, Neil},
title = {Designing Realized Kernels to Measure the ex post Variation of Equity Prices in the Presence of Noise},
journal = {Econometrica},
volume = {76},
number = {6},
pages = {1481-1536},
doi = {10.3982/ECTA6495},
year = {2008}
}

@article{barndorff2009realized,
 ISSN = {13684221, 1368423X},
 author = {O. E. Barndorff-Nielsen and P. Reinhard Hansen and A. Lunde and N. Shephard},
 journal = {The Econometrics Journal},
 number = {3},
 pages = {C1--C32},
 publisher = {[Royal Economic Society, Wiley]},
 title = {Realized kernels in practice: trades and quotes},
 volume = {12},
 doi = {10.1111/j.1368-423X.2008.00275.x},
 year = {2009}
}

@incollection{barndorff2010semivariance,
    author = {Barndorff‐Nielsen, Ole E. and Kinnebrock, Silja and Shephard, Neil},
    isbn = {9780199549498},
    title = {Measuring Downside Risk – Realized Semivariance},
    booktitle = {Volatility and Time Series Econometrics: Essays in Honor of Robert Engle},
    publisher = {Oxford University Press},
    year = {2010},
    month = {03},
    doi = {10.1093/acprof:oso/9780199549498.003.0007},
   }

@article{barndorff2011multivariate,
  title={Multivariate realised kernels: Consistent positive semi-definite estimators of the covariation of equity prices with noise and non-synchronous trading},
  author={Barndorff-Nielsen, Ole E and Hansen, Peter Reinhard and Lunde, Asger and Shephard, Neil},
  journal={Journal of Econometrics},
  volume={162},
  number={2},
  pages={149--169},
  year={2011},
  publisher={Elsevier},
  doi={10.1016/j.jeconom.2010.07.009}
}

@article{bartlett2023market,
    author = {Bartlett, Robert P and McCrary, Justin and O’Hara, Maureen},
    title = {The Market Inside the Market: Odd-Lot Quotes},
    journal = {The Review of Financial Studies},
    volume = {38},
    number = {3},
    pages = {661-711},
    year = {2023},
    month = {09},
    issn = {0893-9454},
    doi = {10.1093/rfs/hhad074},
    }

@article{bollerslev1986generalized,
title = {Generalized autoregressive conditional heteroskedasticity},
journal = {Journal of Econometrics},
volume = {31},
number = {3},
pages = {307-327},
year = {1986},
issn = {0304-4076},
doi = {10.1016/0304-4076(86)90063-1},
author = {Tim Bollerslev},
}

@article{bollerslev2020realized,
  title={Realized semicovariances},
  author={Bollerslev, Tim and Li, Jia and Patton, Andrew J and Quaedvlieg, Rogier},
  journal={Econometrica},
  volume={88},
  number={4},
  pages={1515--1551},
  year={2020},
  publisher={Wiley Online Library},
  doi={10.3982/ECTA17056}
}

@article{bollerslev2016exploiting,
  title={Exploiting the errors: A simple approach for improved volatility forecasting},
  author={Bollerslev, Tim and Patton, Andrew J and Quaedvlieg, Rogier},
  journal={Journal of Econometrics},
  volume={192},
  number={1},
  pages={1--18},
  year={2016},
  doi={10.1016/j.jeconom.2015.10.007},
  publisher={Elsevier}
}

@article{brownlees2006financial,
title = {Financial econometric analysis at ultra-high frequency: Data handling concerns},
journal = {Computational Statistics \& Data Analysis},
volume = {51},
number = {4},
pages = {2232-2245},
year = {2006},
issn = {0167-9473},
doi = {10.1016/j.csda.2006.09.030},
author = {C.T. Brownlees and G.M. Gallo}
}

@incollection{brownlees2012mem,
author = {Brownlees, Christian T. and Cipollini, Fabrizio and Gallo, Giampiero M.},
publisher = {John Wiley \& Sons, Ltd},
isbn = {9781118272039},
title = {Multiplicative {E}rror {M}odels},
booktitle = {Volatility Models and Their Applications},
chapter = {Nine},
pages = {223-247},
doi = {https://doi.org/10.1002/9781118272039.ch9},
year = {2012}
}

@article{cipollini2021realized,
  title={Realized volatility forecasting: Robustness to measurement errors},
  author={Cipollini, Fabrizio and Gallo, Giampiero M and Otranto, Edoardo},
  journal={International Journal of Forecasting},
  volume={37},
  number={1},
  pages={44--57},
  year={2021},
  publisher={Elsevier},
  doi={10.1016/j.ijforecast.2020.02.009}
}

@article{cipollini2025multiplicative20y,
title = {Multiplicative {E}rror {M}odels: 20 years on},
journal = {Econometrics and Statistics},
volume = {33},
pages = {209-229},
year = {2025},
issn = {2452-3062},
doi = {https://doi.org/10.1016/j.ecosta.2022.05.005},
author = {Fabrizio Cipollini and Giampiero M. Gallo}
}

@article{corsi2009simple,
  title={A simple approximate long-memory model of realized volatility},
  author={Corsi, Fulvio},
  journal={Journal of Financial Econometrics},
  volume={7},
  number={2},
  pages={174--196},
  year={2009},
  publisher={Oxford University Press}
}

@article{christensen2007realized,
title = {Realized range-based estimation of integrated variance},
journal = {Journal of Econometrics},
volume = {141},
number = {2},
pages = {323-349},
year = {2007},
issn = {0304-4076},
doi = {10.1016/j.jeconom.2006.06.012},
author = {Kim Christensen and Mark Podolskij},
}

@article{engle1982autoregressive,
 ISSN = {00129682, 14680262},
 author = {Robert F. Engle},
 journal = {Econometrica},
 number = {4},
 pages = {987--1007},
 publisher = {[Wiley, Econometric Society]},
 title = {Autoregressive Conditional Heteroscedasticity with Estimates of the Variance of United Kingdom Inflation},
 urldate = {2024-07-24},
 volume = {50},
 year = {1982}
}

@article{engle2002new,
  title= {New Frontiers for {ARCH} Models},
  author= {Engle, Robert F.},
  year= {2002},
  pages= {425-446},
  volume= {17},
  journal={Journal of Applied Econometrics}
}

@article{engle2006multiple,
  title={A multiple indicators model for volatility using intra-daily data},
  author={Engle, Robert F and Gallo, Giampiero M},
  journal={Journal of econometrics},
  volume={131},
  number={1-2},
  pages={3--27},
  year={2006},
  publisher={Elsevier}
}

@article{garman1980estimation,
  title={On the estimation of security price volatilities from historical data},
  author={Garman, Mark B and Klass, Michael J},
  journal={Journal of business},
  pages={67--78},
  year={1980},
  publisher={JSTOR},
  URL = {http://www.jstor.org/stable/2352358}
}

@book{genccay2001introduction,
  title={An introduction to high-frequency finance},
  author={Gen{\c{c}}ay, Ramazan and Dacorogna, Michel and Muller, Ulrich A and Pictet, Olivier and Olsen, Richard},
  year={2001},
  publisher={Elsevier}
}

@article{hansen2005forecast,
author = {Hansen, Peter R. and Lunde, Asger},
title = {A forecast comparison of volatility models: does anything beat a {GARCH}(1,1)?},
journal = {Journal of Applied Econometrics},
volume = {20},
number = {7},
pages = {873-889},
doi = {https://doi.org/10.1002/jae.800},
year = {2005}
}

@article{hansen2006realized,
author = {Hansen, Peter Reinhard and Lunde, Asger},
title = {Realized Variance and Market Microstructure Noise},
journal = {Journal of Business \& Economic Statistics},
volume = {24},
number = {2},
pages = {127--161},
year = {2006},
publisher = {ASA Website},
doi = {10.1198/073500106000000071}
}

@article{heber2009oxford,
  title={Oxford-{M}an {I}nstitute’s realized library},
  author={Heber, Gerd and Lunde, Asger and Shephard, Neil and Sheppard, Kevin},
  journal={Version 0.1, Oxford-Man Institute, University of Oxford},
  year={2009}
}

@article{johnson2014oddlot,
  title={Odd Lot Trades: The Behavior, Characteristics, and Information Content, Over Time},
  author={Johnson, Hardy},
  journal={Financial Review},
  year={2014},
  volume={49},
  number={4},
  doi = {10.1111/fire.12052},
  pages={669--684}
}

@article{ljung1978measure,
    author = {Ljung, G. M. and Box, G. E. P.},
    title = {On a measure of lack of fit in time series models},
    journal = {Biometrika},
    volume = {65},
    number = {2},
    pages = {297-303},
    year = {1978},
    month = {08},
    issn = {0006-3444},
    doi = {10.1093/biomet/65.2.297},
   }

@article{martens2007measuring,
  title={Measuring volatility with the realized range},
  author={Martens, Martin and Van Dijk, Dick},
  journal={Journal of Econometrics},
  volume={138},
  number={1},
  pages={181--207},
  year={2007},
  publisher={Elsevier},
  doi = {10.1016/j.jeconom.2006.05.019}
}

@article{newey1987simple,
  title={A simple, positive semi-definite, heteroskedasticity and autocorrelation consistent covariance matrix},
  author={Newey, Whitney K and West, Kenneth D},
  journal={Econometrica},
  volume={55},
  number={3},
  pages={703--708},
  year={1987},
  publisher={Wiley}
}

@article{newey1994automatic,
 ISSN = {00346527, 1467937X},
 author = {Whitney K. Newey and Kenneth D. West},
 journal = {The Review of Economic Studies},
 number = {4},
 pages = {631--653},
 publisher = {[Oxford University Press, Review of Economic Studies, Ltd.]},
 title = {Automatic Lag Selection in Covariance Matrix Estimation},
 urldate = {2025-09-09},
 volume = {61},
 year = {1994}
}

@article{ohara2014whats,
author = {O'Hara, Maureen and Yao, Chen and Ye, Mao},
title = {What's Not There: Odd Lots and Market Data},
journal = {The Journal of Finance},
volume = {69},
number = {5},
pages = {2199-2236},
doi = {https://doi.org/10.1111/jofi.12185},
year = {2014}
}

@article{parkinson1980extreme,
  title={The extreme value method for estimating the variance of the rate of return},
  author={Parkinson, Michael},
  journal={Journal of business},
  pages={61--65},
  volume = {53},
  year={1980},
  publisher={JSTOR},
  ISSN = {00219398, 15375374},
  URL = {http://www.jstor.org/stable/2352357}
}

@article{patton2011volatility,
title = {Volatility forecast comparison using imperfect volatility proxies},
journal = {Journal of Econometrics},
volume = {160},
number = {1},
pages = {246-256},
year = {2011},
note = {Realized Volatility},
issn = {0304-4076},
doi = {10.1016/j.jeconom.2010.03.034},
author = {Andrew J. Patton},
}

@article{wu1972oddlots,
 ISSN = {00221090, 17566916},
 author = {Hsiu-Kwang Wu},
 journal = {The Journal of Financial and Quantitative Analysis},
 number = {1},
 pages = {1321--1341},
 publisher = {Cambridge University Press},
 title = {Odd-Lot Trading in the Stock Market and Its Market Impact},
 doi ={10.2307/2330066},
 volume = {7},
 year = {1972}
}

@article{zhang2005tale,
  title={A tale of two time scales: Determining integrated volatility with noisy high-frequency data},
  author={Zhang, Lan and Mykland, Per A and A{\"\i}t-Sahalia, Yacine},
  journal={Journal of the American Statistical Association},
  volume={100},
  number={472},
  pages={1394--1411},
  year={2005},
  publisher={Taylor \& Francis}
}

@article{zhang2011estimating,
  title={Estimating covariation: Epps effect, microstructure noise},
  author={Zhang, Lan},
  journal={Journal of Econometrics},
  volume={160},
  number={1},
  pages={33--47},
  year={2011},
  publisher={Elsevier},
  doi={10.1016/j.jeconom.2010.03.012}
}

@article{zhou1996high,
author = {Bin Zhou},
title = {High-Frequency Data and Volatility in Foreign-Exchange Rates},
journal = {Journal of Business \& Economic Statistics},
ISSN = {07350015},
volume = {14},
number = {1},
pages = {45--52},
year = {1996},
publisher = {[American Statistical Association, Taylor \& Francis, Ltd.]},
doi = {10.1080/07350015.1996.10524628},
}

@article{engle2000econometrics,
  title={The econometrics of ultra-high-frequency data},
  author={Engle, Robert F},
  journal={Econometrica},
  volume={68},
  number={1},
  pages={1--22},
  year={2000},
  publisher={Wiley Online Library}
}

\clearpage

\appendix
\renewcommand{\thetable}{A.\arabic{table}}
\renewcommand{\thefigure}{A.\arabic{figure}}
\setcounter{table}{0}
\setcounter{figure}{0}
\section{Appendix} \label{sec:appendix}
\subsection{Asset list}\label{sec:assetslist}

\begin{table}[h!]
\centering
\resizebox{13cm}{!}{
\begin{tabular}{lllrc}
\toprule
Symbol & Name & Sector & First date\\
\midrule
AAPL * & Apple Inc. & Information Technology &  2015-01-02\\
ADBE & Adobe Inc. & Information Technology & 2015-01-02 \\
AMD & Advanced Micro Devices & Information Technology & 2015-01-02\\
AMGN * & Amgen Inc. & Health Care & 2015-01-02\\
AMZN * & Amazon & Consumer Discretionary  & 2015-01-02 \\
AXP * & American Express & Finance  & 2015-01-02\\
BA *& Boeing & Industrials & 2015-01-02\\
CAT *& Caterpillar Inc. & Industrials & 2015-01-02\\
CRM *& Salesforce Inc. & Information Technology & 2015-01-02\\
CSCO *& Cisco & Information Technology & 2015-01-02 \\
CVX *& Chevron Corporation & Energy & 2015-01-02\\
DIS *& Walt Disney Company (The) & Communication Services & 2015-01-02\\
GE & GE Aerospace & Industrials & 2015-01-02 \\
GOOGL & Alphabet Inc. (Class A) & Communication Services & 2015-01-02 \\
GS *& Goldman Sachs & Finance & 2015-01-02 \\
HD *& Home Depot & Consumer Discretionary & 2015-01-02 \\
HON *& Honeywell International Inc. & Industrials & 2015-01-02 \\
IBM *& IBM & Information Technology & 2015-01-02 \\
JNJ *& Johnson \& Johnson & Health Care & 2015-01-02\\
JPM *& JPMorgan Chase & Finance & 2015-01-02 \\
KO *& Coca-Cola Company (The) & Consumer Staples & 2015-01-02\\
MCD *& McDonald's & Consumer Discretionary &  2015-01-02\\
META & Meta Platforms & Communication Services & 2015-01-02 \\
MMM *& 3M & Industrials & 2015-01-02\\
MRK *& Merck \& Company Inc. & Health Care & 2015-01-02 \\
MSFT *& Microsoft & Information Technology & 2015-01-02\\
NFLX & Netflix, Inc. & Communication Services & 2015-01-02\\
NKE *& Nike, Inc. & Consumer Discretionary & 2015-01-02\\
NVDA *& Nvidia & Information Technology & 2015-01-02 \\
ORCL & Oracle Corporation & Information Technology &  2015-01-02\\
PG *& Procter \& Gamble & Consumer Staples &  2015-01-02 \\
PM & Philip Morris International & Consumer Staples &  2015-01-02\\
SHW *& Sherwin-Williams Company & Consumer Discretionary & 2015-01-02 \\
TRV *& The Travelers Companies Inc. & Finance & 2015-01-02 \\
TSLA & Tesla, Inc. & Consumer Discretionary & 2015-01-02 \\
UNH *& Unitedhealth Group Inc. & Health Care & 2015-01-02 \\
V *& Visa Inc. & Finance &  2015-01-02\\
VZ *& Verizon Communications Inc. & Public Utilities & 2015-01-02 \\
WMT *& Walmart & Consumer Staples &  2015-01-02 \\
XOM & ExxonMobil & Energy &  2015-01-02 \\
\bottomrule
\end{tabular}}
\caption{Stocks Data Availability. It includes the stocks from the Dow 30 (marked with *) along with ten stocks from the S\&P 100.}
\label{tab:stocks_availability}
\end{table}

\begin{table}[h!]
\centering
\begin{tabular}{lll}
\toprule
Symbol & Name & First date \\
\midrule
AUDUSD & Australian dollar / US dollar & 2009-09-25  \\
EURUSD & Euro / US dollar              & 2009-09-25 \\
GBPUSD & British pound / US dollar     & 2009-09-25  \\
USDCAD & US dollar / Canadian dollar   & 2009-09-25 \\
USDJPY & US dollar / Japanese yen      & 2009-09-28  \\
\bottomrule
\end{tabular}
\caption{Exchange Rates Data Availability}
\label{tab:forex_availability}
\end{table}

\begin{table}[H]
\centering
\begin{tabular}{lllrl}
\toprule
Symbol & Name & Sector & Exchange & First date \\
\midrule
CL & Crude Oil              & Energy & NYMEX & 2009-09-28 \\
NG & Natural Gas            & Energy & NYMEX &2009-09-28 \\
GC & Gold                   & Metals & COMEX& 2009-09-28 \\
C  & Corn                   & Agricultural & CME& 2009-09-28 \\
ES & E-mini S\&P 500        & Equity Index & CME &2009-09-28 \\
\bottomrule
\end{tabular}
\caption{Futures Data Availability}
\label{tab:futures_availability}
\end{table}

\subsection{Volatility Analysis across Assets: A Test Run} \label{app:vol_model_analysis}
In this section, we present the summary statistics and univariate estimation results obtained in different asset classes.

As mentioned earlier, we evaluate five model specifications: HAR, HAR-Q, MEM(1,1), AMEM(1,1), and AMEM(2,1). For this empirical analysis, we rely on a larger dataset, comprising 109 stocks, 13 exchange rates, and 13 futures contracts, a dataset that extends beyond the sample currently displayed in the graphical interface. 

The models are estimated using several realized variance (RV) measures, namely \texttt{rv1}, \texttt{rv5}, \texttt{bv1}, \texttt{bv5}, and \texttt{rk}.
The estimates were evaluated using the entire available time series for each asset\footnote{For stocks, the series typically range from 2015-01-02 to 2026-01-30; for exchange rates and futures, from 2009-09-25 to 2026-01-30}.
The discussion below focuses primarily on results based on the \texttt{rv1} measure, followed by a comparison of parameter estimates obtained from the other realized variance measures.

\Crefrange{tab:param_stats_stocks}{tab:param_stats_futures} report the cross-sectional summary statistics for the estimated parameters\footnote{Although t-statistics are reported in these tables for consistency with standard practice, inference should be interpreted with caution for the MEM family. Several parameters in the MEM specifications are subject to non-negativity constraints to ensure positive conditional variances: the scale parameter ($\omega$), the ARCH coefficients ($\alpha_i$), and the GARCH coefficients ($\beta_i$). When a parameter is constrained to be non-negative, the standard two-sided significance test ($H_0: \theta = 0$ vs. $H_1: \theta \neq 0$) is not formally valid, as the parameter space excludes negative values by construction. Under such constraints, the usual asymptotic distribution of the t-statistic no longer holds, and inference based on standard critical values may be misleading. For parameters at or near boundary values, one-sided tests or alternative inference procedures (such as bootstrap methods) would be more appropriate.}. Overall, the estimated parameters display the expected signs and magnitudes across all model classes.
{For the MEM(1,1), both  the $\alpha_1$ and the $\beta_1$ coefficients vary notably across asset classes. {In the equity market, the average $\alpha_1 \approx 0.51$, is greater than the average $\beta_1 \approx 0.39$, suggesting that stock volatility is more responsive to lagged volatility. Regarding exchange rates, the average $\beta_1 \approx 0.69$ is much bigger than the average $\alpha_1 \approx 0.28$, with remarkably low cross-sectional dispersion, indicating homogeneous volatility dynamics across currency pairs. 
The futures market exhibits the highest heterogeneity, with $\beta_1$ ranging from 0.35 to 0.81 and $\alpha_1$ from 0.16 to 0.59, reflecting the diverse nature of underlying commodities and financial contracts. 

{The asymmetric extension (AMEM(1,1)) introduces the leverage parameter ($\gamma_1$), which captures differential responses of volatility to positive and negative shocks. This parameter is generally small and weakly significant across markets: statistically significant in 50\% of stocks, but only in 31\% of exchange rates and 38\% of futures contracts. The stronger asymmetric effect in equities is consistent with the well-documented leverage effect, whereby negative returns tend to increase volatility more than positive returns of the same magnitude, a phenomenon observed across financial markets but particularly pronounced in stock returns.}

{The AMEM(2,1) model adds a short-term component ($\alpha_2$), which enhances the model's ability to capture transient volatility spikes and their rapid decay.}
This parameter is systematically negative across all markets, with mean values of -0.36 for stocks, -0.24 for exchange rates, and -0.20 for futures, and is significant in 99\% , 100\%, and 85\%, and of cases, respectively. The more pronounced mean-reversion in equities reflects the tendency of stock volatility to exhibit sharper but shorter-lived spikes following market shocks.

{For the HAR and HAR-Q models, the daily component ($\alpha_d$) typically dominates the weekly ($\alpha_w$) and monthly ($\alpha_m$) components, reflecting the greater importance of recent volatility in forecasting.}
However, the monthly component ($\alpha_m$) frequently lacks statistical significance, particularly in the equity market where only 9\% of stocks show significant estimates, compared to 77\% for exchange rates and 54\% for futures.  
The realized quarticity term ($\alpha_Q$) in the HAR-Q model is typically negative (mean values:  -11.3 for stocks, -37.9 for exchange rates, -65.1 for futures), implying that higher realized noise, captured by quarticity measures, tends to reduce conditional volatility forecasts, consistent with the view that elevated measurement error should be discounted in forward-looking estimates.

Model adequacy diagnostics are summarized in \autoref{tab:diagnostics}, which reports the percentage of assets passing Ljung–Box and ARCH tests. MEM-type models exhibit the highest pass rates, with 85-95\% of assets passing the Ljung–Box test on squared standardized residuals (LB$^2$) and ARCH tests across all three markets, confirming substantial absence of residual autocorrelation and conditional heteroskedasticity. By contrast, HAR-type models show comparatively weaker performance, with pass rates of only 19\% for stocks, 38\% for both exchange rates and futures, giving clues of some misspecification. Notably, the AMEM(2,1) model demonstrates substantial improvement over simpler MEM specifications in capturing residual dynamics, particularly for equities, where the pass rate for the Ljung–Box test on standardized residuals jumps from 6-8\% (MEM/AMEM) to 76\%, reflecting the model's enhanced ability to capture short-term volatility fluctuations through the $\alpha_2$
 parameter.

The boxplots in \Crefrange{fig:stocks_boxplots_HAR}{fig:forex_boxplots_MEM}, illustrate the cross-sectional distribution of the estimated parameters for the MEM and HAR model families across the three asset classes, using the same realized variance measure (\texttt{rv1}). 
Each boxplot represents the empirical distribution of a given parameter across all assets within an asset class. 
The central line corresponds to the sample median, the box spans the interquartile range (IQR, 25th-75th percentiles), and the whiskers extend to the most extreme data points within 1.5 $\times$ IQR from the lower and upper quartiles. 
Observations outside this range are plotted individually to emphasize values that deviate substantially from the central distribution. 
For the MEM-type specifications, the coefficients $\alpha_1$ and $\beta_1$ are consistently positive and display limited dispersion across assets, confirming the strong and homogeneous persistence of conditional volatility in all markets. 
The leverage parameter $\gamma_1$ in the AMEM specification is close to zero in most cases, with slightly higher variability for stocks, suggesting that asymmetry effects are generally mild and more relevant for equity returns. 
The additional short-term component $\alpha_2$ in the AMEM(2,1) model exhibits the expected negative sign, capturing short-lived mean-reversion dynamics. 
Conversely, the HAR and HAR-Q models show the typical heterogeneous-memory structure: the coefficients associated with the daily component ($\alpha_d$) dominate those of the weekly ($\alpha_w$) and monthly ($\alpha_m$) components, and this pattern holds consistently across stock, exchange rates, and futures. 
Overall, these parameter distributions indicate that volatility reacts promptly to new information but remains highly persistent over time, with only moderate cross-sectional variation.

Finally, \Cref{fig:stocks_stability_persistence_all,fig:forex_stability_persistence_all,fig:futures_stability_persistence_all} show the stability of volatility persistence estimates for MEM-type models across different realized variance measures (\texttt{rv1}, \texttt{rv5}, \texttt{bv1}, \texttt{bv5}, \texttt{rk}). 
The estimated persistence, computed as $\alpha_1 + \beta_1$ for MEM(1,1), $\alpha_1 + \beta_1 + \gamma_1/2$ for AMEM(1,1), and $\alpha_1 + \alpha_2 + \beta_1 + \gamma_1/2$ for AMEM(2,1), remains consistently high across all asset classes and remarkably stable across sampling frequencies and variance estimators.
This robustness is evident in all three markets. 
Among stocks, persistence is strong and lies predominantly in the range $[0.84, 0.98]$, with greater cross-sectional dispersion attributable to idiosyncratic company-specific effects, though the pattern across RV measures remains consistent within each ticker.
In exchange rates, persistence values are tightly clustered in the range $[0.95, 0.99]$, with minimal dispersion across both currency pairs and RV measures, confirming the highly persistent nature of volatility in currency markets.
Futures contracts exhibit persistence values tightly concentrated near unity across almost all contracts and RV measures, typically falling in the range $[0.93, 1.00]$, with only a few isolated exceptions that do not alter the overall picture.
These findings confirm that long-memory behavior in volatility is an intrinsic feature of financial time series, largely independent of the sampling interval (1-minute vs.\ 5-minute returns) or the particular realized-variance estimator used (standard RV vs.\ jump-robust bipower variation vs.\ realized kernel), thereby reinforcing the empirical robustness of the MEM-type specifications implemented within the \textsc{Volare} platform.

\begin{table}[htbp]
\centering
\begin{tabular}{lrrrrrrr}
\toprule
\multicolumn{8}{c}{\textbf{Stocks}} \\
\midrule
Parameter & N & Mean & Std & Min & Median & Max & Signif.(\%) \\
\midrule
\multicolumn{8}{l}{\textbf{MEM(1,1)}} \\
$\omega$ & 109 & 2.134 & 0.776 & 0.621 & 2.004 & 4.450 & 100 \\
$\alpha_1$ & 109 & 0.516 & 0.056 & 0.370 & 0.519 & 0.648 & 100 \\
$\beta_1$ & 109 & 0.392 & 0.069 & 0.228 & 0.390 & 0.580 & 100 \\
\midrule
\multicolumn{8}{l}{\textbf{AMEM(1,1)}} \\
$\omega$ & 109 & 2.099 & 0.753 & 0.538 & 1.990 & 4.450 & 100 \\
$\alpha_1$ & 109 & 0.497 & 0.056 & 0.335 & 0.502 & 0.636 & 100 \\
$\beta_1$ & 109 & 0.402 & 0.069 & 0.247 & 0.399 & 0.598 & 100 \\
$\gamma_1$ & 109 & 0.021 & 0.012 & -0.000 & 0.020 & 0.063 & 50 \\
\midrule
\multicolumn{8}{l}{\textbf{AMEM(2,1)}} \\
$\omega$ & 109 & 0.552 & 0.314 & 0.138 & 0.517 & 2.479 & 97 \\
$\alpha_1$ & 109 & 0.536 & 0.046 & 0.418 & 0.540 & 0.638 & 100 \\
$\alpha_2$ & 109 & -0.360 & 0.083 & -0.551 & -0.365 & -0.106 & 99 \\
$\beta_1$ & 109 & 0.792 & 0.069 & 0.572 & 0.798 & 0.924 & 100 \\
$\gamma_1$ & 109 & 0.014 & 0.008 & -0.004 & 0.013 & 0.037 & 52 \\
\midrule
\multicolumn{8}{l}{\textbf{HAR}} \\
$\omega \cdot 1000$ & 109 & 0.061 & 0.048 & 0.016 & 0.048 & 0.336 & 96 \\
$\alpha_d$ & 109 & 0.410 & 0.169 & 0.065 & 0.426 & 0.853 & 85 \\
$\alpha_w$ & 109 & 0.346 & 0.110 & 0.016 & 0.351 & 0.629 & 88 \\
$\alpha_m$ & 109 & 0.027 & 0.070 & -0.098 & 0.013 & 0.250 & 9 \\
\midrule
\multicolumn{8}{l}{\textbf{HAR-Q}} \\
$\omega \cdot 1000$ & 109 & 0.028 & 0.038 & -0.076 & 0.022 & 0.248 & 33 \\
$\alpha_d$ & 109 & 0.717 & 0.157 & 0.305 & 0.710 & 1.167 & 98 \\
$\alpha_Q$ & 109 & -11.289 & 9.762 & -60.027 & -8.125 & 2.968 & 77 \\
$\alpha_w$ & 109 & 0.217 & 0.097 & -0.101 & 0.211 & 0.498 & 70 \\
$\alpha_m$ & 109 & 0.000 & 0.062 & -0.156 & -0.008 & 0.193 & 11 \\
\bottomrule
\end{tabular}
\caption{Parameter estimates for volatility models on 109 stocks using \texttt{rv1} realized variance measure. For each parameter: N is the number of observations, Mean is the average parameter value across assets, Std is the cross-sectional standard deviation, Min/Median/Max are the distribution quantiles, Signif.(\%) represents percentage of estimates statistically significant at 5\% level.}
\label{tab:param_stats_stocks}
\end{table}

\begin{table}[htbp]
\centering
\begin{tabular}{lrrrrrrr}
\toprule
\multicolumn{8}{c}{\textbf{Exchange Rates}} \\
\midrule
Parameter & N & Mean & Std & Min & Median & Max & Signif.(\%) \\
\midrule
\multicolumn{8}{l}{\textbf{MEM(1,1)}} \\
$\omega$ & 13 & 0.265 & 0.089 & 0.107 & 0.258 & 0.400 & 100 \\
$\alpha_1$ & 13 & 0.285 & 0.064 & 0.153 & 0.289 & 0.391 & 100 \\
$\beta_1$ & 13 & 0.689 & 0.071 & 0.564 & 0.678 & 0.819 & 100 \\
\midrule
\multicolumn{8}{l}{\textbf{AMEM(1,1)}} \\
$\omega$ & 13 & 0.260 & 0.091 & 0.107 & 0.254 & 0.400 & 100 \\
$\alpha_1$ & 13 & 0.279 & 0.066 & 0.153 & 0.285 & 0.385 & 100 \\
$\beta_1$ & 13 & 0.691 & 0.071 & 0.566 & 0.678 & 0.826 & 100 \\
$\gamma_1$ & 13 & 0.008 & 0.010 & -0.000 & 0.003 & 0.030 & 31 \\
\midrule
\multicolumn{8}{l}{\textbf{AMEM(2,1)}} \\
$\omega$ & 13 & 0.093 & 0.040 & 0.020 & 0.086 & 0.179 & 85 \\
$\alpha_1$ & 13 & 0.388 & 0.074 & 0.232 & 0.405 & 0.478 & 100 \\
$\alpha_2$ & 13 &  -0.245 & 0.066 & -0.337 & -0.253 & -0.115 & 100 \\
$\beta_1$ & 13 &  0.849 & 0.022 & 0.812 & 0.846 & 0.879 & 100 \\
$\gamma_1$ & 13 & -0.002 & 0.014 & -0.022 & 0.000 & 0.022 & 62 \\
\midrule
\multicolumn{8}{l}{\textbf{HAR}} \\
$\omega \cdot 1000$ & 13 & 0.013 & 0.011 & 0.002 & 0.009 & 0.044 & 100 \\
$\alpha_d$ & 13 & 0.328 & 0.180 & 0.067 & 0.371 & 0.612 & 92 \\
$\alpha_w$ & 13 & 0.244 & 0.072 & 0.122 & 0.241 & 0.348 & 92 \\
$\alpha_m$ & 13 & 0.178 & 0.092 & 0.011 & 0.176 & 0.337 & 77 \\
\midrule
\multicolumn{8}{l}{\textbf{HAR-Q}} \\
$\omega \cdot 1000$ & 13 & 0.010 & 0.009 & 0.002 & 0.007 & 0.033 & 85 \\
$\alpha_d$ & 13 & 0.495 & 0.157 & 0.127 & 0.519 & 0.705 & 100 \\
$\alpha_Q$ & 13 & -37.993 & 43.502 & -132.750 & -21.437 & -0.199 & 62 \\
$\alpha_w$ & 13 & 0.190 & 0.081 & 0.038 & 0.207 & 0.311 & 92 \\
$\alpha_m$ & 13 & 0.140 & 0.089 & -0.007 & 0.141 & 0.296 & 69 \\
\bottomrule
\end{tabular}
\caption{Parameter estimates for volatility models on 13 exchange rates using \texttt{rv1} realized variance measure. For each parameter: N is the number of observations, Mean is the average parameter value across assets, Std is the cross-sectional standard deviation, Min/Median/Max are the distribution quantiles, Signif.(\%) represents percentage of estimates statistically significant at 5\% level.}
\label{tab:param_stats_forex}
\end{table}

\begin{table}[htbp]
\centering
\begin{tabular}{lrrrrrrr}
\toprule
\multicolumn{8}{c}{\textbf{Futures}} \\
\midrule
Parameter & N & Mean & Std & Min & Median & Max & Signif.(\%) \\
\midrule
\multicolumn{8}{l}{\textbf{MEM(1,1)}} \\
$\omega$ & 13 & 0.590 & 0.402 & 0.075 & 0.570 & 1.232 & 77 \\
$\alpha_1$ & 13 & 0.348 & 0.157 & 0.160 & 0.285 & 0.587 & 100 \\
$\beta_1$ & 13 & 0.622 & 0.170 & 0.347 & 0.675 & 0.812 & 100 \\
\midrule
\multicolumn{8}{l}{\textbf{AMEM(1,1)}} \\
$\omega$ & 13 & 0.584 & 0.394 & 0.075 & 0.570 & 1.232 & 77 \\
$\alpha_1$ & 13 & 0.319 & 0.125 & 0.160 & 0.285 & 0.554 & 100 \\
$\beta_1$ & 13 & 0.639 & 0.147 & 0.404 & 0.675 & 0.812 & 100 \\
$\gamma_1$ & 13 & 0.026 & 0.040 & -0.000 & 0.006 & 0.113 & 38 \\
\midrule
\multicolumn{8}{l}{\textbf{AMEM(2,1)}} \\
$\omega$ & 13 & 0.297 & 0.199 & 0.021 & 0.329 & 0.640 & 77 \\
$\alpha_1$ & 13 & 0.394 & 0.107 & 0.221 & 0.381 & 0.575 & 100 \\
$\beta_1$ & 13 & 0.785 & 0.102 & 0.582 & 0.841 & 0.886 & 100 \\
$\alpha_2$ & 13 & -0.202 & 0.078 & -0.334 & -0.221 & -0.090 & 85 \\
$\gamma_1$ & 13 & 0.017 & 0.034 & -0.010 & 0.007 & 0.090 & 46 \\
\midrule
\multicolumn{8}{l}{\textbf{HAR}} \\
$\omega \cdot 1000$ & 13 & 0.116 & 0.144 & 0.001 & 0.068 & 0.451 & 77 \\
$\alpha_d$ & 13 & 0.284 & 0.230 & -0.000 & 0.285 & 0.649 & 85 \\
$\alpha_w$ & 13 & 0.234 & 0.119 & -0.001 & 0.247 & 0.398 & 85 \\
$\alpha_m$ & 13 & 0.156 & 0.163 & -0.031 & 0.121 & 0.523 & 54 \\
\midrule
\multicolumn{8}{l}{\textbf{HAR-Q}} \\
$\omega \cdot 1000$ & 13 & 0.090 & 0.119 & 0.001 & 0.056 & 0.406 & 54 \\
$\alpha_d$ & 13 & 0.525 & 0.259 & 0.175 & 0.510 & 1.046 & 92 \\
$\alpha_Q$ & 13 & -65.043 & 132.183 & -422.905 & -2.878 & -0.017 & 69 \\
$\alpha_w$ & 13 & 0.177 & 0.094 & -0.001 & 0.188 & 0.348 & 85 \\
$\alpha_m$ & 13 & 0.126 & 0.144 & -0.037 & 0.083 & 0.436 & 54 \\
\bottomrule
\end{tabular}
\caption{Parameter estimates for volatility models on 13 futures using \texttt{rv1} realized variance measure. For each parameter: N is the number of observations, Mean is the average parameter value across assets, Std is the cross-sectional standard deviation, Min/Median/Max are the distribution quantiles, Signif.(\%) represents percentage of estimates statistically significant at 5\% level.}
\label{tab:param_stats_futures}
\end{table}

\begin{table}[htbp]
\centering
\begin{tabular}{lrrrrrrr}
\toprule
\multicolumn{5}{c}{\textbf{Stocks}} \\
\midrule
 Model & N& LB Test& LB$^2$ Test & ARCH Test\\
\midrule
MEM(1,1)  &109     & 8\%     &95\%    &  95\%          \\
  AMEM(1,1)  &109    &  6\%   &  96\%    &  96\%         \\
AMEM(2,1)&  109   &  76\% &    94\%  &    94\%         \\
   HAR & 109  &  10\%     & 19\%     &  19\%         \\
  HAR-Q & 109  & 14\%    &  35\%    &   35\%         \\
\midrule
\multicolumn{5}{c}{\textbf{Exchange Rates}} \\
\midrule
 Model & N& LB Test& LB$^2$ Test & ARCH Test\\
\midrule
MEM(1,1)  &13     & 0\%     &85\%    &  85\%          \\
  AMEM(1,1)  &13    &  0\%   &  85\%    &  85\%         \\
AMEM(2,1) & 13   &  8\% &    85\%  &    85\%         \\
   HAR  &13 &  77\%     & 38\%     &  38\%         \\
  HAR-Q  &13    & 38\%    &  46\%    &   46\%         \\
\midrule 
\multicolumn{5}{c}{\textbf{Futures}} \\
\midrule
 Model & N& LB Test& LB$^2$ Test & ARCH Test\\
\midrule
MEM(1,1)  &13     & 23\%     &92\%    &  92\%          \\
  AMEM(1,1)  &13    &  23\%   &  92\%    &  92\%         \\
AMEM(2,1)  &13   &  54\% &    92\%  &    92\%         \\
   HAR & 13  &  38\%     & 38\%     &  38\%         \\
  HAR-Q  &13    & 38\%    &  38\%    &   38\%         \\
\bottomrule
\end{tabular}
\caption{Percentage of assets passing diagnostic tests for model adequacy (\texttt{rv1} measure). Ljung-Box test for autocorrelation in standardized residuals (LB Test); Test Ljung-Box test for autocorrelation in squared standardized residuals (LB$^2$ test); ARCH test for remaining conditional heteroskedasticity. Pass rates indicate the percentage of assets where the null hypothesis (no misspecification) is not rejected at 5\% significance level. }
\label{tab:diagnostics}
\end{table}

\begin{figure}[htbp]
    \centering
    \includegraphics[width=0.95\textwidth]{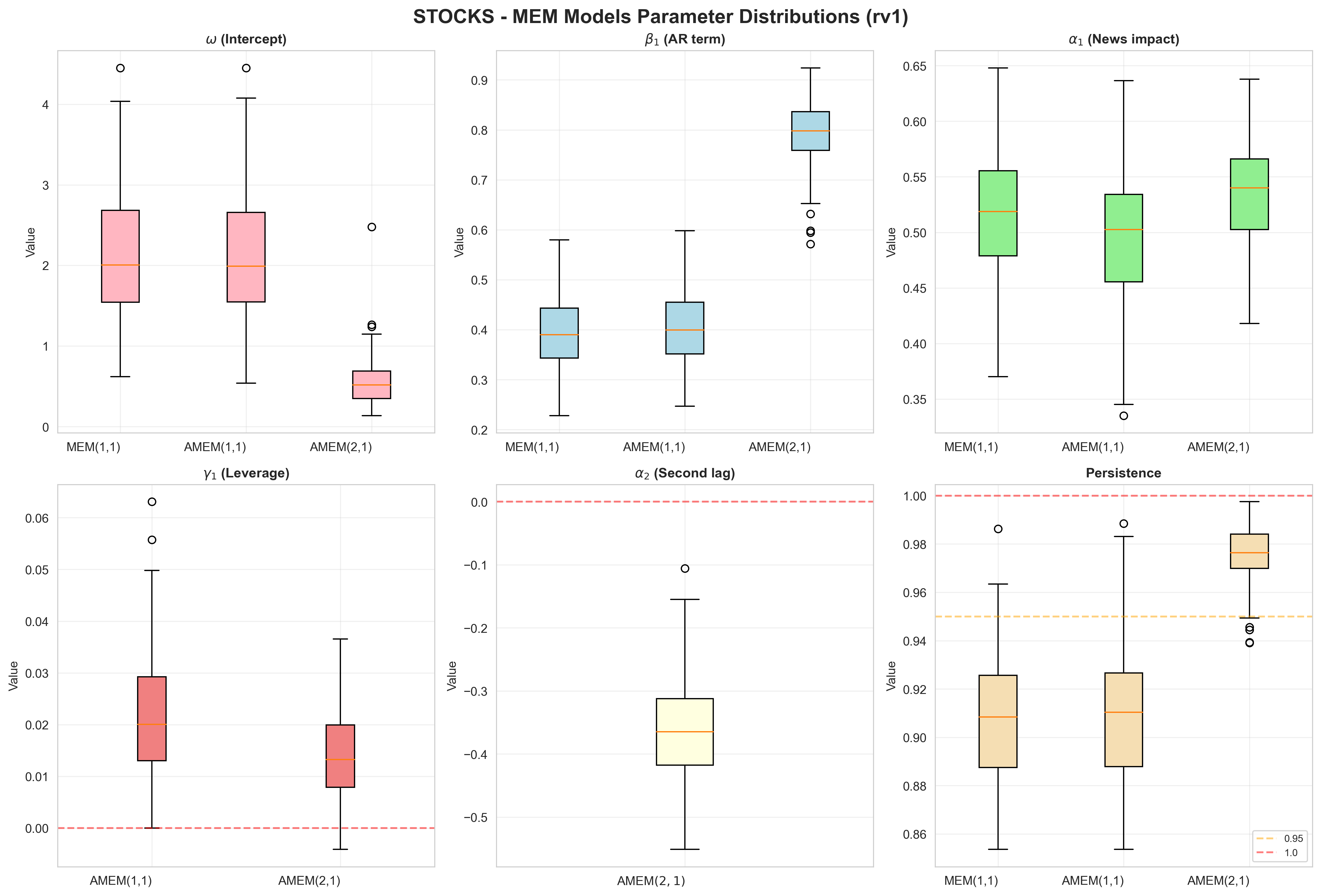}
    \caption{Distribution of parameter estimates for MEM family models across 109 stocks using \texttt{rv1} measure. Boxes display the interquartile range (IQR) with median line; whiskers extend to 
$\pm1.5\times$IQR; observations beyond this range shown as circles. }
    \label{fig:stocks_boxplots_MEM}
\end{figure}

\begin{figure}[htbp]
    \centering
    \includegraphics[width=0.95\textwidth]{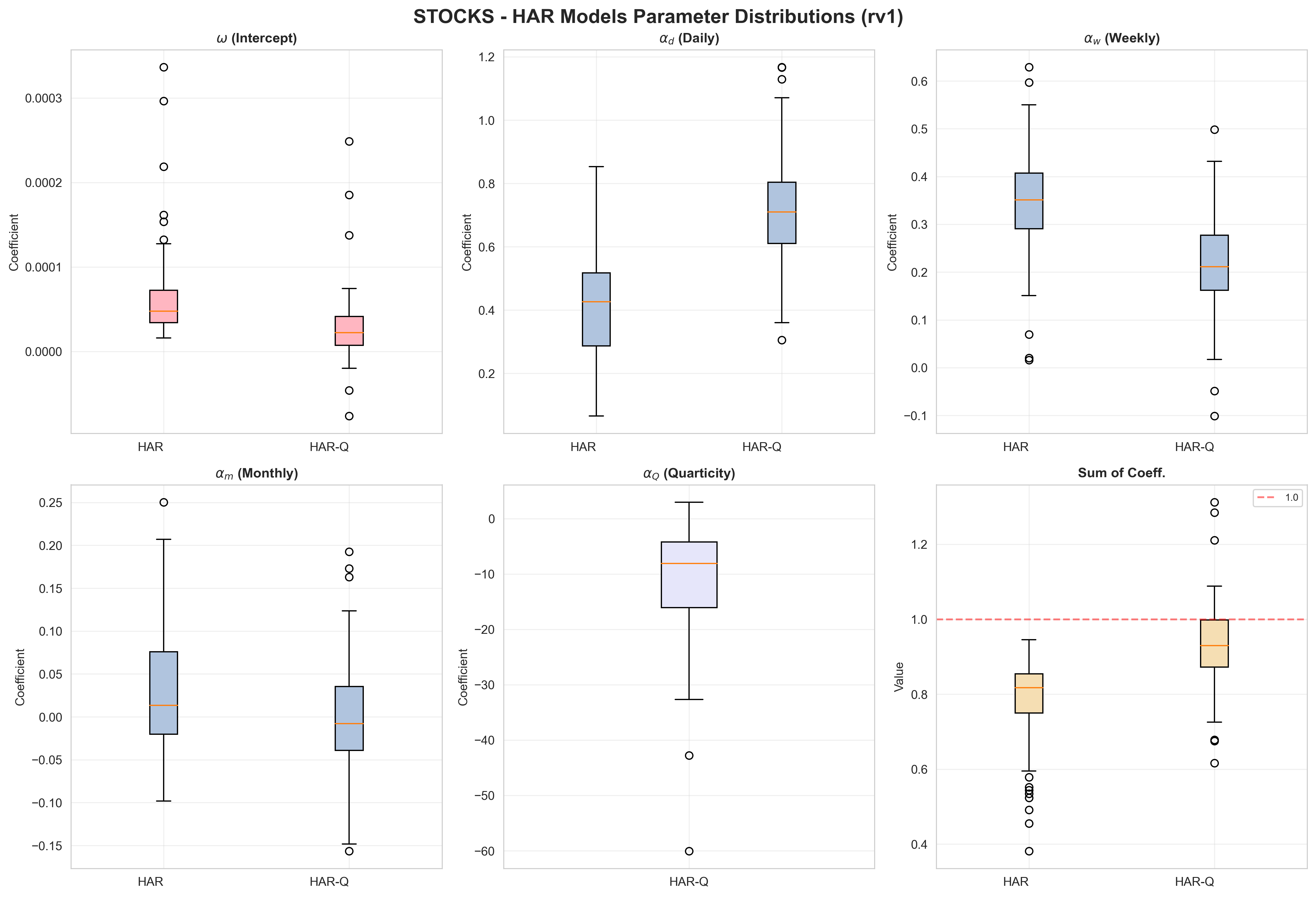}
    \caption{Distribution of parameter estimates for HAR family models across 109 stocks using \texttt{rv1} measure. Boxes display the interquartile range (IQR) with median line; whiskers extend to 
$\pm1.5\times$IQR; observations beyond this range shown as circles.}
    \label{fig:stocks_boxplots_HAR}
\end{figure}

\begin{figure}[htbp]
    \centering
    \includegraphics[width=0.95\textwidth]{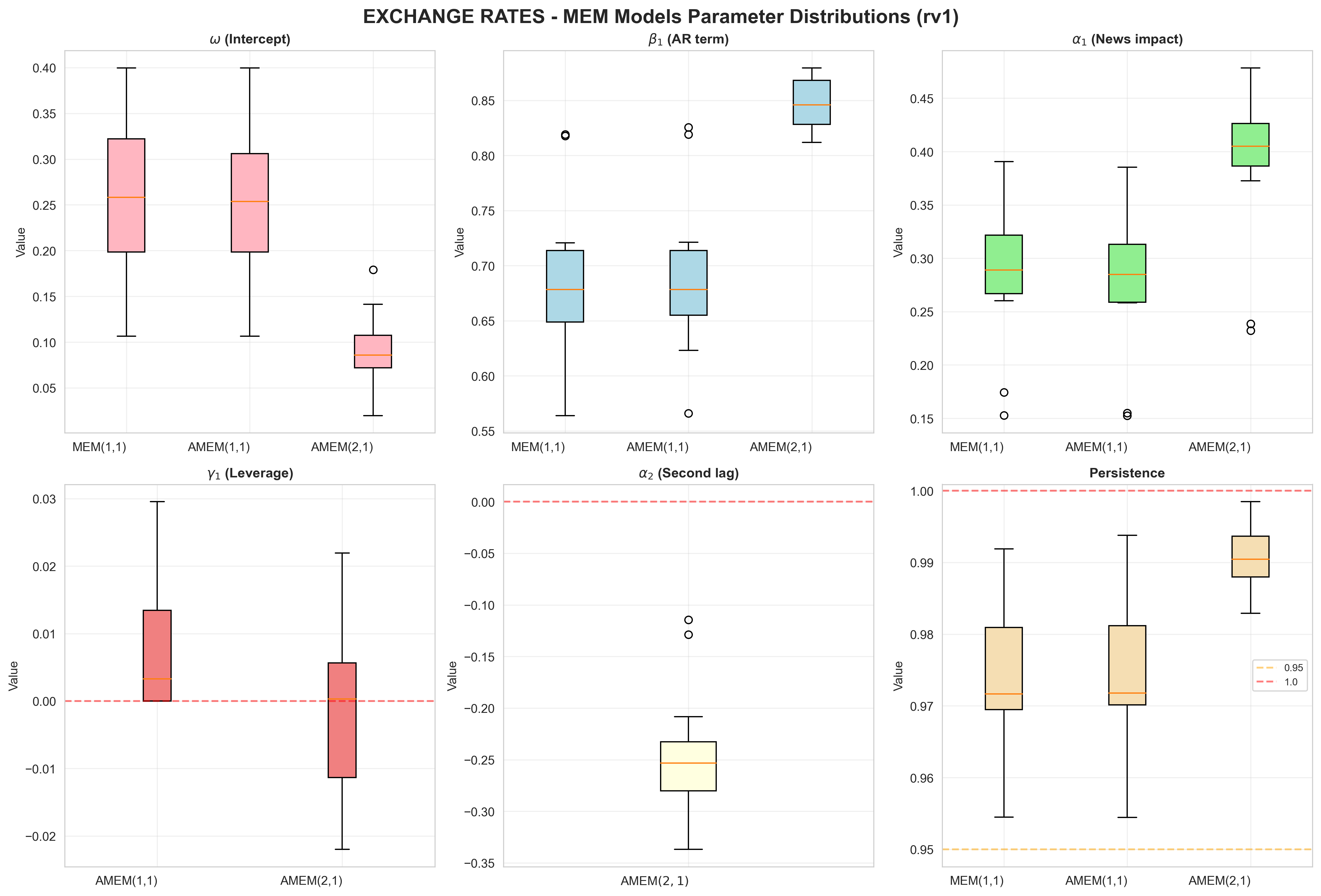}
    \caption{Distribution of parameter estimates for MEM family models across 13 exchange rates using \texttt{rv1} measure. Boxes display the interquartile range (IQR) with median line; whiskers extend to 
$\pm1.5\times$IQR; observations beyond this range shown as circles.}
    \label{fig:forex_boxplots_MEM}
\end{figure}

\begin{figure}[htbp]
    \centering
    \includegraphics[width=0.95\textwidth]{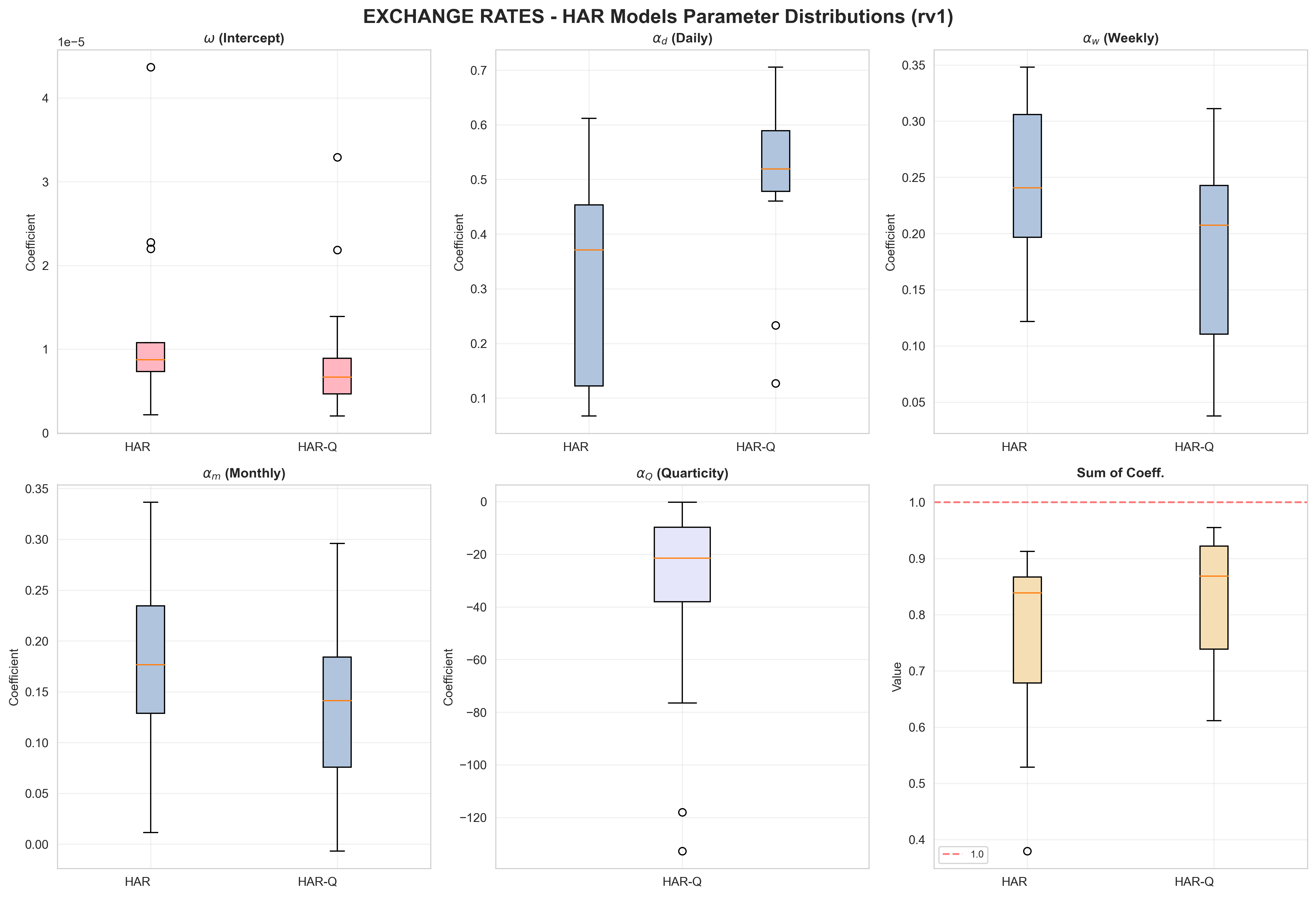}
    \caption{Distribution of parameter estimates for HAR family models across 13 exchange rates using \texttt{rv1} measure. Boxes display the interquartile range (IQR) with median line; whiskers extend to 
$\pm1.5\times$IQR; observations beyond this range shown as circles. }
    \label{fig:forex_boxplots_HAR}
\end{figure}

\begin{figure}[htbp]
    \centering
    \includegraphics[width=0.95\textwidth]{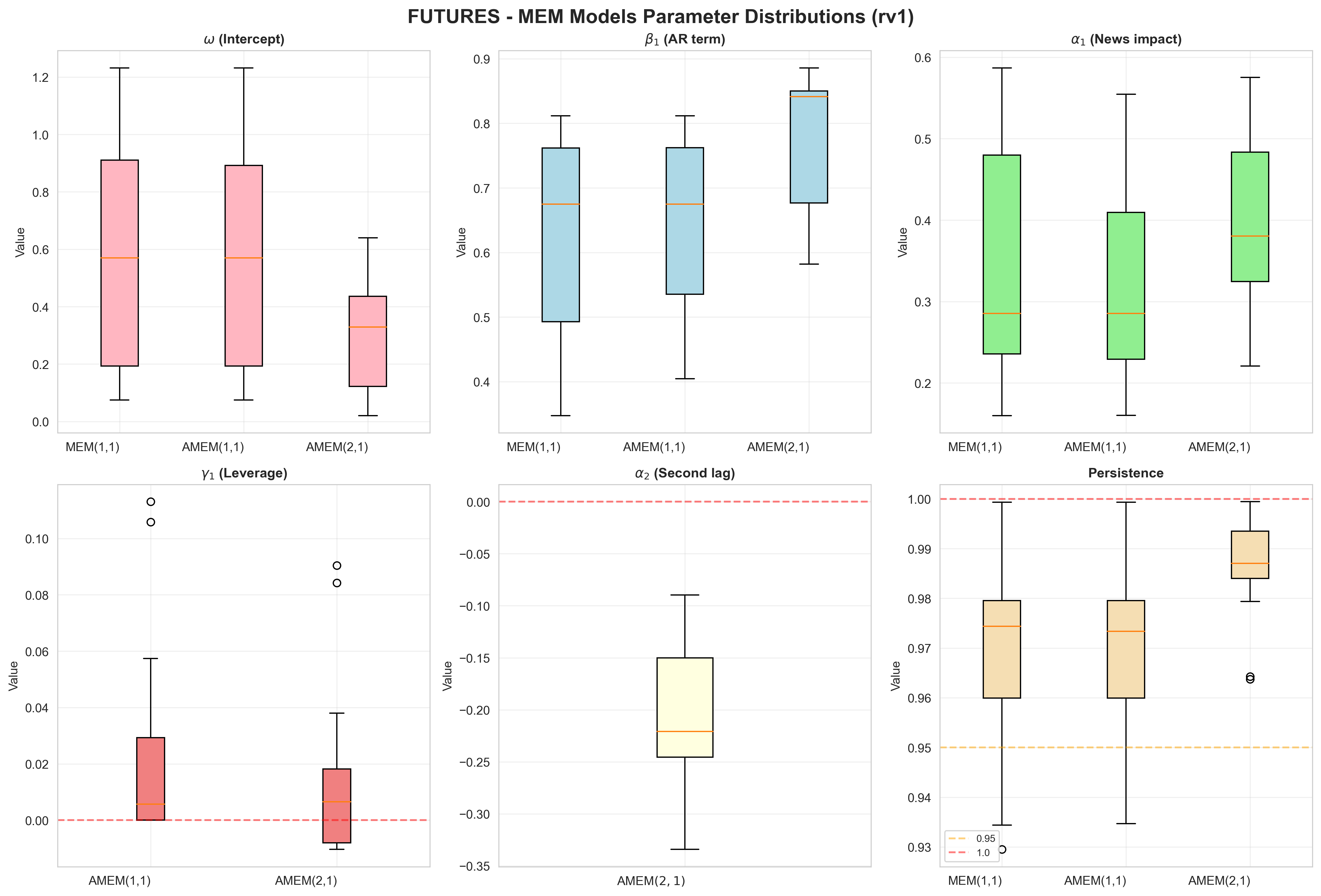}
    \caption{Distribution of parameter estimates for MEM family models across 13 futures using \texttt{rv1} measure. Boxes display the interquartile range (IQR) with median line; whiskers extend to 
$\pm1.5\times$IQR; observations beyond this range shown as circles. }
    \label{fig:futures_boxplots_MEM}
\end{figure}

\begin{figure}[htbp]
    \centering
    \includegraphics[width=0.95\textwidth]{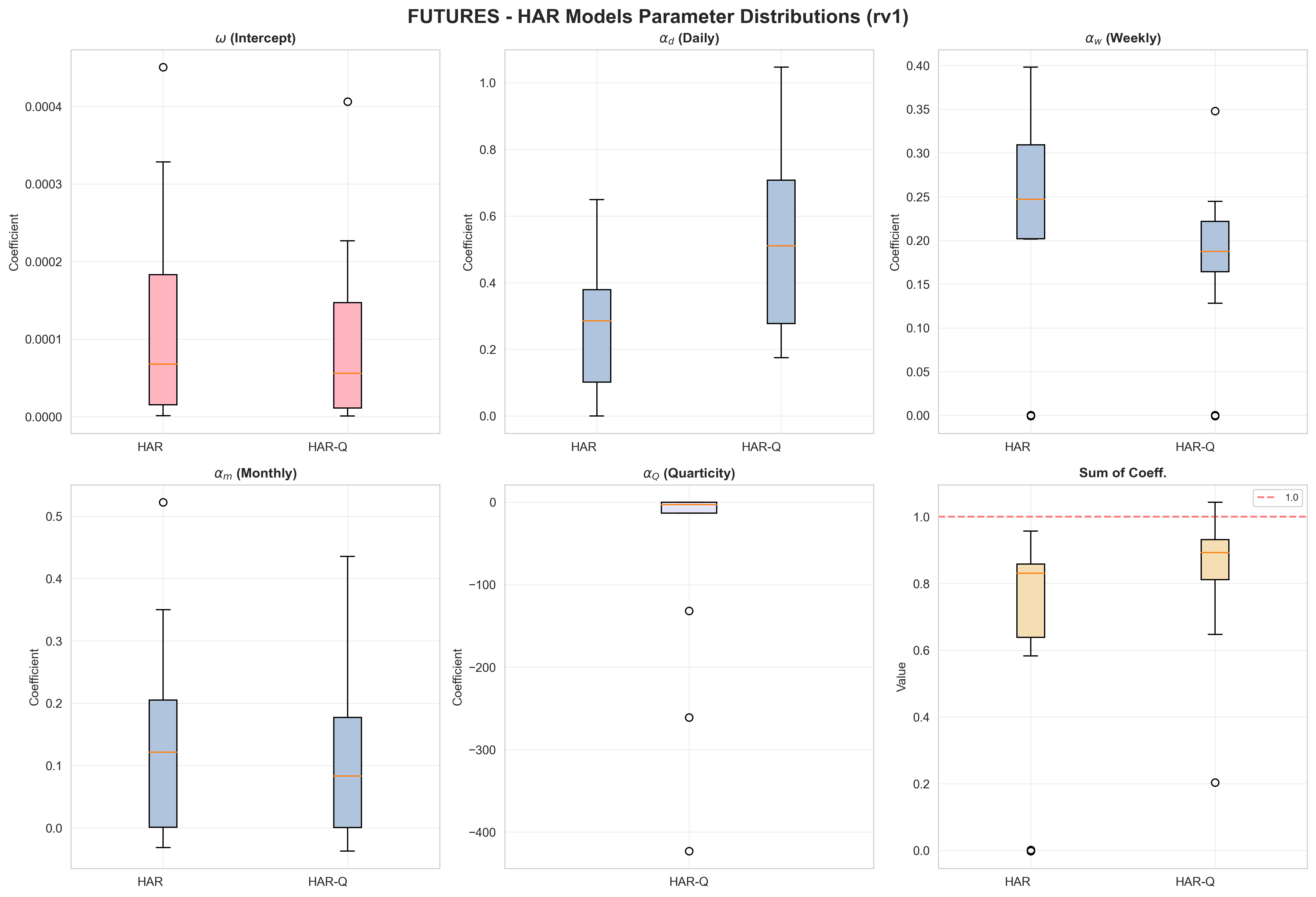}
    \caption{Distribution of parameter estimates for HAR family models across 13 futures using \texttt{rv1} measure. Boxes display the interquartile range (IQR) with median line; whiskers extend to 
$\pm1.5\times$IQR; observations beyond this range shown as circles. }
    \label{fig:futures_boxplots_HAR}
\end{figure}


\begin{figure}[htbp]
    \centering
    \includegraphics[width=\textwidth]{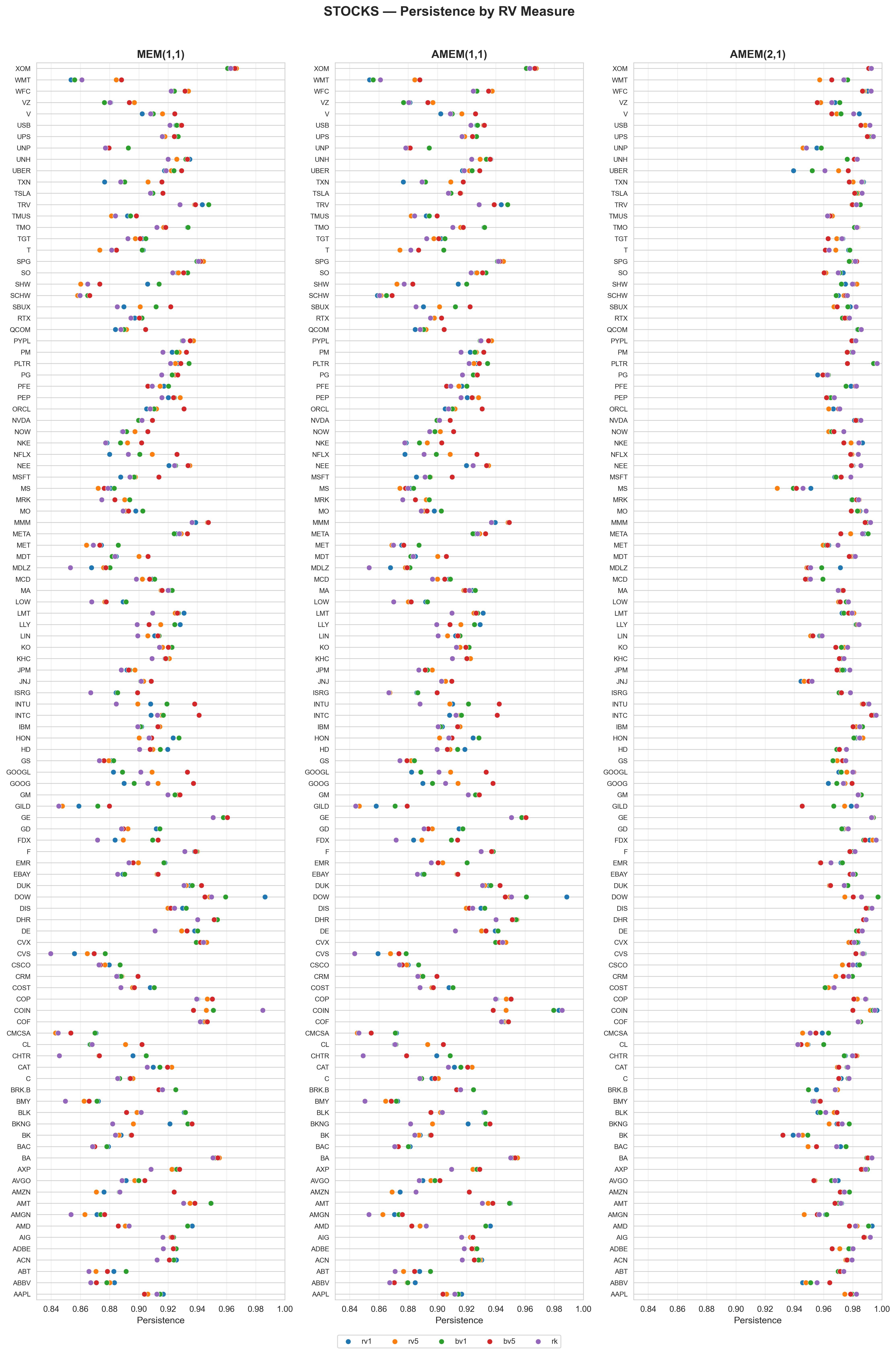}
            
    \caption{Robustness of total persistence across RV measures for stocks. Persistence for MEM(1,1) $= \alpha_1 + \beta_1 $, for AMEM(1,1) $= \alpha_1 + \beta_1 +  \gamma_1/2$, AMEM(2,1) $= \alpha_1 + \alpha_2 +\beta_1 + \gamma_1/2$. Each dot represents the persistence estimate for one asset using a specific RV measure (\texttt{rv1}, \texttt{rv5}, \texttt{bv1}, \texttt{bv5}, \texttt{rk}), with colors distinguishing the five estimators.}
    \label{fig:stocks_stability_persistence_all}
\end{figure}

\begin{figure}[htbp]
    \centering
    \includegraphics[width=\textwidth]{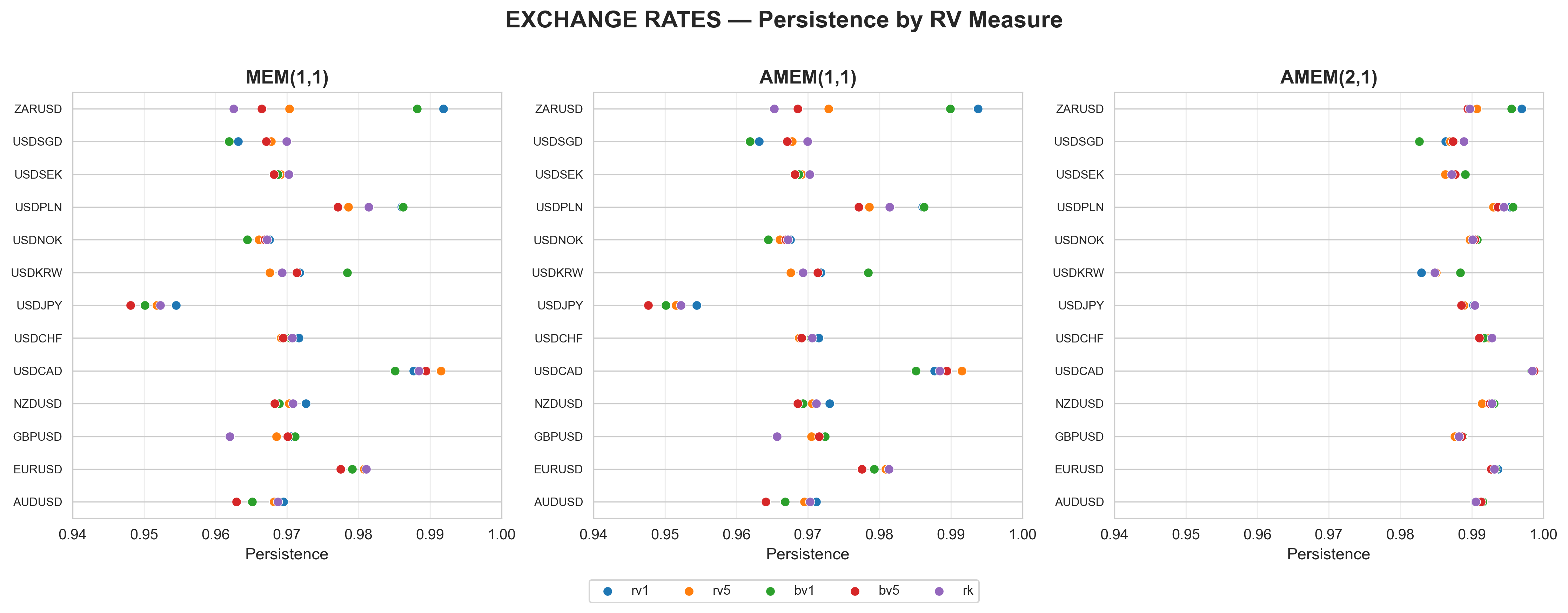}
            
    \caption{Robustness of total persistence across RV measures for echange rates. Persistence for MEM(1,1) $= \alpha_1 + \beta_1 $, for AMEM(1,1) $= \alpha_1 + \beta_1 +  \gamma_1/2$, AMEM(2,1) $= \alpha_1 + \alpha_2 +\beta_1 + \gamma_1/2$. Each dot represents the persistence estimate for one asset using a specific RV measure (\texttt{rv1}, \texttt{rv5}, \texttt{bv1}, \texttt{bv5}, \texttt{rk}), with colors distinguishing the five estimators.}
    \label{fig:forex_stability_persistence_all}
\end{figure}

\begin{figure}[htbp]
    \centering
    \includegraphics[width=\textwidth]{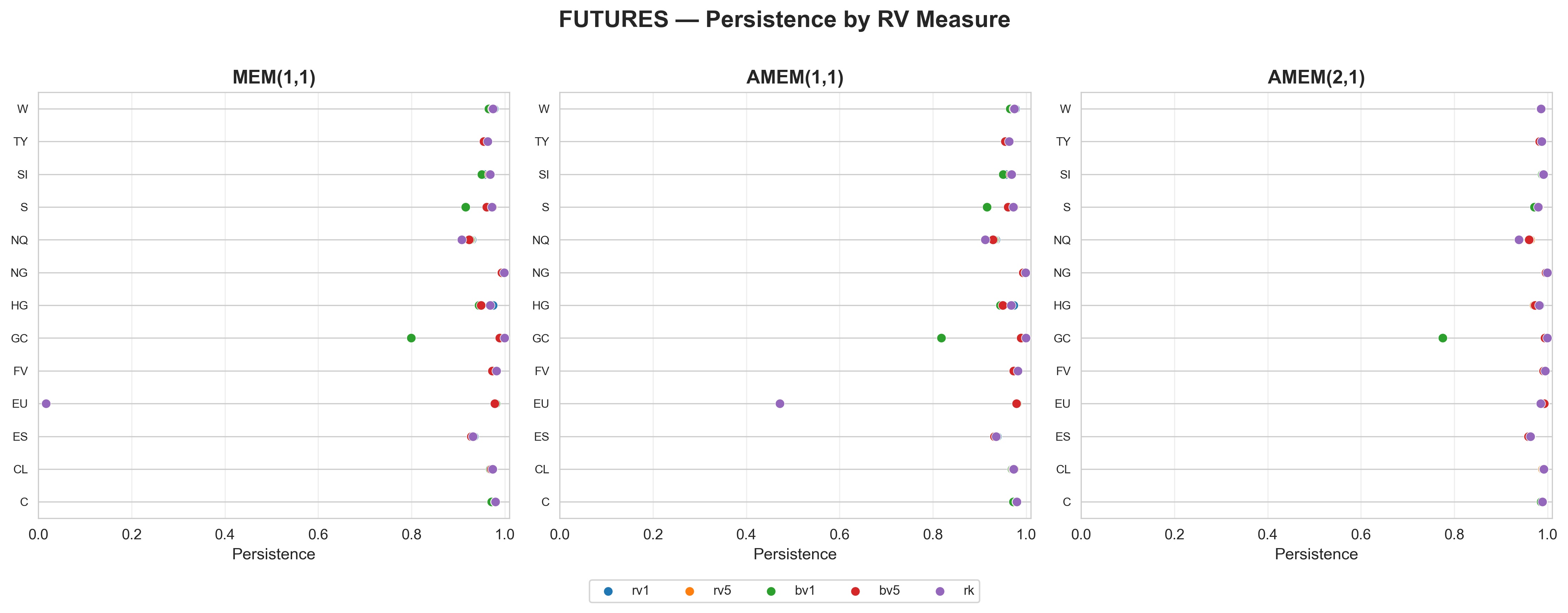}
            
    \caption{Robustness of total persistence across RV measures for futures contracts. Persistence for MEM(1,1) $= \alpha_1 + \beta_1 $, for AMEM(1,1) $= \alpha_1 + \beta_1 +  \gamma_1/2$, AMEM(2,1) $= \alpha_1 + \alpha_2 +\beta_1 + \gamma_1/2$. Each dot represents the persistence estimate for one asset using a specific RV measure (\texttt{rv1}, \texttt{rv5}, \texttt{bv1}, \texttt{bv5}, \texttt{rk}), with colors distinguishing the five estimators.}
    \label{fig:futures_stability_persistence_all}
\end{figure}

\end{document}